\documentclass[twoside,11pt]{article}
\usepackage{LAMB}
\usepackage{hyperref}

\begin{document}

\begin{title}
{\Large\bf Latent Association Mining for Binary Data\\
}
\end{title}
\author{Carson Mosso$^{* \, \S}$, Kelly Bodwin$^{* \, \natural}$, Suman Chakraborty$^{\dagger}$, Kai Zhang$^{\S}$, Andrew B.\ Nobel$^{\S}$\\
$*:$ These authors contributed equally to this work.\\
$\S:$ University of North Carolina at Chapel Hill\\
$\natural:$ California Polytechnic State University\\
$\dagger:$ University of Michigan at Ann Arbor}
\date{\today}
\maketitle

\begin{abstract}
	We consider the problem of identifying stable sets of mutually 
	associated features in moderate or high-dimensional binary data. 
	In this context we develop and investigate a method called 
	Latent Association Mining for Binary Data (LAMB).  The LAMB method 
	is based on a simple threshold model in which the observed 
	binary values represent a random thresholding of a latent 
	continuous vector that may have a complex association structure.  
	We consider a measure of latent association that quantifies 
	association in the latent continuous vector without bias due 
	to the random thresholding. The LAMB method uses an 
	iterative testing based search procedure to identify 
	stable sets of mutually associated features.  We compare the 
	LAMB method with several competing methods on 
	artificial binary-valued datasets and two real count-valued datasets. 
	The LAMB method detects meaningful associations in these datasets. 
	In the case of the count-valued datasets, associations
	detected by the LAMB method are based only on 
	information about whether the counts are zero or non-zero, 
	and is competitive with methods that have access to the full count data.
%%%	We consider the problem of identifying sets of mutually associated features
%%%	in moderate or high-dimensional binary data. 
%%%%%	Pearson correlation provides
%%%%%	useful information concerning the association between
%%%%%	features in many cases. For binary data, however, Pearson correlation
%%%%%	may not be appropriate. 
%%%	We develop and investigate a method called Latent Association
%%%	Mining for Binary Data (LAMB) %. The LAMB method 
%%%	that is based on a simple
%%%	threshold model: binary observations represent a random 
%%%	thresholding of a latent continuous vector that may have a complex
%%%	association structure. In particular, we consider 
%%%	a measure called latent association
%%%	designed to measure the association 
%%%	in the latent continuous vector without 
%%%	bias due to the random thresholding. The LAMB
%%%	method uses an iterative testing based search 
%%%	procedure to identify sets of mutually associated features.
%%%	We analyze the LAMB method on artificial 
%%%	datasets and two real datasets.
%%%%%	Results from the LAMB method are compared
%%%%%	to results from the Nonnegative Matrix Factorization
%%%%%	and Latent Dirichlet Allocation methods.
%%%	The LAMB method is shown to detect meaningful 
%%%	associations in the artificial and real datasets.
\end{abstract}

\providecommand{\keywords}[1]{\textbf{\textit{Keywords: }} #1}

\keywords{binary data, binary association, data mining, unsupervised learning, latent variable models}

\newpage

\section{Introduction}
\label{sec: intro}

%This sentence is too long in my opinion. changing it some.-CM
%A common problem in exploratory data analysis is to identify relational 
%structure among a moderate to large set of features based on a modest 
%set of samples for which measurements may, in some cases be missing 
%or restricted in their values.  

A common problem in exploratory data analysis is to 
identify relational structure among a set of features
based on a set of samples. It is often the case
in real datasets
that the set of features or the set of samples or both
are of moderate to large size.
In some cases measurements may
be missing or restricted in their values.
This problem, which has points of contact 
with market basket analysis, recommender systems, and unsupervised 
learning, falls under the broad umbrella of association mining.

%I think this paragraph is better in the related works section.-CM
%In some cases, association mining problems can be addressed by model- or
%dissimilarity-based clustering methods, or by search procedures, often exhaustive, that
%identify patterns of interest among sets of features.  Clustering methods typically partition the 
%available features into disjoint groups; search procedures can produce unwieldy outputs when
%many instances of the desired pattern are present; both can be prone to errors when
%samples exhibit substantial inhomogeneity.sample

%In this paper we introduce a statistical method for association mining that is focused 
%on finding sets of features with strong latent association in binary data.   
% latent association needs to be defined before it is used in a sentence like this,
% I don't think it's clear without more context. -CM
In this paper we introduce a statistical method for association mining that is focused 
on finding sets of features with strong positive association in binary-valued data. 
The method, Latent Association Mining for Binary Data (LAMB), is based on
a simple threshold model and association measure.
In our threshold model observed binary samples are obtained by 
randomly thresholding the entries of a continuous random vector. 
The continuous random vector and the random
thresholds are assumed to be independent, and neither are observed in this model.
%%The random thresholds are modeled by %truncated 
%%%AN: Replaced factorization distribution by model.  Is this the right terminology?
%%%Poisson factorization %Miheer's suggestion to not say this yet
%%distributions that rely on a single parameter for each sample and
%%a single parameter for each feature.
The model for the random thresholds is of limited complexity, but is flexible enough to capture 
heterogeneity in the frequency of 1s between samples and between features in the dataset.

Association between two binary features is measured indirectly
in our statistical framework:
it is not a function of the features alone, but of the features in conjunction with the 
unobserved thresholds. We call this measure latent association. 
In particular, the latent association between two features
is the expected value of their conditional correlation given the random thresholds.
%In practice, unobserved random thresholds are estimated and 
%then plugged-in to obtain sample estimates of latent association.

To identify sets of features having strong positive latent association,
we employ a novel iterative testing based search procedure.  
%
%%The third component is the use of an iterative hypothesis testing 
%%search procedure to identify sets of features that
%%are positively associated. 
%
%Like most association mining
%methods, the LAMB method's search procedure 
%searches the lattice of sets 
%of features to detect associated sets of features. 
Our search procedure is computationally efficient, so it can be applied 
to high-dimensional datasets.
In contrast to many conventional association mining methods,
the LAMB method is carried out in a 
statistical framework and based on 
hypothesis tests. The LAMB method's
statistical framework is able to
account for sample heterogeneity,
and the testing based search procedure moderates false discoveries in both the
real and artificial datasets discussed in this paper.

%%In the next section we give an example motivating our consideration of a model for binary data with
%%sample heterogeneity, and exhibiting the usefulness of latent association.
%%%Then in Section~\ref{subsec: notation} we briefly
%%%describe applications of the LAMB
%%%method to real datasets. Finally, i
%%Organization of the rest of the paper is discussed in Section~\ref{subsec: organization}.

\subsection{Motivating Example}
\label{subsec: motivating example}

%In order to motivate the LAMB method, we consider a toy market basket data set.
%
%Conventional association mining methods were 
%originally applied to 
%binary-valued market basket data.\footnote{Binary-valued 
%data was due to both barcode and data storage 
%technology at the time. The datasets 
%did not necessarily reflect purchases made all at once.
%(CITE AGARWAL and 2006 association mining survey)}

Binary-valued market basket data motivated a number of important association
mining methods in the data mining literature.
In market basket data, features correspond to items available for purchase and samples
correspond to transactions carried out by a buyer.  Binary measurements
indicate whether or not a particular buyer purchased a particular item.\footnote{Binary-valued
data was due to barcode technology at the time and could
include purchases made at different times \citep{agrawalMiningAssociationRules-1993}.}
Market basket data often exhibits sample heterogeneity, reflecting the fact that a
buyer's decision to purchase an item can depend on the buyer's valuation of that item
relative to other items, as well as the buyer's needs, 
financial resources, time, et cetera.

Figure~\ref{tab: toy dataset} contains a toy market basket
dataset with 12 buyers and 14 items.
The $(i,j)$ entry of the table is 1 if item $i$ was purchased by buyer $j$ and 0 otherwise.
%A 1 means that buyer $j$ purchased item $i$,
%and a 0 means that buyer $j$ did not purchase item $i$.
Note the heterogeneity among the samples:
samples 1 through 5 correspond to high volume buyers, samples 6 and 7 to medium volume buyers, 
and samples 8 through 12 to low volume buyers.
Consider the item sets \{1, 2\} and \{3, 4\}, which
show identical behavior up to a permutation of the samples.  
The correlation between the two items in each set is the same ($r_{12} = r_{34} = 0.667$), and 
the items in each set are also equally far apart in both 
$\ell_1$ and $\ell_2$ distance ($d_{12}^1 = d_{34}^1 = 2$ and 
$d^2_{12} = d^2_{34} = \sqrt{2}$, respectively). 
Differences emerge, however, when one considers these pairs in the context of buyer behavior. 
The association between items 3 and 4 (and other pairs of items 3 through 9) is driven by
high volume buyers who are purchasing the majority of items for sale.  By contrast, the association between 
items 1 and 2 is driven by low volume buyers who purchase relatively few items, but buy items
1 and 2 together.  It is reasonable, then, to treat the reported association measures of these two item
sets differently.

%Nevertheless, it is not clear that there is any intrinsic association between items 3 and 4, 
%aside from an overall pattern in buyer behavior
%To see this, note that buyers 1-5 bought most available items, 
%buyers 8-12 bought very few items,
%and buyers 6 and 7 bought an intermediate number of items.  
%
%However, the association between items 3 and 4 (and other pairs of items 3 through 9) reflects an
%overall pattern in buyer behavior:
%most of these items were purchased by high and medium volume buyers, none were purchased by
%low volume buyers.
%%rather than a true underlying association between these items.  
%%Items 1 and 2, on the other hand, show similar buying patterns that can {\it not} be explained by buyer differences.  
%%%%% [I think this is a false sentence and so am removing it. The fact that only low volume 
%%%%% consumers buy these two items in fact means the pattern is potentially due to buyer difference!]
%By contrast, low volume buyers, who buy relatively few items, tend to purchase items 1 and 2 together, an 
%indicator of an underlying association between these features that is of interest in market basket analysis.
%

\begin{figure}
\centering
\small
Buyers\\
\vspace{.025cm}
\begin{tabular}{ccccccccccccc}
  \hline
 & 1 & 2 & 3 & 4 & 5 & 6 & 7 & 8 & 9 & 10 & 11 & 12\\
  \hline
 Item 1 &   0 &   0 &   0 &   0 &   0 &   1 &   0 &   1 &   1 &   1 &   1 &   1 \\
  Item  2 &   0 &   0 &   0 &   0 &   0 &   0 &   1 &   1 &   1 &   1 &   1 &   1 \\
  Item  3 &   1 &   1 &   1 &   1 &   1 &   1 &   0 &   0 &   0 &   0 &   0 &   0 \\
  Item  4 &   1 &   1 &   1 &   1 &   1 &   0 &   1 &   0 &   0 &   0 &   0 &   0 \\
  Item  5 &   1 &   1 &   1 &   1 &   0 &   1 &   0 &   0 &   0 &   0 &   0 &   0 \\
  Item  6 &   1 &   1 &   1 &   0 &   1 &   0 &   1 &   0 &   0 &   0 &   0 &   0 \\
 Item   7 &   1 &   1 &   0 &   1 &   1 &   1 &   0 &   0 &   0 &   0 &   0 &   0 \\
 Item   8 &   1 &   0 &   1 &   1 &   1 &   0 &   1 &   0 &   0 &   0 &   0 &   0 \\
  Item  9 &   0 &   1 &   1 &   1 &   1 &   1 &   0 &   0 &   0 &   0 &   0 &   0 \\
  Item  10 &   1 &   1 &   1 &   1 &   1 &   0 &   1 &   0 &   0 &   0 &   0 &   1 \\
  Item  11 &   1 &   1 &   1 &   1 &   1 &   1 &   0 &   0 &   0 &   0 &   1 &   0 \\
  Item  12 &   1 &   1 &   1 &   1 &   1 &   0 &   1 &   0 &   0 &   1 &   0 &   0 \\
   Item 13 &   1 &   1 &   1 &   1 &   1 &   1 &   0 &   0 &   1 &   0 &   0 &   0 \\
   Item 14 &   1 &   1 &   1 &   1 &   1 &   0 &   1 &   1 &   0 &   0 &   0 &   0 \\
   \hline
\end{tabular}
\caption{Toy market basket dataset.}
\label{tab: toy dataset}
\end{figure}

\begin{figure}
\begin{subfigure}[b]{0.32\textwidth}
\centering
\includegraphics[width = .9\textwidth]{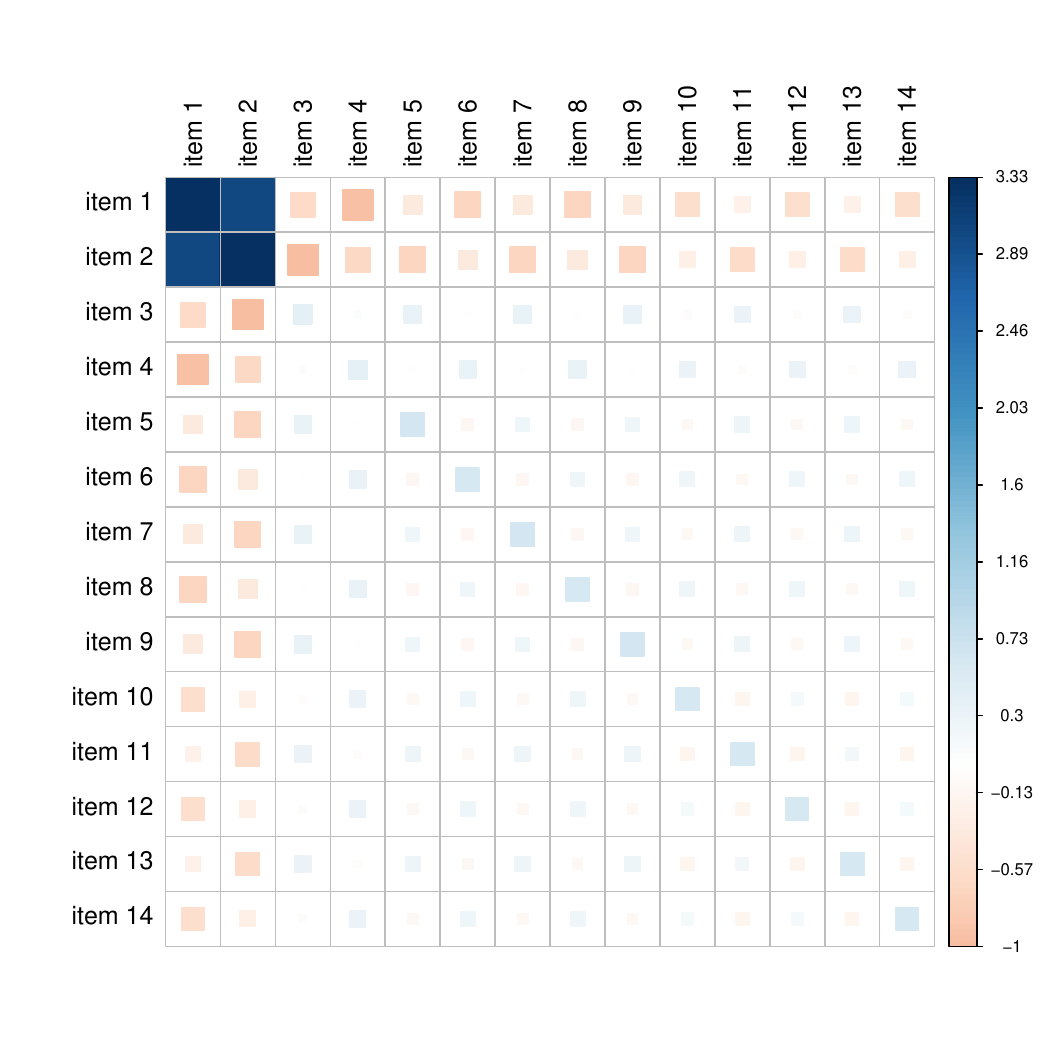}
\caption{Latent association in the toy market basket dataset.}
\end{subfigure}
~
\begin{subfigure}[b]{0.32\textwidth}
\centering
\includegraphics[width = .9\textwidth]{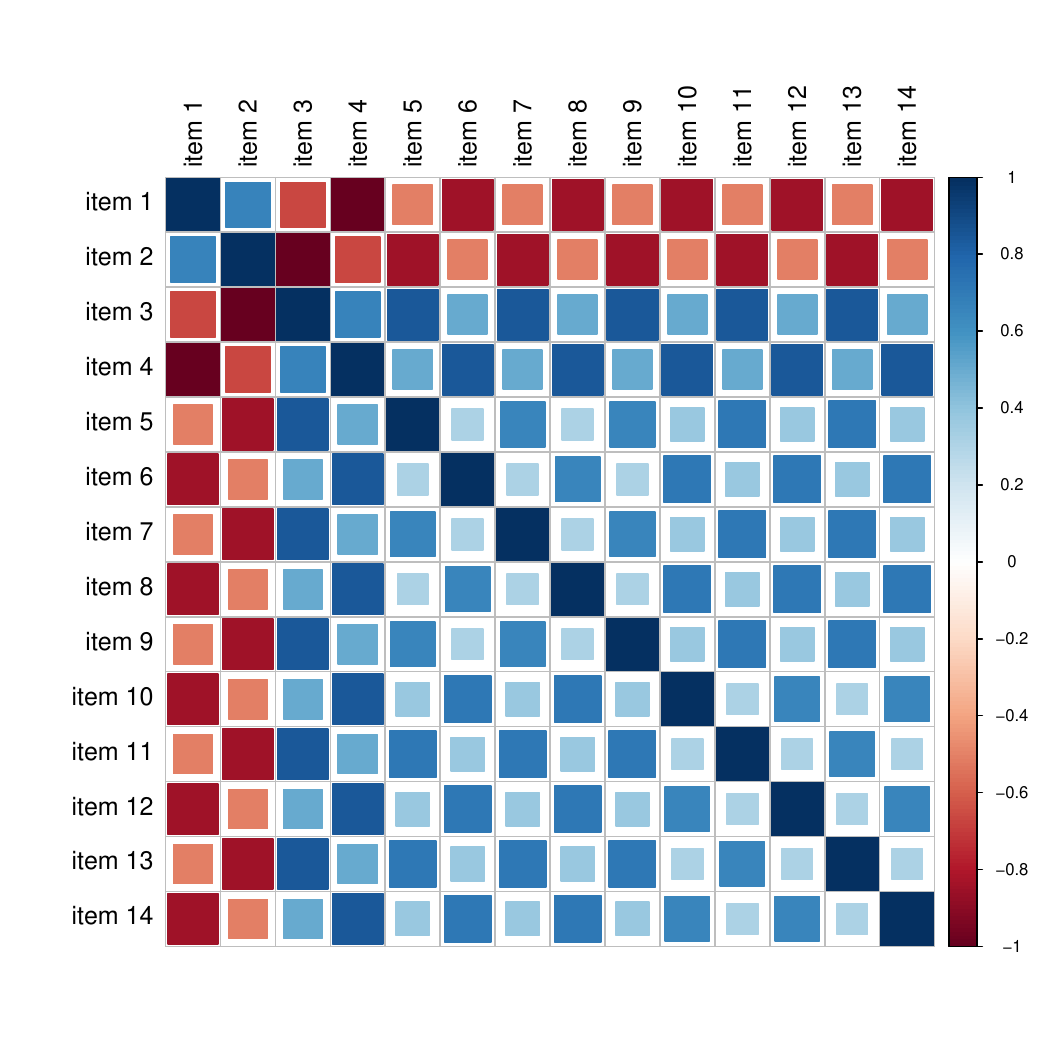}
\caption{Correlation in the toy market basket dataset.}
\end{subfigure}
~
\begin{subfigure}[b]{0.32\textwidth}
\centering
\includegraphics[width = .9\textwidth]{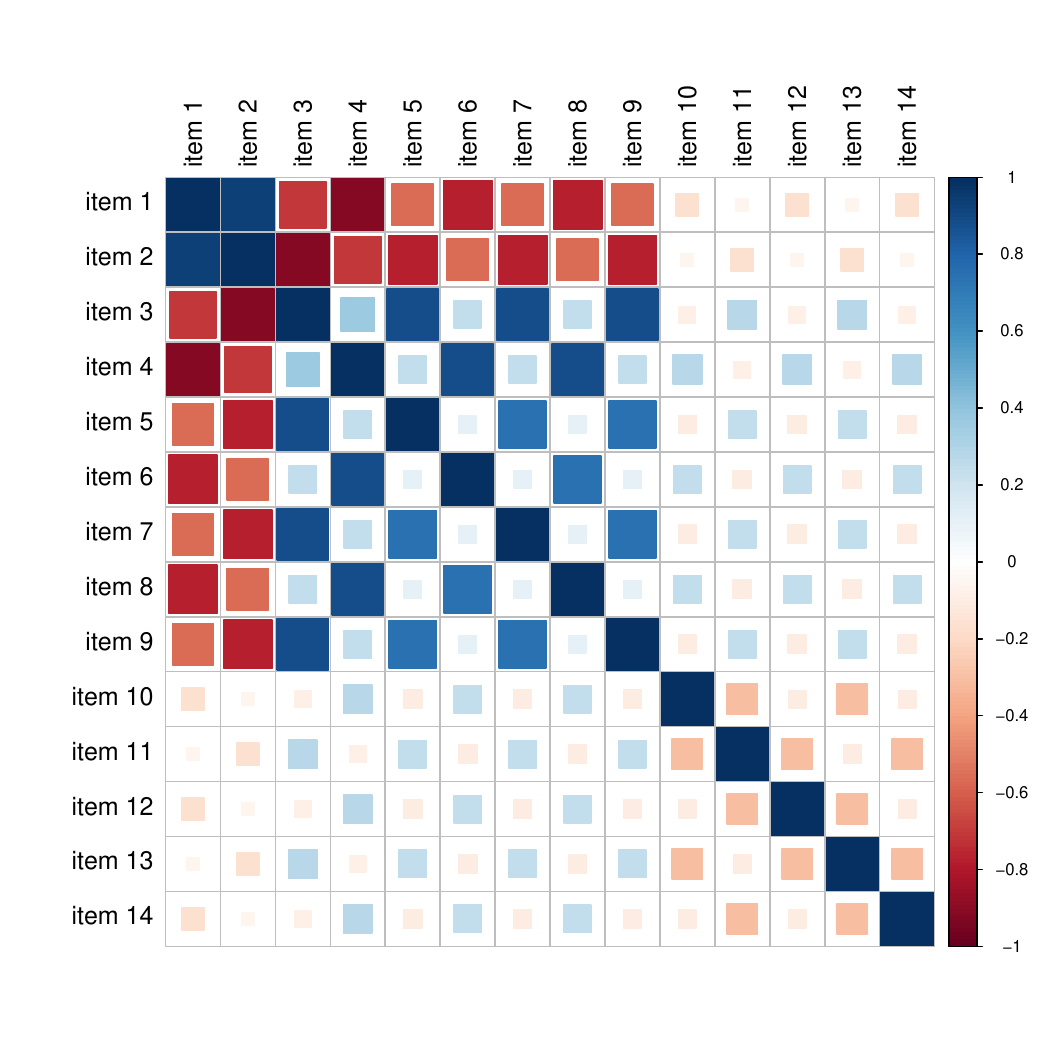}
\caption{Correlation in the ``buyer normalized'' dataset.}
\end{subfigure}

\caption{Item association heat maps.}
%%Figures (a) and (b) correspond to latent association and correlation,
%%respectively, in the original toy dataset.
%%Figure (c) corresponds to correlation in the ``buyer normalized'' dataset.}
\label{fig: app toy coh}
\label{fig: toy coh}
\end{figure}

Figure~\ref{fig: toy coh} illustrates the difference 
between latent association
(calculated using the methods of Section~\ref{sec: model}) 
and correlation
for items in the toy market basket data.
Note that in Figure \ref{fig: toy coh}(a) the item set \{1, 2\} has strong latent association, 
but other associations are attenuated since they are not distinguishable from the pattern among high volume buyers.
The item set \{1, 2\} is the only set of items that the LAMB method
declares to be associated.
%when it is applied to the toy market basket dataset. 
This result is in contrast with those of conventional methods.
For example, the item sets 
$\{1, 2 \}$, $\{3, 5, 7, 9, 11, 13\}$,
and $\{4, 6, 8, 10, 12, 14\}$ are often 
considered associated by other methods.
See Appendix~\ref{app: toy data results} for details.

A simple way of addressing the effects of high volume buyers 
is to divide each sample by its Hamming weight, which is the total number of purchased items.
We will refer to data transformed in this way as ``buyer normalized''.  Note that buyer normalized
data will not, in general, be binary, and that the sign of the correlation between items can change
after this normalization.
Figure~\ref{fig: app toy coh}(c) illustrates the sample correlation matrix of the buyer normalized dataset.
We note that there are still strong correlations between items 3 through 9, and that the correlations of
of items 10 through 14 have undergone a sign change.

As mentioned in the introduction, the 
LAMB method handles heterogeneity among both samples and features 
(buyers and items in this context) 
through a statistical framework.
%and meaningful hypothesis tests. 
The LAMB method's statistical framework acts as a de facto normalization
of the observed binary samples. 
The hypothesis tests used in the LAMB
method's search procedure help to account for 
sampling artifacts, as opposed to population quantities.
As a result of this, the LAMB method
is less prone to making spurious claims of association
among features,
e.g., the association between items 3 and 4 
(and other pairs of items 3 through 9)
in the toy market basket data.

%%In conventional association mining methods,
%%sampling artifacts may lead to spurious claims of association among features,

%\begin{figure}
%\begin{subfigure}[b]{0.32\textwidth}
%\centering
%\includegraphics[width = .9\textwidth]{./img/toy_cor_plot.pdf}\\
%\hspace{1em}
%\caption{Sample correlation}
%\end{subfigure}
%~
%\begin{subfigure}[b]{0.32\textwidth}
%\centering
%\includegraphics[width = .9\textwidth]{./img/toy_coh_plot.pdf}\\
%\hspace{1em}
%\caption{Estimated latent association}
%\end{subfigure}
%
%\caption{Item association matrix comparison for toy market basket dataset.
%\red{REDO (b) using new LAMB}}
%\label{fig: toy coh}
%\end{figure}

\subsection{Related Work}
\label{subsec: rel work}

%%\blue{
%%In some cases, association mining problems can be addressed by model- or
%%dissimilarity-based clustering methods, or by search procedures, often exhaustive, that
%%identify patterns of interest among sets of features.  Clustering methods typically partition the 
%%available features into disjoint groups; search procedures can produce unwieldy outputs when
%%many instances of the desired pattern are present; both can be prone to errors when
%%samples exhibit substantial inhomogeneity.}

\subsubsection{Itemset Mining Methods}

The LAMB method is 
related to Itemset Mining (IM) in the data mining literature. 
%%a class of methods
%%originally developed for market basket datasets to discover sets of items that are typically
%%bought together \citep{agrawalMiningAssociationRules-1993, apriori-1996}.
%%IM methods have also been applied to text and census datasets \citep{brinGeneralizingAssociationRules-1997}.
As noted earlier, IM methods
were originally motivated by market basket datasets that were
often binary-valued \citep{agrawalMiningAssociationRules-1993, apriori-1996}.
In general, most IM methods can be applied
to undirected bipartite graphs \citep{zakiScalableAssociationMining-2000, 
nobelBinarySignificanceOfBlockStructures-2006}.\footnote{There 
is a bijection between binary data matrices 
and undirected bipartite graphs.}
See \citet{goethalsFPMsurvey-2003,
ceglarAssociationMining-2006, hanFPMsurvey-2007, aggarwalFPM-2014}, 
and \citet{SurveyOfItemsetMining-2017}
for recent surveys of IM methods.

%%Both the LAMB and IM methods discover associated features using 
%%\red{redundant:} search procedures.
The LAMB method uses a novel iterative testing 
based search procedure that is
similar to the IM method of \citet{brinGeneralizingAssociationRules-1997},
which explicitly tests for dependence between features. 
In most cases, however, IM methods use
exhaustive search procedures to discover sets of features
that co-occur at a given frequency threshold
\citep{brinDynamicItemsetCounting-1997, zakiScalableAssociationMining-2000}.
%Most IM method search procedures are based on the frequency 
%of an observed set of features.

%Many IM methods differ from the LAMB method 
%in that they (implicitly) assume homogeneous samples. 
As noted by \citet{yangErrorTolerantItemsets-2001, nobelMiningApproximateItemsetsNoisyData-2005,
nobelBinarySignificanceOfBlockStructures-2006,
alvesFPMGeneAnalysis-2010} and
\citet{naulaertsPrimerFrequentItemsetMiningBioinformatics-2015},
the results of many IM methods
are not always robust to noisy datasets or datasets with sample heterogeneity.
IM methods may struggle with discovering associations in
sparse datasets or datasets with a moderate to large amount of features,
due to modeling assumptions or exhaustive search procedures or both
\citep{brinGeneralizingAssociationRules-1997, moensFIMBigData-2013, tattiRobustItemsets-2014}.

\subsubsection{Iterative Hypothesis Testing}

The iterative testing framework of the LAMB method's search procedure 
has been employed in community detection \citep{wilsonESSC-2014},
differential correlation mining \citep{bodwinDCM-2018}, 
and biclustering of multi-view data \citep{miheerBimodules-2020}.
\citet{liuFalsePositivesAssociationMining-2011} discuss the 
importance of considering false positives
in IM methods, and they propose multiple 
approaches to account for this.
The LAMB method's iterative testing based search procedure 
moderates false discoveries
by using a multiple testing procedure per iteration.
In \citet{boltonIterativeTesting-2003} a statistical model for generating 
data is iteratively tested, and this is called an 
iterative hypothesis testing strategy. 
%%, however this is 
%%different than the iterative hypothesis testing done in the LAMB method's
%%search procedure. 
In contrast to the iterative hypothesis testing strategy,
in the LAMB method's search procedure
the statistical model is fixed and a test statistic is calculated per iteration.

\subsubsection{Clustering Methods}

In some cases, association mining problems can be addressed by model- or
dissimilarity-based clustering methods that
identify patterns of interest among sets of features.  
See \citet{kriegelClusterReview-2009}, \citet{everittClustAnalysis-2011},
and \citet{hastieESL-2017} for reviews of clustering methods.
A collection of binary distance and dissimilarity measures
is provided in \citet{choiBinarySurvey-2010},
and clustering binary data is surveyed in
\citet{liGeneralModelBinaryClustering-2005}
and
\citet{liUnifiedViewClusteringBinaryData-2006}.
Clustering methods typically partition the 
available features into disjoint groups, and 
these methods can be prone to spurious results when
samples exhibit substantial inhomogeneity.

\subsubsection{Topic Model Methods}

Topic Model (TM) methods postulate a
latent variable model that generates observed data.\footnote{TM
methods are usually applied to count-valued data and were
originally motivated by text data.}
In contrast to the LAMB method, the latent variables in
TM methods are discrete-valued and each value 
corresponds to a generative probability distribution (called a topic)
over the set of features.
TM methods are related to IM and clustering methods. 
Whereas IM methods use a search procedure based on counting the frequency 
of occurrences of a set of features, 
TM methods usually estimate probability distributions.
The latent variables and estimated probabilities %, and the corresponding probability distributions,
in TM methods allow for a soft clustering of features or samples.
%%Topic Models are often applied to count-valued data
%%but can also be applied to binary-valued data. 

Commonly used Topic Model methods include Nonnegative Matrix Factorization (NMF)
\citep{leeNMF-1999, leeNMFAlgorithm-2000} and Latent Dirichlet Allocation (LDA) \citep{bleiLDA-2003}.
%%There are also many extensions of the NMF and LDA methods.
See \citet{steyversTopicModels-2007} and \citet{bleiTopicModels-2009}
for surveys of Topic Model methods and extensions of the LDA method.
Recent surveys of the NMF method include \citet{huangNMFSurvey-2012}
and \citet{wangNMFReview-2013}.
Work on binary matrix factorization methods and algorithms
includes \citet{zhangBinaryMatrixFactorization-2007,
kumarFasterAlgorithmsBinaryMatrixFactorization-2019},
and \citet{luWeightedRankOneBMF-2020}.

%%In contrast to most IM methods, the NMF and LDA methods can be used
%%to discover associations among both features and samples.\footnote{In 
%%some situations the LAMB method can be successfully 
%%employed to discover associations among both features
%%and samples. However, this is not automatically done, 
%%as it is in the NMF and LDA methods, because the data matrix 
%%must be transposed and the LAMB method rerun.
%%Transposing a data matrix reverses the roles of features and samples,
%%so this does not always make sense to do in practice.
%%However, for the real datasets discussed in this paper
%%it makes sense to apply the LAMB method to both the original 
%%and the transposed data matrix.}

%%There exist binary matrix constraint versions of the NMF method.
%%Work on binary matrix factorization methods and algorithms
%%includes \citet{zhangBinaryMatrixFactorization-2007,
%%kumarFasterAlgorithmsBinaryMatrixFactorization-2019},
%%and \citet{luWeightedRankOneBMF-2020}.
%%In this paper we do not directly compare the LAMB method
%%to binary matrix constraint versions of NMF.
%%One reason is because the NMF method is 
%%successful at discovering associated sets of features in binary-valued 
%%data matrices in our simulation study (see Section~\ref{sec: sim study}).
%%Another reason is that the LAMB method is closer to 
%%the NMF method in theory,
%%since the latent vectors in our threshold model are continuous.
%There is also a lack of binary constraint NMF package support in R.

\subsubsection{Bernoulli Mixture Models}

Another class of methods that
can be used for association mining includes the
Bernoulli Mixture Model (BMM) and its various extensions 
\citep{govaertBlockClusteringBernoulliMixtureModels-2008,
saeedMachineLearningUsingBernoulliMixtures-2013,
tangModelBasedClusteringBinaryData-2015,
yamamotoClusteringMultivariateBinary-2015,
yeClusteringBinaryDataHierarchicalBernoulli-2018}.
A BMM is a latent variable model and the association between 
features is usually modeled through hyperparameters. 
The Expectation-Maximization (EM) algorithm and its extensions are often used for 
estimating the hyperparameters to provide clustering information
\citep{bishopML-2006}.
Compared to BMM methods, 
the LAMB method assumes a more general threshold model. 
Furthermore, to improve the computational efficiency, instead of 
applying an EM algorithm globally to all of the hyperparameters, 
the LAMB method employs an efficient search procedure
to discover associations between features.
%%iterative testing 
%%based search procedure to discover sets of associated features. 

%%This search procedure 
%%framework has been employed in community detection \citep{wilsonESSC-2014},
%%differential correlation mining \citep{bodwinDCM-2018}, 
%%and biclustering \citep{miheerBimodules-2020}.

%%The LAMB method includes a novel iterative hypothesis
%%testing based search procedure. This search procedure
%%framework has been employed in community detection \citet{wilsonESSC-2014}, 
%%differential correlation mining \citet{bodwinDCM-2018}, and biclustering (CITE Miheer).

%\citet{zhangBET}
%
%\citet{everittClustAnalysis}
%
%\citet{szekelyDistCorr}

\subsection{Organization}
\label{subsec: organization}

The remainder of the paper is organized as follows.
The threshold model and latent association measure for
the LAMB method are formalized in Section~\ref{sec: model}.
In Section~\ref{sec: search procedure} we describe the 
LAMB method and outline the
iterative testing based search procedure.
%including a high-level discussion of the Central Limit Theorem
%that is discussed in detail in Section~\ref{sec: CLT}.
We review a simulation study in Section~\ref{sec: sim study}
that demonstrates how the LAMB, NMF, and LDA methods
perform under controlled conditions.
The LAMB, NMF, and LDA methods are also
applied to two real (count-valued) datasets
in Section~\ref{sec: applications}.
%%Sections~\ref{sec: CLT} and \ref{sec:params} present 
%%supporting theoretical results. 
%%Section~\ref{sec: CLT} goes through the central limit
%%theorem that is used in the LAMB method's search
%%procedure. Section~\ref{sec:params} goes through
%%both the theory and implementation for
%%estimating the random thresholds 
%%in the LAMB method's threshold model.
Section~\ref{sec: summary} summarizes the LAMB method
and the results in this paper. Supporting theory, details,
and results are included in the appendices 
and is referenced throughout the paper.

%%\section{Related Work}
%%\label{sec: related work}
%%\input{./tex/rel_work.tex}

\section{Threshold Model and Latent Association}
\label{sec: model}

In this section we describe the statistical model %framework 
and the latent association measure 
used in the LAMB method. 
We explore and prove some basic properties of 
the threshold model and latent association 
in Appendix~\ref{app: supporting theory}.

\subsection{Threshold Model}
\label{subsec: threshold model}

The LAMB method is based on a simple threshold model in which
an observed binary-valued vector indicates whether or not
the components of a latent continuous vector lie above or below the corresponding components of a 
(transformed) threshold vector.\footnote{Note that $F_i^{-1}(\theta_i)$ 
in Equation~\eqref{eq:thresh} is simply the $\theta_i$ percentile of $F_i$, and
that $X_i$ records whether the realized value of $V_i$ is less than this number.}
We assume that the continuous and threshold vectors are random and independent of one another, though the components
of each may be dependent. 
The formal definition of the model follows.

\begin{definition}[\textbf{Threshold Model}]
\label{def:mod}
Let $\bm{\alpha} = (\alpha_1, \ldots, \alpha_d)^T$ be a fixed vector of positive constants
and let $\tau$ be a positive-valued random variable with density $\pi$.  Define the $d$-dimensional 
random threshold vector $\bm{\theta} = (\theta_1, \ldots, \theta_d)^T \in (0,1)^d$ by 
\begin{align}
	\label{eq:theta}
	\theta_i 
		&:=  1 - \exp(- \alpha_i \tau) \ \ \ \ i \in \{1, \ldots, d\}.
\end{align}
Let $\bm{V}  = (V_1, \ldots, V_d)^T \in \mathbb{R}^d$ be a latent random vector
with continuous distribution function $F$ that is independent of $\bm{\theta}$. 
We assume that the threshold vector $\bm{\theta}$ and continuous vector $\bm{V}$ are unobserved.  
The observable random vector 
$\bm{X} = (X_1, \ldots, X_d)^T \in \{0, 1\}^d$ is defined by
\begin{align}
	\label{eq:thresh}
	X_i 
		&:= \mathbb{I}\left( V_i \, \leq \, F_i^{-1}(\theta_i) \right)
			\ \ \ \ i \in \{1, \ldots, d\}
\end{align}
where $F_i$ is the marginal distribution function of $V_i$.
In what follows, we will describe the model above as the {\em threshold model} for $\bm{X}$ 
generated by $(\bm{\theta}, \bm{V})$, and use the shorthand
$\bm{X} = \mathbb{I}\left[ \bm{V} \leq \bm{F}^{-1}( \bm{\theta}) \right]$.
\end{definition}

As the model for the threshold
$\bm{\theta}$ is similar to the Poisson factorization approach of 
\citet{gopalanNonparametricPF-2014, gopalanScalablePF-2015}
and \citet{huTruncatedPoissonFactorization-2015},
we refer to it as a truncated Poisson factorization model (TPFM). 
For a given sample, the value of $\theta_i$ depends only on 
%the popularity of feature $i$ and the overall resources allocated to the sample through 
the product $\alpha_i \tau$.
Randomness in $\bm{\theta}$ arises from the (shared) random variable $\tau$.
In the case of market basket data, the fixed parameters $\bm{\alpha}$ 
account for the intrinsic popularity or utility of different features (items),
and the resources or budget allocated to each sample (buyer)
is captured by the value of $\tau$.
In this way, the threshold model is able
to capture heterogeneity among the features and the samples without the need for normalization
or preprocessing steps that might not be suitable for binary data.  

%%In practice, we estimate the popularity vector $\bm{\alpha}$ and the resources $\tau$ of different samples 
%%from the available data. See Section~\ref{sec:tildes} for details about estimating 
%%$\bm{\alpha}$ and $\tau$ for different samples.  
%%Identifiability of the TPFM is discussed in Proposition~\ref{prop: identifiable}.

\subsection{Latent Association}
\label{subsec: latent coh}

Under the threshold model dependence among the components of the binary vector 
$\bm{X}$ arises from two sources: dependence among the components of the continuous
vector $\bm{V}$ and dependence among the components of the threshold vector
$\bm{\theta}$.  Dependence among the components of $\bm{\theta}$ is due to the
shared value of $\tau$ in the truncated Poisson factorization model.
In our analysis, we regard dependence in $\bm{X}$ arising from 
the dependence in $\bm{\theta}$ as a nuisance parameter;
we seek to capture dependence in $\bm{X}$ induced by 
dependence in the continuous vector $\bm{V}$. 
We make use of a simple measure called latent association
to quantify the dependence in $\bm{X}$ that arises from $\bm{V}$.

\begin{definition}[\textbf{Latent Association}]
\label{def:lat_cor}
Let the binary random vector $\bm{X} = \mathbb{I}\left[ \bm{V} \leq \bm{F}^{-1}( \bm{\theta}) \right]$
follow the threshold model of Definition~\ref{def:mod}.
The \emph{latent association} between $X_i$ and $X_k$ is defined by
\begin{align}
\label{eq:lat_cor}
\psi(i, k)
	&:= \Ex \left( \frac{\left( X_i - \theta_i \right) \left( X_k - \theta_k \right) }{ \sqrt{ \theta_i (1-\theta_i) \, \theta_k (1-\theta_k) } } \right) \,.
\end{align}
Here, and in what follows, all expectations are taken with respect to the joint distribution of $(\bm{\theta}, \bm{X})$ 
inherited from the distribution of $(\bm{\theta}, \bm{V})$.
\end{definition}

Conditional on the threshold vector $\bm{\theta}$, the components $X_1,\ldots, X_d$ 
of $\bm{X}$ are Bernoulli random variables, with $X_i \given \bm{\theta} \sim \mbox{Bern}(\theta_i)$. 
The variables $X_i \given \bm{\theta}$ will exhibit
dependence arising from dependence among the 
components of $\bm{V}$.  
It is easy to see that 
\begin{align*}
	\psi(i, k) = \Ex [ \rho(X_i,X_k \given \bm{\theta}) ],
\end{align*}
where $\rho(X_i,X_k \given \bm{\theta})$ denotes the
conditional correlation of $X_i$ and $X_k$ given $\bm{\theta}$.
%which arises from dependence between the latent 
%variables $V_i$ and $V_k$ with fixed thresholds $\theta_i$ and $\theta_k$.   
In particular, $\psi(i, k) = 0$ if $V_i$ and $V_k$ are independent, regardless of the distribution of $\bm{\theta}$.

In Section~\ref{sec: intro} we provided an 
illustration of the importance of measuring latent association 
rather than correlation when sample
heterogeneity is present in a binary-valued dataset. %on a toy market basket dataset.  
Now we give an explicit example in which $\bm{X}$ has standard correlation
induced by $\bm{\theta}$, despite independence in $\bm{V}$.

\begin{example}
\label{ex1}
	Let the continuous vector $\bm{V} \sim \mathcal{N}_d(\bm{0}, \mathbf{I})$, the threshold $\bm{\theta}$ be
	such that
	\begin{align}
		\theta_1 = \cdots = \theta_d \, = \, \begin{cases}
      		\epsilon & \text{ with probability } \, \frac{1}{2}, \\
    		1 - \epsilon & \text{ with probability } \, \frac{1}{2},
  	 \end{cases}
	\end{align}
	for some fixed $\epsilon \in\left(0, \frac{1}{2}\right)$,
	and $\bm{X} = \mathbb{I}\left[ \bm{V} \leq \bm{F}^{-1}( \bm{\theta}) \right]$. 
	Then it is easy to see that, for $i \neq k$,
	\begin{align*}
		\mathbb{E}(X_i X_k) - \mathbb{E}(X_i) \mathbb{E}(X_k)
			&= %\left( \frac{\epsilon^2}{2} + \frac{ (1-\epsilon)^2 }{2} \right) - \frac{1}{4} =
				 \frac{1}{4} - \epsilon (1 - \epsilon) \, .
	\end{align*}
	Thus, the correlation $\rho(X_i, X_k)$ is positive, and tends to one as $\epsilon$ tends to zero.
	Dependence arises from simultaneously thresholding the variables $V_i$ and $V_k$ at a very high or 
	a very low percentile. Consequently, $X_i$ and $X_k$ are not constant, but 
	are equal with high probability for every distinct pair $(i, k)$.  
	The dependence between the components of $\bm{X}$ arises from the threshold vector
	$\bm{\theta}$, and does not reflect dependence between the components of $\bm{V}$, which are 
	independent.  In contrast to $\rho(X_i, X_k)$, 
	the latent association $\psi(i, k)$ is equal to 0 for any distinct pair $(i,k)$,
	which accurately reflects the lack of dependence between the components of $\bm{V}$.
\end{example}

%%%\red{Maybe the following should be moved up?}
%%%We also remark here that one common measure of association between binary variables $X_i$ and $X_k$ is the odds ratio
%%%\[
%%%{\prob{ X_i = 1,  \, X_k =  1 } \prob{ X_i = 0,  \, X_k =  0 } \over \prob{ X_i = 0,  \, X_k =  1 } \prob{ X_i = 1,  \, X_k =  0 }}.
%%%\] 
%%%However, we decide to focus on latent association, 
%%%since the Threshold Model of Definition~\ref{def:mod} 
%%%is reasonable for many practical problems.
%%%Furthermore, latent association incorporates the 
%%%generative model for handling heterogeneity,
%%%and is more related to our goal of recovering the dependence structure in $\bm{V}$ 
%%%while mitigating the effects from the dependence structure in $\bm{\theta}$.
%%%

\section{Latent Association Mining in Binary Data (LAMB)}
\label{sec: search procedure}
In this section we present the details of how the LAMB method
%In particular, the iterative hypothesis testing search procedure 
detects sets of mutually associated features in binary data.  
%%In Appendix~\ref{app: additional lamb method details}
%%we give additional details that are helpful
%%for using the LAMB method.
%The search procedure used in the LAMB method
%is based on iterative hypothesis testing.
%Given a binary-valued data matrix 
%$\mathbb{X} = [\bm{X}_{\cdot 1}, \ldots, \bm{X}_{\cdot n}]\in \{0,1\}^{d\times n}$, 
%the goal of the iterative testing procedure is to
%identify sets of features with positive association.\footnote{Here 
%association is short for average pairwise latent association.}
%%These discovered sets are subsets of $[d]$ 
%%that represent an estimate of intrinsic association among features,
%%mitigated against heterogeneity among samples.
%%%We begin by specifying the target sets, called coherent sets, 
%%%that the iterative testing algorithm is designed to identify.
%%%We then briefly discuss the details of the multiple testing update process, 
%%%including the statement of a central limit theorem that justifies our hypothesis testing approach.
%In what follows we assume binary datasets 
%arise as independent replicates 
%%random samples 
%from a fixed, but unknown, threshold model 
%$\bm{X} = \mathbbm{1}\left[ \bm{V} \leq \bm{F}^{-1}(\bm{\theta}) \right]$.  
%
%In Section~\ref{subsec: coh sets} we introduce the targets
%of the LAMB method's search procedure:
%associated sets of features under the threshold model.
%The LAMB method's search procedure
%involves hypothesis testing that we formulate
%in Section~\ref{sec:test}. Finally, in Section~\ref{sec:lamb}
%we describe the LAMB method's iterative hypothesis
%testing search procedure in detail.

\subsection{Coherent Sets}
\label{subsec: coh sets}

Definition~\ref{def:lat_cor} provides a measure 
of association between two features under the threshold model 
of Definition~\ref{def:mod}.  
%%This statistic is supposed to approximately capture the 
%%dependence among the features of interest, represented through $\bm{V}$,
%%while mitigating the dependence among the observations, 
%%represented through $\bm{\theta}$.
The goal of the LAMB method, however, 
is to identify sets, rather than just pairs, 
of associated features. %that we call coherent sets.  
To this end we state the following two definitions.

%%\blue{
%%[I made a change to the definition, so that $\psi(j,j)$ 
%%can occur if $j\in A$. Then we make a slight change to 
%%the definition of coherent sets, which is the simplest 
%%thing to do to fix our issue of $j\in A$ in the CLT 
%%section. Recall that there was inconsistencies in this paper 
%%arising from sometimes summing over $A\setminus\{j\}$ 
%%and other times summing over $A$.]}
%%

\begin{definition}[\textbf{Average Latent Association}]
\label{def: set lat asc}
Given $i \in [d]$ and $A \subseteq [d]$, let
\begin{align}
	\label{eq:avg_coh}
	\psi(i, A) 
		&:= \frac{1}{|A|} \sum_{k \in A}  \psi(i, k)
\end{align}
be the \emph{average latent association} between $X_i$ and $\{ X_k : k \in A\}$.
\end{definition}

%%\begin{definition}
%%\label{def: set lat asc}
%%{\bf (Average Latent Coherence)}
%%Given $j \in [d]$ and $A \subseteq [d]$ let
%%\begin{equation}
%%\label{eq:avg_coh}
%%\psi(j,A) \, := \,\frac{1}{|A|} \sum_{k \in A \backslash \{ j \}} \! \psi(j, k)
%%\end{equation}
%%be the \emph{average latent coherence} between $X_j$ and $\{ X_k : k \in A\}$.
%%\end{definition}

%\noindent\blue{[Do we need $A\setminus\{j\}$ in the definition? I need to carefully check this. 
%I do not think this is consistent throughout the paper. For now I think it is good to have 
%for how we define Coherent Sets. However, in the CLT section we run into issues
%for the cases $j\in A$ because of the difference in the two sections definitions. Must rectify this.]}

\begin{definition}[\textbf{Coherent Set}]
\label{def:coherent_set}
%Let $\psi(\cdot, \cdot)$ be defined as in Definition~\ref{def: set lat asc}.  
A subset $A \subseteq [d]$ with at least two elements
is a \emph{coherent set} with respect to average latent association if
\begin{enumerate}
	\item $\psi(i, A_{-i}) \, > \, 0$ for each $i \in A$, and
	\item $\psi(i, A) \, \leq \, 0$ for each $i \notin A$,
\end{enumerate}
where $A_{-i}:= A\setminus\{i\}$.
A coherent set is \emph{minimal} if no proper subset is a coherent set.
\end{definition}

% Side Note from Carson: the structure of this definition 
% seems to imply a tendency towards larger sets of features,
% because the definition only requires a sign, as opposed to
% a threshold value. I believe in real data this will allow 
% large sets to arise from a few significant features that are 
% positively associated with many other features.
% An introduction of a threshold value to this definition
% seems interesting both theoretically and for applications.
% Alternatively, instead of implicitly using additivity,
% considering an interaction effect for the average latent association
% might make this sign definition more interesting for future work.

A set is coherent if each element in the set has positive 
average latent association with the other elements in the set, 
and no element outside of the set satisfies this property.  
This definition ensures that if we add or remove a single feature to 
a coherent set, then it is no longer a coherent set.
Note that this does not exclude the possibility of adding or removing more than 
one feature from a set and maintaining the 
coherent set definition.\footnote{See 
Proposition~\ref{prop: coh set greedy} 
in Appendix~\ref{app: lamb greedy search proc} for an explicit example of such behavior.}
%\red{I think this following sentence is better somewhere else, if said at all.} 
%In this way, the definition of coherent sets makes 
%the LAMB method's search procedure (discussed in Section~\ref{sec:lamb}) to be a Greedy algorithm.
For this reason the LAMB method seeks to discover minimal coherent sets.

%Like clusters of objects in exploratory data analysis
Analogous to a block of positive correlations in a covariance matrix, 
a coherent set of features is mutually positively associated.
For binary-valued data, coherent sets offer a narrative advantage over other measures
for associated sets, since the threshold model allows latent association to be interpreted as
mutual dependence in an unobserved continuous measure.
Coherent sets can 
be estimated in a computationally efficient search procedure
based on iterative hypothesis testing
that is formalized in Sections~\ref{sec:test} and \ref{sec:lamb}.
%%Hypothesis testing of latent association is discussed in 
%%Section~\ref{sec:test}. The search 
%%procedure is discussed in Section~\ref{sec:lamb}.

\subsection{Hypothesis Testing}
\label{sec:test}

The LAMB method's search procedure is based on 
iterative hypothesis testing. %See Equation \eqref{eq:hypo}.  
To carry out hypothesis tests we construct a test statistic for latent association, 
and then appeal to a central limit theorem to calculate approximate p-values.

Binary data $\bm{X}_{\cdot 1}, \ldots, \bm{X}_{\cdot n}$ 
is assumed to arise from 
independent replicates  from the threshold model of Definition \ref{def:mod}.
Suppose that the random threshold vectors 
$\bm{\theta}_{\cdot 1}, \ldots, \bm{\theta}_{\cdot n}$ are observed along with
the binary vectors $\bm{X}_{\cdot 1}, \ldots, \bm{X}_{\cdot n}$.  
%$\mathbb{X} = [\bm{X}_{\cdot 1}, \ldots, \bm{X}_{\cdot n}] = [x_{ij}]$.
In this case, a straightforward estimator for the latent association
of Definition \ref{def:lat_cor} is the corresponding sample average
\begin{align}
\label{eq:sam_coh}
	\widehat{\psi}_n(i,k)
		&:=  \frac{1}{n} \sum_{j = 1}^n U_{ij} U_{kj}\,, \quad\text{where}\quad 
			U_{ij} := \frac{X_{ij} - \theta_{ij} }{ \sqrt{ \theta_{ij} (1-\theta_{ij})  } } \, .
\end{align}
%$X_{ij}$ is the $i^{\textnormal{th}}$ feature in $\bm{X}_{\cdot j}$, and $\theta_{ij}$ is defined similarly.
%\blue{[Why not use $A\setminus\{j\}$ here?]
%Note that based on our null hypothesis we assume the condition $j\notin A$ 
%when we estimate $\psi(j, A)$ to utilize the condition $C_j$ independent of $\{C_k\colon k\in A\}$,
%see section \ref{sec: CLT}. }
For $A\subseteq[d]$ we then define
%\begin{align*}
	$\widehat{\psi}_n(i,A) 
		:= |A|^{-1} \sum_{k \in A} \widehat{\psi}_n(i,k)$.
%\end{align*}
Note that the sample quantities $\widehat{\psi}_n(i,k)$ and  $\widehat{\psi}_n(i,A)$ are not guaranteed to fall between -1 and 1.
%Under mild conditions, however, their values will converge to the interval $[-1,1]$ as $n$ tends to infinity.
See Proposition~\ref{prop:u_consist} in Appendix~\ref{sec: CLT} for details.

Suppose that $V_i$ is independent of $\{V_k \colon k\in A\}$
for some $i\in [d]$ and $A\subset [d]\setminus\{i\}$.
Theorem~\ref{thm:csm_clt} of Appendix~\ref{sec: CLT} establishes that, 
for a suitable variance estimator\footnote{There are two consistent
estimators for the variance under the same conditions. However,
one is more computationally efficient. 
See Appendix~\ref{sec: CLT} for details.}
$\widehat{\sigma}_n^2(i, A)$ and under some conditions\footnote{These
conditions provide an upper bound that guarantees
the Lindeberg condition is satisfied
and the Lindeberg-Feller Central Limit Theorem can be applied.}
on $\bm{\theta}_{\cdot 1}, \ldots, \bm{\theta}_{\cdot n}$, the quantity
$\sqrt{n} \,  \widehat{\psi}_n(i, A) / \widehat{\sigma}_n(i, A) $
is approximately standard normal, i.e.,
\begin{align*}
 	\frac{\sqrt{n}\, \widehat{\psi}_n (i, A)} {\widehat{\sigma}_n(i, A)}
		&\overset{d}{\longrightarrow}  \mathcal{N}(0,1) \quad\textnormal{as  $n \to \infty$} .
\end{align*}

In the iterative testing based search procedure, we wish to test 
the following hypotheses for a fixed $A\subset [d]$ and each $k\in [d]$:
\begin{align}
	\textnormal{H}_0 (k, A): \psi(k, A_{-k})  \leq  0 \quad\text{vs}\quad  
		\textnormal{H}_1 (k, A): \psi(k, A_{-k})  >  0 \label{eq: search proc hypothesis tests}
\end{align}
where $A_{-k} := A\setminus\{k\}$.
Large positive values of $\widehat{\psi}_n(k, A_{-k})$ provide 
strong evidence for positive association, i.e.,
strong evidence against the null hypothesis $\textnormal{H}_0(k, A)$.
Therefore, we define approximate p-values by
\begin{align}
\label{eq:pvs}
	\textnormal{pv}(k, A) 
		&:= 1 - \Phi^{-1} \left( \frac{ \sqrt{n} \widehat{\psi}_n(k, A_{-k}) }{ \widehat{\sigma}_n(k, A_{-k}) } \right)\,.
\end{align}
%Empirical p-values are passed to a given multiple testing procedure in order to perform a set update
%step in the search procedure that is discussed in Section~\ref{sec:lamb}.
The approximate p-values defined by Equation~\eqref{eq:pvs} 
are used to perform a set update step in the LAMB
method's search procedure that is discussed
in the next section.

Recall that above it is assumed
the random threshold vectors 
$\bm{\theta}_{\cdot 1}, \ldots, \bm{\theta}_{\cdot n}$ are observed along with
the binary vectors $\bm{X}_{\cdot 1}, \ldots, \bm{X}_{\cdot n}$. 
In practice the thresholds 
$\bm{\theta}_{\cdot 1}, \ldots, \bm{\theta}_{\cdot n}$ 
are not observable,
but they can be consistently
estimated 
%from the binary samples 
%$\bm{X}_{\cdot 1}, \ldots, \bm{X}_{\cdot n}$
under suitable assumptions.
Supporting theory for estimating 
$\bm{\theta}_{\cdot 1}, \ldots, \bm{\theta}_{\cdot n}$ is detailed in 
Appendix~\ref{sec: est thresholds}.
The computational approach to estimating 
$\bm{\theta}_{\cdot 1}, \ldots, \bm{\theta}_{\cdot n}$ 
from binary data is discussed in Section~\ref{sec:tildes}.
The LAMB method uses estimates $\widehat{\bm{\theta}}_{\cdot j}$ of $\bm{\theta}_{\cdot j}$ 
%to define 
and plugs in these values to calculate the statistics
$\widehat{\psi}_n(i,A)$ and $\widehat{\sigma}_n(i, A)$ defined above.
%and the empirical p-values $\text{pv}(i, A)$.

\subsection{Iterative Testing Search Procedure}
\label{sec:lamb}

We now discuss the LAMB method's search procedure
for discovering coherent sets.
The pseudocode for the LAMB method's
iterative testing based search procedure is 
contained in Algorithm~\ref{alg: lamb search proc}.
Initializing all possible singleton sets is recommended, and
we consider this an ``exhaustive'' search in 
practice.\footnote{Algorithm~\ref{alg: lamb search proc} 
allows for initializing sets of features other than singletons,
and the search procedure 
can be performed in parallel across multiple seeds $A^0$
and across the for  $k\in [d]$ loop
(see Appendix~\ref{app: lamb parallel search}).
Multiple testing procedures
other than \citet{yekutieli-2001} could be used
(see Appendix~\ref{app: multiple testing procedures}).}
Intuitively, initializing the search
procedure with a singleton set increases the 
chance of estimating a minimal coherent set.
However, Algorithm~\ref{alg: lamb search proc} is 
not guaranteed to estimate 
minimal coherent sets.

\begin{algorithm}[h]
\SetAlgoLined
\SetKwInOut{Input}{input}\SetKwInOut{Output}{output}

\Input{$i \in [d]$, $T\in \mathbb{N}$, $N \geq 2$, and $\delta \in (0,1)$}
	Initialize $t:= 0$ and $A^0 := \{i\}$\;
	\While{$t\leq T$} {
		\For{$k\in[d]$}{
%			Let $A^t_{-k}:= A^t\setminus\{k\}$\; 
			Test $\textnormal{H}_0(k, A^t)$ using the approximate p-value from Equation~\eqref{eq:pvs}\;
		}
		Calculate Benjamini-Yekutieli adjusted approximate p-values $\textnormal{BY}(1, A^t), \ldots, \textnormal{BY}(d, A^t)$\;
		Define $A^{t+1} := \{k\in [d] \colon \textnormal{BY}(k, A^t) \leq \delta \}$\;
		\eIf{$A^t = A^{t+1}$ or $A^{t+1} = \emptyset$}{
			\eIf{$|A^{t+1}| \geq N$}{
				\Output{$A^* := A^t$}
			}{
				\Output{$A^* := \emptyset$}
			}
		}{
			$t := t + 1$\;
		}
	}
\caption{Iterative Testing Search Procedure \label{alg: lamb search proc}}
\end{algorithm}

%%\begin{enumerate}
%%	\item[(1)] Choose an initial set 
%%		$A^0 := \{i\}$ for some $i \in [d]$.
%%	
%%	\item[(2)] Given $A^t$ for $t\in \mathbb{N}_0$
%%		and notation $A^t_{-i}:= A^t\setminus\{i\}$, test
%%		\begin{align} \label{eq:hypo}
%%			H_0 (i, A^t):\quad \psi(i,A^t_{-i})  \leq  0 \quad\text{vs}\quad  H_1 (i, A^t):\quad \psi(i,A^t_{-i})  >  0, \quad \text{for $i \in [d]$}.
%%		\end{align}
%%		
%%	\item[(3)] Let $A^{t+1}$ be the set of $i\in [d]$ such that 
%%		$H_0(i, A^t)$ is rejected by a chosen multiple testing 
%%		procedure. %\footnote{We recommend the FDR control of \citet{yekutieli-2001}.}
%%		
%%	\item[(4)] Repeat steps (2) and (3) until $A^{t} = A^{t'}$ for some 
%%			$t>t'$.\footnote{If $t$ reaches a maximum number of allowed iterations
%%			and this does not occur, the search procedure terminates and does 
%%			not return any set.}
%%	
%%	\item[(5)] Output set $A^* := A^{t}$ if $t = t' + 1$ and $|A^*| \geq 2$ 
%%		or some other chosen minimum set size.
%%\end{enumerate}

The output set $A^*$ in Algorithm~\ref{alg: lamb search proc} 
is a fixed point of the search procedure;
further set updates will not change $A^*$.
Note that $\emptyset$ is a fixed point 
of the search procedure that represents an unsuccessful search.
There is a close relationship between 
nonempty fixed points and coherent sets. %\ref{def:coherent_set}.
By definition, a nonempty fixed point $A^*$ of the 
search procedure has the properties: 
$\textnormal{H}_0(k, A^*)$ is rejected for all $k \in A^*$, and
$\textnormal{H}_0(k, A^*)$ is accepted for all $k \notin A^*$.\footnote{We reject
or accept the null hypotheses using a multiple testing procedure.}
Therefore a nonempty fixed point  $A^*$ satisfies Definition~\ref{def:coherent_set} 
up to a level of statistical significance.
Consequently, nonempty fixed points of the 
search procedure are natural estimates of 
coherent sets.\footnote{Fixed points are not the only thing
we observe from this search procedure in practice.
Cycles do occur and are discussed in 
Appendix~\ref{app: search procedure details}.}

In practice, many or all of the LAMB method's searches may degenerate to $A^* = \emptyset$,
indicating lack of evidence of a true underlying signal in the features of a dataset.  
Other searches might result in overlapping or identical estimated 
coherent sets.  Multiple instances
of the same set are considered a single estimated coherent set.  
In cases where substantial overlap is present, a variety of heuristic methods
may be employed to form a representative set from an overlapping class of 
sets.\footnote{See Appendix~\ref{app: post process} for details about 
post-processing the LAMB method's estimated coherent sets.}

\subsection{Estimating Random Thresholds}
\label{sec:tildes}

In Section~\ref{sec:test}, we assumed that the random thresholds
$\bm{\theta}_{\cdot 1}, \ldots, \bm{\theta}_{\cdot n}$
were observed to define estimators of latent association.
However, as noted earlier, in practice
we must estimate the random thresholds and
plug in the estimates 
$\widehat{\bm{\theta}}_{\cdot 1}, \ldots, \widehat{\bm{\theta}}_{\cdot n}$
into the estimators of latent association defined in Section~\ref{sec:test}.
%%The approximation and optimization approach that is outlined below
%%was robust in both the artificial and real data settings
%%discussed in this paper. 

First we note that the threshold model of Definition~\ref{def:mod}
underlying the LAMB method
is not a true generative model.
Recall that we  do not impose any assumptions 
for the underlying distribution of $\bm{V}$,
other than assuming it has a continuous joint distribution function $F$.
We also do not assume the density $\pi$
of the random variables $\tau_1, \ldots, \tau_n$.
%%hence we do not
%%assume the distribution that generates
%%the random thresholds $\bm{\theta}_{\cdot 1}, \ldots, \bm{\theta}_{\cdot n}$.
Consequently, more distribution assumptions for $\bm{V}$ and $\bm{\theta}$
are needed in our threshold model 
in order to use the popular Expectation-Maximization (EM) 
or Variational Inference (VI) optimization algorithms 
\citep{jordanGraphicalModels-2008, bleiVariationalInferenceCopulas-2015}.
However, the marginal distributions of $\bm{X} \given \bm{\theta}$
are Bernoulli distributions, which
allows us to implement a pseudo-likelihood approach.

To estimate the random thresholds $\theta_{ij}$
it is necessary and sufficient to estimate the parameters
in the truncated Poisson factorization model
(see Equation~\eqref{eq:theta} in Definition~\ref{def:mod}
and Proposition~\ref{prop: identifiable}).
In particular, instead of estimating $d\cdot n$ 
random thresholds $\{\theta_{ij}\}_{i\in [d], j\in [n]}$, 
we estimate $d + n - 1$ parameters\footnote{Identifiability 
of the truncated Poisson factorization
model requires $\alpha_1 := 1$. See Proposition~\ref{prop: identifiable}.}
$\bm{\alpha} := (1, \alpha_2, \ldots, \alpha_d)^T$
and $\bm{\tau}^0 := (\tau_1^0, \ldots, \tau_n^0)^T$,
where $\tau_j^0$ is the realized value of the
random variable $\tau_j$, for each $j\in [n]$.

Consistent estimators of $\bm{\alpha}$ and $\bm{\tau}^0$
are derived in Appendix~\ref{sec: est thresholds}
(see Proposition \ref{prop:theta_mean} and 
Theorem \ref{thm:taus}).
These consistent estimators rely on knowing the density $\pi$ 
for the random sample $\tau_1, \ldots, \tau_n$.
The results in Appendix~\ref{sec: est thresholds}
demonstrate that the problem of 
estimating random thresholds from binary data is 
a well-posed inference problem.
%Despite the theoretical advantages of the estimators using a known 
%distribution $\pi$, in practice we generally do not know 
%nor wish to assume the distribution $\pi$.
However, estimating $\bm{\alpha}$ and $\bm{\tau}^0$ is still
a computationally difficult problem.
The computational goal, therefore, is to 
efficiently approximate the consistent 
estimators discussed in Appendix~\ref{sec: est thresholds}.
We use two different approximations for the density 
$\pi$ and the likelihood $f(\bm{X} \given \bm{\alpha}, \tau)$
to accomplish this goal. 
For the first approximation, we replace the 
unknown density $\pi$ with an uninformative prior
$\widehat{\pi} \sim \textnormal{Unif}(0, 8)$.\footnote{The 
interval $(0, 8)$ is arbitrary, and any number larger
than 8 could be chosen. As the values $\alpha_i$ or $\tau_j^0$
get larger than 8, the random threshold gets exponentially close 
to the value 1. In practice, the interval $(0, 8)$
is large enough for robust estimation.}
%%an empirical distribution function
%%\begin{align*}
%%%	\label{eq:prior}
%%	\hat{\pi}_n(t)
%%		&= \begin{cases}
%%			\frac{1}{n}, &\text{if} \tab t \in \{ \tau_1^0, \ldots, \tau_n^0 \} \\
%%			0, & \textnormal{otherwise.}
%%		\end{cases}
%%\end{align*}
%Then the estimation approach suggested by Proposition \ref{prop:theta_mean} and 
%Theorem \ref{thm:taus} is to maximize an approximation to the posterior likelihood for $\tau$ 
%under an appropriate ``moment-of-moments'' estimation of $\bm{\alpha}$.  
For the second approximation we use, for each $j\in[n]$,
\begin{align*}
	\hat{f}(\bm{X}_{\cdot j} \given \bm{\alpha}, \tau^0_j)
		&:= \prod_{i = 1}^d  \left(1 - e^{ - \alpha_i \tau^0_j }\right)^{X_{ij}} \left(e^{- \alpha_i \tau^0_j }\right)^{1- X_{ij}}\,.
\end{align*}
%i.e., a factorized form across the variables that share $\tau_j^0$.
This approximation is sometimes called a pseudo-likelihood
\citep{besagLatticeInteractions-1974, besagNonLatticeInteractions-1975, 
amini-2013}.\footnote{This approximation resembles the conditional independence assumption
used in naive Bayes classification \citep{domingosNaiveBayes-1997}.} 
Note that the distribution for $\bm{V}$
is not approximated.
%%; we do not make a distribution
%%assumption for $\bm{V}$, because we are not interested in estimating 
%%the joint distribution function $F$ of $\bm{V}$.

To approximate the consistent estimators of Appendix~\ref{sec: est thresholds},
we want to find $\bm{\alpha}$ and 
$\bm{\tau}^0$ that maximize the pseudo-likelihood 
subject to a method of moments constraint:
\begin{align}
	\hat{f}(\mathbb{X}, \bm{\tau}^0 \given \bm{\alpha})
		&:= \prod_{j = 1}^n \hat{f}(\bm{X}_{\cdot j} \given \bm{\alpha}, \tau^0_j)\, \hat{\pi}(\tau^0_j)
%			= \frac{1}{8} \prod_{j=1}^n \prod_{i = 1}^d
%		 	 \left(1 - e^{ - \alpha_i \tau_j }\right)^{X_{ij}} \left(e^{- \alpha_i \tau_j }\right)^{(1- X_{ij})}
%			 \mathbb{I}_{(0, 8)}(\tau_j)
		\label{eq: tau pseudolikelihood}\\
	\overline{X}_i
		:= \frac{1}{n} \sum_{j\in [n]} X_{ij} 
			&\approx \frac{1}{n} \sum_{j\in [n]} (1 - e^{- \alpha_i \tau_j^0})
				=: \overline{\theta}_i \quad\textnormal{for each $i\in [d]$}\,.\label{eq: alpha MoM}
\end{align}
Equation~\eqref{eq: alpha MoM} is based on the fact that under the threshold model we have
$\mathbb{E}(X_{ij}) = \mathbb{E}\left( \mathbb{E}[X_{ij} \given \bm{\theta}_{\cdot j}]\right) = \mathbb{E}(\theta_{ij})$, 
for each $i\in [d]$ and $j\in [n]$.\footnote{This approach
resembles a constrained MLE estimator in the frequentist
perspective, and a constrained MAP estimator in the Bayesian perspective.}
Although this optimization problem does not have a closed form solution, 
it can be approximately computed. 

\begin{algorithm}[h]
\SetAlgoLined
\SetKwInOut{Input}{input}\SetKwInOut{Output}{output}

\KwData{binary data matrix $\mathbb{X} \in \{0, 1\}^{d \times n}$}
\Input{$\epsilon \in (0, 1)$ and $M \in \mathbb{N}$}
	Let $\alpha_1 := 1$, $m := 0$, and $\Delta := 1$\;
	Initialize $\alpha_2, \ldots, \alpha_d$ randomly or using row means of $\mathbb{X}$\;
	\While{$\Delta \geq \epsilon$ and  $m\leq M$}{
		\For{$j \in [n]$}{
			$\tau_j^0 \given \bm{\alpha} := \argmin_{t\in (0, 8)} \sum_{i\in [d]} \big[\alpha_i t - X_{ij}\big( \alpha_i t + \ln[1 - \exp\{-\alpha_i t\} ]\big) \big]$\;
		}
		\For{$i \in [d]\setminus\{1\}$}{
			$\alpha_i \given \bm{\tau}^0 \in \argmin_{a\in (0, 8)} \left(1 - \overline{X}_i - \frac{1}{n} \sum_{j\in [n]} \exp\{- a \tau_j^0\} \right)^2$\;
		}
		$m := m + 1$\;
		\If{$m > 1$}{
			$a^* := \textnormal{median}\left( ||\bm{\alpha} - \bm{\alpha}_{\textnormal{old}}||_1 \right)$\;
			$t^* := \textnormal{median}\left( ||\bm{\tau}^0 - \bm{\tau}^0_{\textnormal{old}} ||_1 \right)$\;
			$\Delta := \textnormal{max}\{a^*, t^*\}$\;
		}
		$\bm{\alpha}_{\textnormal{old}} := \bm{\alpha}$ and $\bm{\tau}^0_{\textnormal{old}} := \bm{\tau}^0$\;
	}
\caption{Estimate Random Thresholds \label{alg: est rand thresh}}
\end{algorithm}

Algorithm~\ref{alg: est rand thresh} contains the
pseudocode for estimating the random thresholds.
Note that Algorithm~\ref{alg: est rand thresh}
is a coordinate-wise optimization procedure with 
two alternating steps.
%When does the coordinate-wise optimization 
% converge to the global optimization? - Kai
Given values for $\bm{\alpha}$,
we minimize the negative logarithm\footnote{This is 
commonly done. In this case, it protects the optimization procedure
from numerical underflow.}
of each sample in 
the pseudo-likelihood given by Equation~\eqref{eq: tau pseudolikelihood}.
Then, given values for $\bm{\tau}^0$, we minimize a smoothed version of the constraint in
Equation~\eqref{eq: alpha MoM}.
We note here that it is possible to 
optimize $\alpha_i \given \bm{\tau}^0$ in Algorithm~\ref{alg: est rand thresh}
subject to a constraint that ensures convexity.
See Appendix~\ref{app: conv opt theory} for details.

\section{Simulation Study}
\label{sec: sim study}
The LAMB method was applied
to artificial binary datasets to 
establish the effectiveness of the method
at discovering associated sets of features
under controlled conditions.
%%To establish the effectiveness of the LAMB method
%%at discovering associated sets of features
%%we analyzed its results under the controlled conditions
%%of artificial binary datasets.
In Section~\ref{subsec: high level artificial datasets} we discuss
the high-level details of generating 
these artificial binary datasets. 
The full details of generating the artificial datasets 
is discussed in Appendix~\ref{subsec: artificial datasets}.
%%In Section~\ref{subsec: artificial datasets}
%%we discuss how the artificial datasets used in this simulation study
%%were created. 
Analysis of the results obtained from different association mining methods
applied to the artificial datasets is discussed
in Section~\ref{subsec: sim study analysis}.
Additional figures are included in Appendix~\ref{subsec: additional sim study figures}.

\subsection{Artificial Datasets}
\label{subsec: high level artificial datasets}

A total of 450 artificial binary datasets $\mathbb{X}\in \{0, 1\}^{d \times n}$
were created in accordance with
the threshold model of Definition~\ref{def:mod}.
In particular, binary data contains a randomly thresholded 
version of a continuous latent random vector with nontrivial 
covariance between its components. 

To resemble the high-dimension low-sample size setting
of most modern datasets, we used $n = 200$ and $d = 2,000$.
Analyzing datasets with larger values of $n$ or $d$ is possible, 
and we expect the results to be robust to an increased value of $n$ or 
an increased value of $d$ or both. 
However, because of the approximate p-values used 
in the iterative testing search procedure (see Section~\ref{sec:test}),
using $n < 200$ could lead to spurious results
from the LAMB method.

Continuous vectors $\bm{V}_{\cdot 1}, \ldots, \bm{V}_{\cdot n}\in \mathbb{R}^d$
with nontrivial association structure were generated 
from a $d$-multivariate normal distribution.
Five disjoint sets of associated features $A_1, \ldots, A_5 \subset [d]$,
all of size 200,
were embedded into the artificial binary datasets.\footnote{Five 
disjoint sets were embedded 
into the artificial datasets so that half of the features in 
the datasets are white noise.}\footnote{The 
results obtained from the LAMB method do not depend
on knowing the number nor the size of the embedded sets 
of associated features. Using 
the same number of sets and the same size for each set makes 
comparing the results obtained from the LAMB method to the 
results obtained from the NMF and LDA 
methods more systematic and informative.}
Embedding associated sets of features is accomplished by using
blocks of correlation in the covariance matrix for the
$d$-multivariate normal distribution that
generates $\bm{V}_{\cdot 1}, \ldots, \bm{V}_{\cdot n}$.
The sets of associated features $A_1, \ldots, A_5$ 
were independently given (population)
correlation values $\rho_1, \ldots, \rho_5 \in [0, 1]$.
Note that features belonging 
to different sets $A_p$ and $A_q$, for $p,q\in [5]$,
are independent in their population covariance 
because of the block correlation
structure used to generate $\bm{V}_{\cdot 1}, \ldots, \bm{V}_{\cdot n}$.
%%Finally, the underlying (population) correlations
%%$\rho_1, \ldots, \rho_5$ used to generate $A_1, \ldots, A_5$
This process for generating $\bm{V}_{\cdot 1}, \ldots, \bm{V}_{\cdot n}$ 
was done a total of 50 times; values of the (population) correlation
$\rho_1, \ldots, \rho_5$ corresponding to the sets
of features $A_1, \ldots, A_5$
varied significantly across the interval $[0,1]$.

Random thresholds $\bm{\theta}_{\cdot 1}, \ldots, \bm{\theta}_{\cdot n}$
were generated independently of $\bm{V}_{\cdot 1}, \ldots, \bm{V}_{\cdot n}$.
Recall that the random thresholds %$\bm{\theta}_{\cdot j}$
are completely specified by the values of
$\bm{\alpha} := (\alpha_1,  \ldots, \alpha_d)^T$
and $\bm{\tau}^0 := (\tau_1^0, \ldots, \tau_n^0)^T$ (see Equation~\eqref{eq:theta} in Definition~\ref{def:mod}).
The values of $\bm{\alpha}$ were independently generated using a (scaled) beta distribution
and the values of $\bm{\tau}^0$ were independently generated using 
a gamma distribution.
Three different pairs of parameters for the beta distribution
were used to generate three different 
vectors $\bm{\alpha}_{\textnormal{low}}, \bm{\alpha}_{\textnormal{med}}, \bm{\alpha}_{\textnormal{high}}$.
Similarly, three different pairs of parameters for the gamma distribution
were used to generate three different 
vectors $\bm{\tau}^0_{\textnormal{low}}, \bm{\tau}^0_{\textnormal{med}}, \bm{\tau}^0_{\textnormal{high}}$.
See Figures~\ref{app fig: alphas histogram} and \ref{app fig: taus histogram}
in Appendix~\ref{subsec: artificial datasets}
for histograms of the corresponding values of $\bm{\alpha}$
and $\bm{\tau}^0$, respectively. 
Using all combinations of $\bm{\alpha}_{\textnormal{low}}, \bm{\alpha}_{\textnormal{med}}, \bm{\alpha}_{\textnormal{high}}$ and $\bm{\tau}^0_{\textnormal{low}}, \bm{\tau}^0_{\textnormal{med}}, \bm{\tau}^0_{\textnormal{high}}$ generated a total of 9 different 
threshold matrices $\mathbf{\Theta}:= [\bm{\theta}_{\cdot 1}, \ldots, \bm{\theta}_{\cdot n}]$.

\subsection{Analysis}
\label{subsec: sim study analysis}

We now discuss the results of the LAMB method applied to the (450)
artificial binary datasets that were generated in 
accordance to the threshold model of Definition~\ref{def:mod}.
The Nonnegative Matrix Factorization (NMF) and
Latent Dirichlet Allocation (LDA) methods discussed
in Section~\ref{subsec: rel work} were also applied 
to the artificial datasets. 
The LAMB, NMF, and LDA methods are 
further compared on two real (count-valued) dataset applications
in Section~\ref{sec: applications}. 
%The NMF and LDA methods were also applied to these 
%artificial binary datasets, because these two 
%methods are commonly used and they are
%related to the LAMB method.

Precision and recall statistics were used to 
compare the results of the different association mining
methods.
Suppose an estimated set of associated
features is denoted by $E$ (for any method),
and assume that the discovered set $E$ best 
estimates\footnote{Here ``best'' refers to
the set $A_j$ with the smallest Jaccard distance from $E$ \citep{jaccard-1901}.
In particular, we are not interested in the precision or recall 
statistics of ``noisy'' sets or clusters from the methods.
For example, methods such as NMF, LDA, and hierarchical clustering
benefit from have more topics or clusters than known sets
because at least one of the clusters produced by these methods
can then contain the ``noisy'' features. See Appendix~\ref{subsec: additional sim study figures}
for additional figures for each method.}
the embedded set of features $A_j$, for some $j\in [5]$.
Then we define the precision and recall statistics
for the set $E$ estimating the set $A_j$ by
\begin{align*}
	\textbf{precision}(E, A_j)
		&:= \frac{|E \cap A_j|}{|E|}
		\quad\textnormal{and}\quad
		\textbf{recall}(E, A_j)
			:= \frac{|E \cap A_j|}{|A_j|}\,.
\end{align*}
The embedded associated sets 
of features $A_1, \ldots, A_5$
were all of the same size, 
and the corresponding features 
of $\bm{V}_{\cdot 1}, \ldots, \bm{V}_{\cdot n}$
varied in their (population) correlation independently.
It is therefore reasonable to take the median values of 
the precision and recall statistics for a fixed value of (population)
correlation. 
Due to the manner in which the 
associated sets of features $A_1, \ldots, A_5$
were independently given (population) correlation
values $\rho_1, \ldots, \rho_5$, we expect the
median precision and recall statistics to be robust estimates
of the results obtained from all of the association mining methods that we considered.
%%Across the 50 generations of the random vectors
%%$\bm{V}_{\cdot 1}, \ldots, \bm{V}_{\cdot n}$,
%%the values of the
%%(population) correlation $\rho_1, \ldots, \rho_5$ for the embedded
%%sets of associated features $A_1, \ldots, A_5$
%%were occasionally identical, or similar in value, 
%%for two or more different sets. 
%%This makes the results of the median precision and recall statistics more robust,
%%because in principle the association mining methods should
%%have a more difficult time discovering the (disjoint) 
%%sets of associated features embedded
%%in the artificial datasets.

\begin{figure}[h]
\begin{subfigure}[b]{0.49\textwidth}
\centering
\includegraphics[width = \textwidth]{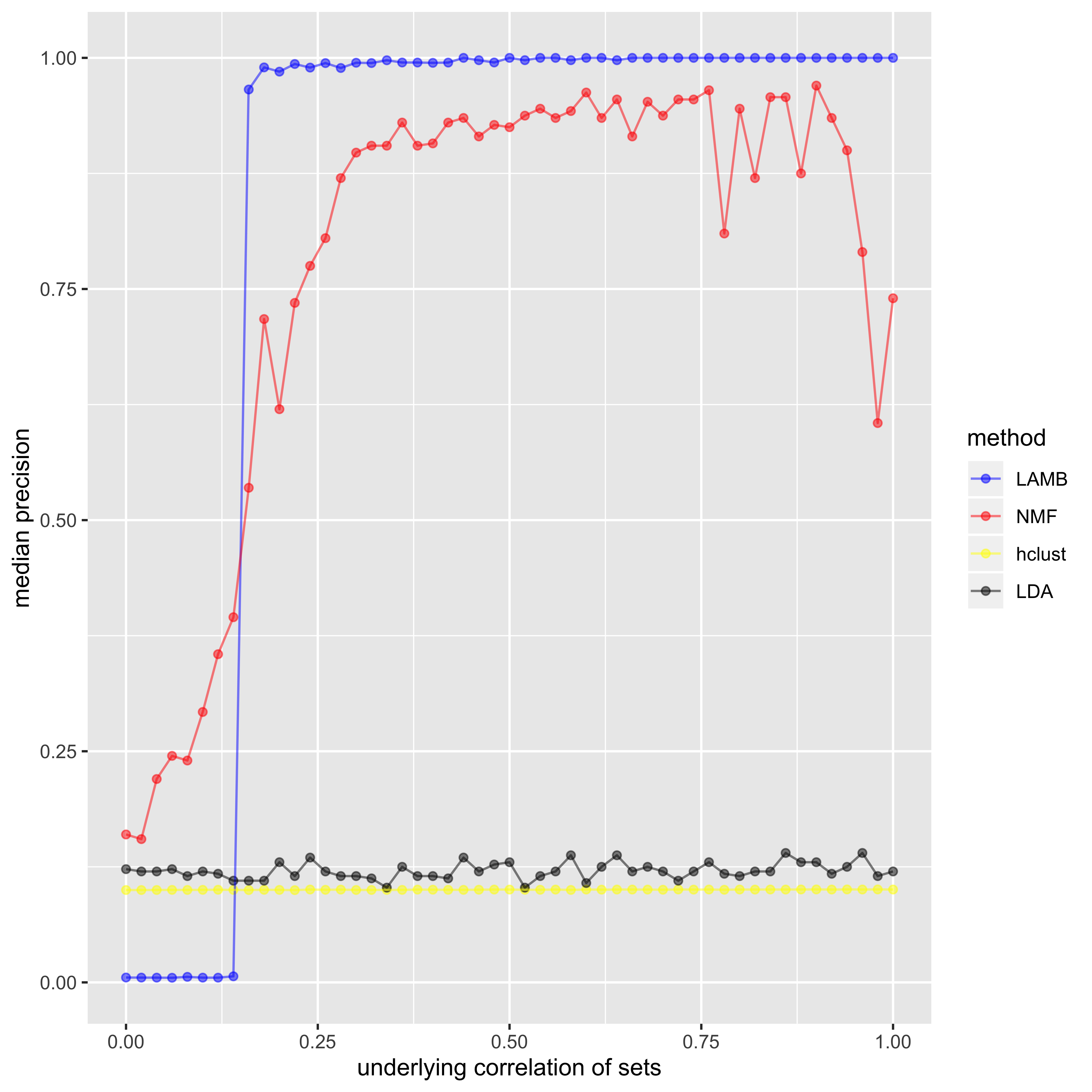}%\\
\end{subfigure}
~
\begin{subfigure}[b]{0.49\textwidth}
\centering
\includegraphics[width = \textwidth]{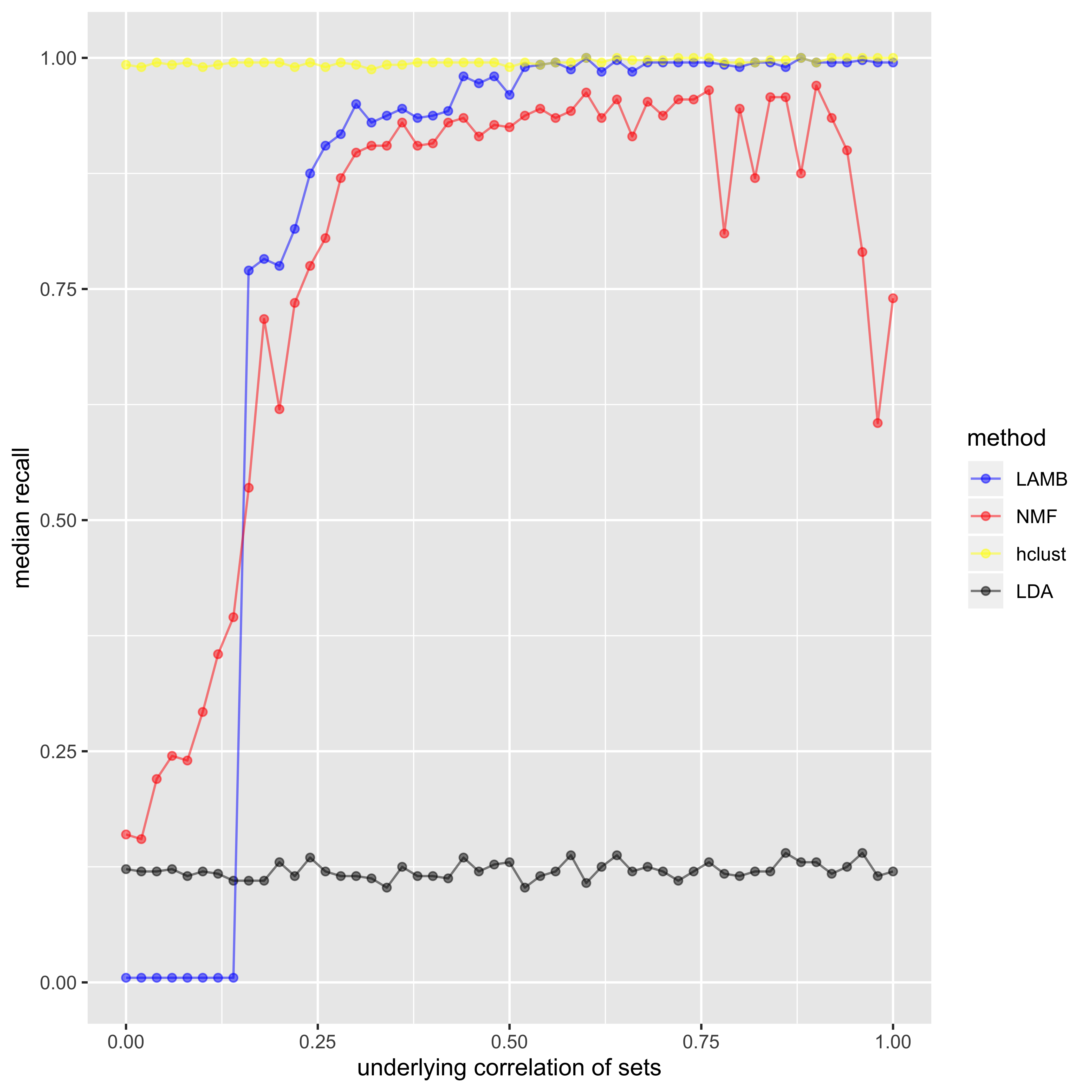}%\\
\end{subfigure}

\caption{Comparison of the LAMB, NMF, and LDA methods on the artificial datasets of Section~\ref{sec: sim study} using the (median) precision  and recall of their estimated associated sets. 
%%See Section~\ref{subsec: sim study analysis} for a detailed analysis.
%%The LAMB method produces estimated coherent sets while
%%the NMF and LDA methods produce soft clusters  
%%of all the features per topic. Only the ``top 200'' features
%%in the soft clusters of the NMF and LDA methods are used
%%since the embedded sets all contained 200 features.
For this figure we used $\delta = .05$ for the LAMB method
and fit 6 latent topics for the NMF and LDA methods.
As a baseline reference we include results obtained from hierarchical clustering 
with binary distance and average linkage 
using 7 clusters.
Additional figures for different values of these parameters 
are included in Appendix~\ref{subsec: additional sim study figures}.}
\label{fig: sim study comp}
\end{figure}

Results from the LAMB, NMF, and LDA methods are
visualized in Figure~\ref{fig: sim study comp}.
The LAMB method is successful at discovering
truly associated sets of features whenever the 
underlying correlation in the corresponding
features of the continuous latent
vectors $\bm{V}_{\cdot 1}, \ldots, \bm{V}_{\cdot n}$
is above around $.125$.
Soft clusters of features arising from the NMF and LDA
methods were restricted to the ``top 200'' features
to better visualize the results of the different methods.\footnote{In 
general, the size and the number of the soft clusters of features generated
by the NMF and LDA methods is essentially another parameter to consider.}
Note that since only the top 200 features are
used for both the NMF and LDA methods,
and the embedded sets of associated features
were all of size 200, both the precision and
recall statistics are exactly the same for these two methods.\footnote{That is to say,
for the estimated sets $E$ from the NMF and LDA methods
we have $|E| = |A_j|$, in the notation used to define 
$\textbf{precision}(E, A_j)$ and $\textbf{recall}(E, A_j)$.
Hence, $\textbf{precision}(E, A_j) = \textbf{recall}(E, A_j)$
for the NMF and LDA methods across all estimated sets and artificial datasets.}
In contrast to the NMF and LDA methods,
the number of sets and the sizes of the sets 
discovered by the LAMB method is not fixed. %, 
%%and so 
%%the precision and recall statistics corresponding
%%to the LAMB method's results are different.

The NMF method also successfully discovers truly associated
sets of features (see Figure~\ref{fig: sim study comp}). Note that
the NMF and LDA methods are Topic 
Model methods for count-valued data.
Restriction to binary-valued data negatively impacted
the LDA method more than the NMF method.\footnote{It 
appears that the binary-valued artificial datasets do not provide enough information for 
the inference problem used in the LDA method.}
These results are not an indictment
of the LDA method. Rather, the results
of this simulation study indicate the difficulty 
that binary-valued data can present for inference
and association mining problems.
On the other hand, we see in 
Section~\ref{sec: applications} that the LAMB method
is able to discover meaningful associations 
in count-valued data that is ``binarized''.
Additional results are included in Appendix~\ref{subsec: additional sim study figures}.

\section{Real Dataset Applications}
\label{sec: applications}
Applying association mining methods
to real datasets is similar to Exploratory Data Analysis \citep{tukeyEDA-1977}
and Feature Selection \citep{guyonFeatureSelection-2003, khalidFeatureSelection-2014}.
%%In this section we discuss applications of 
%%association mining method to real datasets.
%%As mentioned in Section~\ref{sec: intro},
%%%conventional 
%%association mining methods were 
%%originally motivated by discovering associated
%%items in market basket data to better predict buyer behavior.
Binary-valued datasets are still encountered
in modern settings \citep{needellBinaryClassification-2018, quonSCbinarize-2019}.
The purpose of this section, however, is to demonstrate how the LAMB method 
performs relative to the well-known Nonnegative Matrix Factorization (NMF) and 
Latent Dirichlet Allocation (LDA) methods 
on count-valued datasets that contain important sample heterogeneity.
%%However, there is no simple way of directly 
%%comparing results from the three methods.
%%We will now discuss the application of
%%the LAMB, Nonnegative Matrix Factorization (NMF), 
%%and Latent Dirichlet  Allocation (LDA) methods 
%%to count-valued datasets that contain sample heterogeneity. 
%%There are important similarities and distinctions between 
%%the results of applying the LAMB, NMF, and LDA methods to the data
%% in this section.
%%The LAMB method performs well
%%compared to the other methods considered.
%%However, binary-valued datasets are still encountered
%%in modern settings \citep{needellBinaryClassification-2018, quonSCbinarize-2019}.

%%Recall that the LAMB method takes
%%a binary-valued data matrix as input.
In this section, count-valued data matrices were ``binarized''
into co-occurence data matrices
for the LAMB method. Unlike the LAMB method, the
NMF and LDA methods 
were applied to the count-valued data matrices.
In principle, the count data matrices
contain sufficiently more information than the binary
co-occurence matrices. This section demonstrates
that the LAMB method is able to detect meaningful
associations among features without information of the raw counts.

%%Throughout this section we use 
%%a statistic called the effective number of sets
%%that is defined in Definition~\ref{def: eff num} in Appendix~\ref{app: eff num sets}.
In Section~\ref{sec: text} 
we study a text dataset of famous works in
Western literature. Then in Section~\ref{sec: lastfm}
we study a bipartite graph dataset.
Section~\ref{sec: other lamb apps} discusses
another suitable application of the LAMB method 
to gene expression data.
Additional details, figures, and tables
are included in Appendices~\ref{app: text add results}
and \ref{app: lastfm add results}.

\subsection{Text Data}
\label{sec: text}

Text analysts are often interested in 
classifying texts by finding sets 
of terms that appear together frequently. See
\citet{bagofwords-1986} and \citet{manningInformationRetrieval-2008}
for introductions to text analysis and information retrieval,
and see \citet{forman-2003} and \citet{aggarwalTextClassification-2012}
for surveys of text classification methods.
%Term usage in documents presents an ideal data source for this paradigm.  
In text data samples are documents within a work
(e.g., chapters in a novel) 
and features are unique terms.\footnote{It is possible
to use $n$-grams instead of distinct terms
as features.}
Different text documents can vary significantly 
in length, content, and style. 
Text data is therefore an ideal source for 
detecting associated sets of features based
on heterogeneous samples.
%%Failing to account for sample heterogeneity 
%%can therefore lead to spurious association results.

In this section we consider a text dataset that
consists of famous works in Western literature. 
Text sources were restricted to Western literature only to 
facilitate interpreting associated sets of terms. 
Nonetheless, the chosen works are
very heterogeneous in terms of
document length, authors, and themes.
The online database \verb!gutenberg.org! and the R package 
\verb!gutenbergr! from \citet{gutenbergr-package} were used to create this text dataset.
See Tables~\ref{tab: west lit non-shake docs}
and \ref{tab: west lit shake docs} in Appendix~\ref{app: text add results} for
the different works used in this text dataset.

%\begin{table}[H]
%\caption{Estimated coherent sets for text data from LAMB using $\alpha = .05$.}
%\label{tab: large shake house coh set}
%\footnotesize
%\centering
%\fbox{ \begin{minipage}{0.8\textwidth}
%
%\begin{enumerate}
%
%	\item $\{$    $\}$
%
%\end{enumerate}
%
%\end{minipage} }
%
%\end{table}

Text data is naturally sequence-valued.
However, term sequences within a document can be ``tokenized''
using a so-called ``bag-of-words'' assumption to create
a count-valued term-document matrix.
In particular, if the $(i, j)$-entry of the term-document matrix 
has value $c$, then the term $i$ was used $c$ times in document $j$.

Individual works such as Moby Dick or Romeo and Juliet were broken up by chapter 
or scene, respectively.\footnote{Individual Shakespeare sonnets 
were treated as one document
to add more heterogeneity among the documents in this text dataset.}
%instead of considering an entire work
%as a single document.
For example, Moby Dick was broken up into
135 documents, each document
representing one chapter in the novel.
After removing stop words\footnote{Including 
Shakespearean stop words such as ``thou''. 
This filtering step was not perfect. For example,
we failed to combine the counts of
``england's'' and ``england''. 
%%However, for the purpose
%%of our analysis in this section, 
%%we believe the preprocessing done was sufficient to illustrate similarities 
%%and differences between the LAMB, LDA, and NMF
%%methods for this text dataset.
}, 
the resulting term-document matrix
consisted of 40,561 unique terms
and 971 documents. 
This text data matrix was preprocessed to further reduce it to 
11,757 unique terms and 971 documents by removing
terms that occurred fewer than 
three times across all documents.\footnote{All documents
contained at least 6 distinct terms among the remaining 11,757 terms.}
The original $40,561\times 971$ dimensional term-document matrix is $99\%$
sparse, i.e., $99\%$ of the
term-document matrix entries are 0.
The reduced $11,757\times 971$ dimensional term-document matrix is $97\%$
sparse. Both the final term-document matrix size and the amount of sparsity 
is common for a text dataset.

Typically, term-document data matrices use either a 
term-frequency (tf) weight or term-frequency inverse-document-frequency
(tf-idf) weight for each matrix entry. 
The NMF and LDA methods take as input a tf-weighted term-document matrix.
%The tf-idf-weighted term-document matrix was provided to HC.
As mentioned earlier, a binary co-occurrence (or binary-weighted)
term-document matrix is used as input 
to the LAMB method. In particular,
a 1 in the $(i, j)$-entry of the co-occurrence term-document matrix indicates
that term $i$ appeared at least once in document $j$,
and a 0 indicates that term $i$ was never used in document $j$.

Recall that the LAMB method outputs estimated coherent sets.
The NMF and LDA methods can produce soft clusters of terms, 
and the elements of these soft clusters can essentially be ranked.\footnote{For 
the NMF method the rank can be based on the latent nonnegative value in
the matrix factorization. For the LDA method there are two 
ways to rank terms: one rank is based on 
estimated probabilities and another is based on so-called ``term-scores''
that resemble tf-idf weighting (see \citep{bleiTopicModels-2009}).}
To demonstrate the soft clustering by the NMF and LDA methods
we considered different sizes of soft clusters
and different numbers of latent topics. To form
soft clusters of different sizes we selected ``top'' 
cluster elements based on the corresponding method's rankings.

Some of the effective numbers (see 
Definition~\ref{def: eff num} in Appendix~\ref{app: eff num sets}) 
of distinct term sets  
produced by the LAMB, NMF, and LDA methods is presented
in Table~\ref{tab: lamb text effective numbers}.
The results in Table~\ref{tab: lamb text effective numbers}
and Appendix~\ref{app: text add results}
demonstrate that the LAMB method
detects sets of associated terms that are as 
discriminative as the results obtained from the NMF and LDA methods,
but usually contain many more terms in the associated sets. 
Appendix~\ref{app: text add results} contains 
more detailed results obtained from the
LAMB, NMF, and LDA methods.

\begin{figure}[h]
\centering
\includegraphics[width = \textwidth]{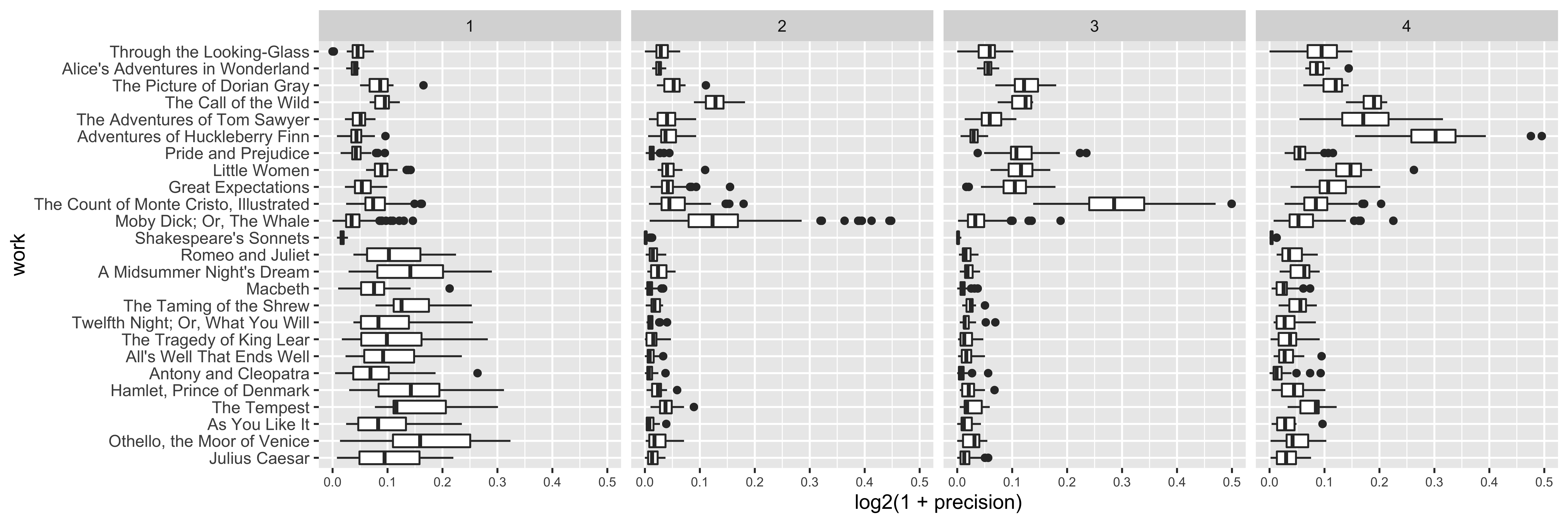}

\caption{Box plots of precision per work for four estimated coherent sets
obtained from the LAMB method
with $\delta = .01$ on the text dataset in Section~\ref{sec: text}.
Each data point within a box plot row corresponds to a chapter or scene in a given
work. Higher values of precision correspond to the estimated
coherent sets containing more terms in the work.}
\label{fig: reduced lamb coh set box plots alpha 01}
\end{figure}

Figure~\ref{fig: reduced lamb coh set box plots alpha 01} provides
a simple numerical description of some of
the LAMB method's estimated coherent sets.
%%Figure~\ref{fig: reduced lamb coh set box plots alpha 01}
%%demonstrates that the sets of terms discovered
%%by the LAMB method are meaningful.
For example, the second estimated coherent set
in Figure~\ref{fig: reduced lamb coh set box plots alpha 01}
contains many terms present in Moby Dick 
such as characters, sailing terms, and hunting terms.
Based on the plots and settings of these two novels, it is not very surprising that The Call of the Wild and
Moby Dick share a decent amount of terms.
In particular, the LAMB method's estimated 
coherent sets of terms for this text dataset
contain important terms such as characters, but the sets
also contain more general terms that belong to many works. 
It is important to note that the LAMB method is able to
detect these associated terms with access
to only binary co-occurence data,
as opposed to the raw count data used by the NMF and LDA 
methods.

\begin{table}[h]
\centering
\begin{tabular}{ || c || c | c | c || } 
 \hline
 \textbf{Method} & \textbf{Parameter} & \textbf{Set Filtering} & \textbf{Effective Number} \\          
 \hline \hline
 LAMB & $\delta = .05$ & $\geq 25$ terms & 7.510 \\
 \hline
 LAMB & $\delta = .01$ & $\geq 25$ terms & 10.004 \\
 \hline
 NMF & 25 topics & top 50 terms & 12.42 \\
 \hline
 NMF & 25 topics & top 1000 terms & 7.11 \\
 \hline
 LDA & 25 topics & top 50 terms & 11.12 \\
 \hline
 LDA & 25 topics & top 1000 terms & 6.40 \\
 \hline
\end{tabular}
\caption{Effective number of distinct term sets obtained
from applying the LAMB, NMF, and LDA methods
to the text dataset in Section~\ref{sec: text}. 
The LAMB method estimates coherent sets
while the NMF and LDA methods can produce soft clusters of terms.}
\label{tab: lamb text effective numbers}
\end{table}

\subsection{Last.fm  Artist-User Data}
\label{sec: lastfm}

Music streaming services such as Pandora (\verb!pandora.com!), 
Last.fm (\verb!last.fm!), and Spotify (\verb!spotify.com!)
offer their users the opportunity to 
discover new  artists or bands based on the user's musical preferences.  
Companies usually develop complex algorithms 
for finding similar artists or bands based on era, genre, user ratings, et cetera. 
In this section we study a Last.fm dataset
provided by \citet{celmaLastfm-2010} and accessible from the \verb!last.fm! public API
through the author's website.\footnote{This 
Last.fm data was collected during the aughts, up until May 2009.} 
It is natural to apply association mining
methods to the Last.fm dataset in order to discover associated artists
or bands. 
%%In this context,
%%results from association mining
%%methods applied to the Last.fm dataset 
%%can be treated as feature selection
%%\citep{guyonFeatureSelection-2003, khalidFeatureSelection-2014}.
However, one can also analyze this Last.fm dataset 
using a recommender system method,
e.g., \citet{daiSmoothRecommenderSystems-2019} have recently
studied a similar Last.fm dataset using such a method.

The Last.fm dataset presents another ideal source
for detecting associated sets of features in heterogeneous samples. 
For this dataset, individual Last.fm users are treated as samples
and individual artists or bands are treated as features.
Heterogeneity among the samples in the dataset arises 
from the variety in the demographics of the Last.fm users, as well as
different users listening to different amounts of artists or bands.
Note that the Last.fm dataset can be represented as a bipartite graph
where one set of vertices corresponds to artists or bands, and another set
of vertices corresponds to Last.fm users.

The corresponding artist-user data matrix
is count-valued; if the $(i,j)$-entry of the artist-user matrix 
has value $c$, then Last.fm user $j$ listened to $c$
songs by artist $i$.\footnote{Not 
necessarily a count of distinct songs. The original dataset includes the individual songs that users
listened to. However, for the purposes of this section, it makes sense to focus on associated artists
or bands rather than associated songs. Therefore, we agglomerate
the counts of individual songs per artist or band.}
The Last.fm dataset contains
153,898 unique artists or bands available
to listen to, the listening history of 
924 anonymized Last.fm users, and the matrix is $99\%$ sparse.\footnote{This dataset 
contains a lot of metadata that is not utilized for this application.}
A basic preprocessing step filtered out artists or bands and Last.fm users below 
a certain listening threshold. The filtering done 
was such that all remaining artists or bands were listened to by at least 11 different 
Last.fm users,
and all remaining Last.fm users listened to at least 11 different artists or bands.
This data preprocessing step reduced the artist-user data matrix to
10,957 distinct artists or bands, 913 distinct Last.fm users,
and is 94\% sparse.

The NMF and LDA  methods were applied
to the count-valued artist-user data matrix.
However, the LAMB method was applied to the corresponding 
binary-valued artist-user co-occurence matrix.
In this co-occurence matrix, a value of 
1 in the $(i, j)$-entry indicates that Last.fm user $j$ listened  to artist or band $i$ at least once,
and a 0 indicates user $j$ never listened to artist or band $i$.

Latent association and coherent sets provide 
a novel means of matching  artists or bands when 
Last.fm users
exhibit diverse listening behavior.  
The LAMB method detects large and diverse 
sets of associated artists and bands
in the Last.fm dataset (see Table~\ref{tab: lamb lastfm effective numbers} 
and Appendix~\ref{app: lastfm add results}).
Results from the LAMB method
are not heavily skewed towards popular music.
In particular, the LAMB method consistently detects 
the most popular artists and bands in the Last.fm dataset
in one of its estimated coherent sets, while the other estimated
coherent sets detect more diverse sets of artists and bands.
Additional results are included in Appendix~\ref{app: lastfm add results}.
%%Differences in Last.fm user behavior effectively mask associations
%%among artists or bands in the NMF and LDA methods.

\begin{table}[h]
\centering
\begin{tabular}{ || c || c | c | c || } 
 \hline
 \textbf{Method} & \textbf{Parameter} & \textbf{Set Sizes} & \textbf{Effective Number} \\          
 \hline \hline
 LAMB & $\delta = .05$ & $\geq 25$ & 4.710 \\
 \hline
 LAMB & $\delta = .01$ & $\geq 25$ & 5.166 \\
 \hline
  NMF & 12 topics & top 50 & 8.840 \\
 \hline
 NMF & 12 topics & top 1000 & 4.784 \\
 \hline
 LDA & 12 topics & top 50 (probabilities) & 8.820 \\
 \hline
 LDA & 12 topics & top 1000 (probabilities) & 6.157 \\
 \hline
\end{tabular}
\caption{Effective number of distinct artist sets
from applying the LAMB, NMF, and LDA methods
to the Last.fm dataset in Section~\ref{sec: lastfm}. The LAMB method outputs estimated coherent sets
while the NMF and LDA methods can produce soft clusters of artists and bands.}
\label{tab: lamb lastfm effective numbers}
\end{table}

\subsection{Further Applications}
\label{sec: other lamb apps}

%%The real dataset applications in Sections~\ref{sec: text}
%%and \ref{sec: lastfm}, as well as the 
%%results under the controlled conditions
%%of artificial datasets in Section~\ref{sec: sim study}, demonstrate
%%that the LAMB method is capable of discovering 
%%sets of associated features in high-dimensional datasets
%%that contain heterogeneous samples.
%%In particular, the results from the LAMB method
%%in Sections~\ref{sec: sim study}, \ref{sec: text},
%%and \ref{sec: lastfm} lead us to believe
%%that the LAMB method will be useful
%%in similar data settings such as gene expression data.

Gene interaction networks and gene expression data in general
are an important application of association mining methods
\citep{alvesFPMGeneAnalysis-2010,
naulaertsPrimerFrequentItemsetMiningBioinformatics-2015,
aamriCoOccurrenceGeneAssociations-2019}.
One important future application of the LAMB method
is to single-cell data known colloquially as scRNA-seq and scATAQ-seq
data.
See \citet{chenSCRNAreview-2019} and
\citet{kiselevSCRNAchallenges-2019} for general reviews of scRNA-seq data,
and see \citet{chenSCATAQassessment-2019} and 
\citet{yanSCATAQreads-2020} for general reviews of scATAQ-seq data.
Usually both scRNA-seq and scATAQ-seq data are 
high-dimensional, contain heterogeneous samples (e.g., cells),
and usually have a low signal-to-noise ratio.
\citet{quonSCbinarize-2019}
demonstrate the utility of binary methods for analyzing scRNA-seq and scATAQ-seq data.

\section{Summary}
\label{sec: summary}

The LAMB method is based on a simple 
threshold model and latent association measure for binary data.
We have shown that the LAMB method
is able to detect meaningful sets of mutually associated features
in artificial and real datasets.
The LAMB method's iterative testing based search procedure 
moderated false discoveries when applied to the artificial datasets.
For the two real dataset applications that we considered,
the data is naturally count-valued. However, we have demonstrated
that there is sufficient information present
in the corresponding binary co-occurence data
for the LAMB method to detect
meaningful sets of associated features.
In particular, the LAMB method is able to find large 
sets of associated features that are as discriminative as the 
best soft clusters produced by the NMF and LDA methods.

In principle, methods developed for count-valued data
do not have sufficient information from binary-valued data 
to detect meaningful sets of associated features.
For example, this occurs for the Latent Dirichlet Allocation (LDA) method
on the artificial binary datasets in Section~\ref{sec: sim study}.
However, the Nonnegative Matrix 
Factorization (NMF) method is able to discover 
meaningful sets of associated features in binary data.
On the other hand, the LAMB method is able to discover
meaningful sets of associated features in binary data in a less 
supervised manner than the NMF method.

\section*{Acknowledgement}
This work was partly supported by the National Institutes of Health under
award number R01HG009125-01. %AN NIH grant
The content is solely the responsibility of the authors and 
does not necessarily represent the official views of the National Institutes of Health.
This work was also partly supported by the National Science Foundation
under Grants DMS-1613072, DMS-1613261, %AN's NSF grants
DMS-1613112, IIS-1633212, and DMS-1916237. %KZ's NSF grants
Any opinions, findings, and conclusions or recommendations
expressed in this material
are those of the authors and do not necessarily 
reflect the views of the National Science Foundation.
The authors thank Miheer Dewaskar and 
Kevin O'Connor for helpful discussions about 
the LAMB method's search procedure and 
post-processing results obtained from the LAMB method.

%%Code for the LAMB method was written in the \verb!R!
%%language. The following packages were helpful in writing the code
%%and for the analyses within this paper: 
%%\citet{R-stats-package, tidyverse-package, 
%%reshape2-R-package, R-utils-package, here-R-package, nmf-R-package,
%%tm-R-package, topicmodels-R-package, tidytext-R-package, gutenbergr-package,
%%pracma-R-package}, and 
%%\citet{corrplot-R-package}.

%\Bibliography
%\clearpage
%\newpage
%\input{./citations.bib}
\bibliography{citations.bib}
\bibliographystyle{plainnat}

\newpage

\appendix

\section{Supporting Theory}
\label{app: supporting theory}

\subsection{Properties of the Threshold Model}
\label{app: prop of thresh mod}

An important feature of the threshold model
of Definition~\ref{def:mod} is that it imposes
no structure on the latent vector $\bm{V}$
beyond the assumption that its joint distribution function $F$ is continuous.
%This allows the LAMB method to 
%focus on detecting the joint association structure of $\bm{V}$. 
While we wish to assess latent association among the components of $\bm{X}$ 
arising from dependence in $\bm{V}$, the latent associations do not determine $F$, 
and we do not seek to estimate $F$ nor the marginal distribution functions $F_1, \ldots, F_d$.  
Indeed, if $\gamma: \mathbb{R}^d \to \mathbb{R}^d$ is defined by
$\gamma(v_1, \ldots, v_d) := (\gamma_1(v_1), \ldots, \gamma_d(v_d))^T$
where each function $\gamma_i: \mathbb{R} \to \mathbb{R}$ is continuous
and strictly increasing, then the threshold model generated by $(\bm{\theta}, \bm{V})$
is the same as the model generated by $(\bm{\theta}, \gamma(\bm{V}))$.

The threshold model is also invariant to more general, multivariate monotone transformations.
This can be described by viewing the threshold model
in terms of copulas instead of quantiles.
Continuity of the joint distribution function $F$ of $\bm{V}$
ensures that of the marginal distribution functions $F_1, \ldots, F_d$.  
Thus, $F_1(V_1), \ldots, F_d(V_d)$ 
are (potentially dependent) Unif$(0, 1)$ random variables.  The joint distribution function 
\begin{align*}
	C(u_1,\ldots, u_d) 
		&:= \mathbb{P}(F_1(V_1) \leq u_1, \ldots, F_d(V_d) \leq u_d)
\end{align*}
is the copula for the joint distribution function $F$
with marginals $F_1, \ldots, F_d$ \citep[Chapter~2]{nelsenCopulas-2010}.
If $\bm{V}' \in \mathbb{R}^d$ has a continuous distribution function 
$\tilde{F}$ and marginals $\tilde{F}_1, \ldots, \tilde{F}_d$ with the same copula $C$, 
then the threshold model generated
by $(\bm{\theta}, \bm{V})$ is the same as that generated by $(\bm{\theta}, \bm{V}')$.
The converse, however, is not the case: 
it can be shown that under the threshold model two random vectors 
$\bm{V}$ and $\bm{V}'$ having different copulas can yield identical (in distribution) 
binary random vectors. 
% Trying to find an explicit example to demonstrate this. 
% Intuitively this makes sense. Kai came up with a simple univariate example
% to try and generalize to two dimensional case. -CM
%%\red{I don't think this is an accurate statement:
%%As consequence of this invariance, once the threshold vector $\bm{\theta}$ has been specified,
%%the distribution of $\bm{X}$ %and the latent associations $\psi(i, k)$ (is not random, so I don't understand this statement)
%%depends only on the copula $C$ of $\bm{V}$, and not
%%on the marginal distributions of the latent random variables $V_1, \ldots, V_d$. }
As a result, different copulas for $\bm{V}$ can lead to identical latent associations.
%in particular, they do not depend on the location, scale, or moments of the latent variables $V_i$.

We will now show
how Example~\ref{ex1} in Section~\ref{subsec: latent coh} is not exceptional.
In \citet{zhangBET-2019} truncated binary expansions
are used for a nonparametric test
of independence. Our threshold model conditioned on 
$\bm{\theta}$ can be considered a 
binary expansion truncated at the first resolution.
A simple, but important,
point used in \citet{zhangBET-2019} is that 
uncorrelated binary variables implies
independent binary variables.
% \blue{[Insert law of total covariance calculation here.]}
Note that under the threshold model the Law of Total Covariance implies that 
\begin{align}
	\text{Cov}(X_i, X_k)
%		&= \mathbb{E}\left(\text{Cov}[X_i, X_k| \bm{\theta}]\right)
%			+ \text{Cov}\left(\mathbb{E}[X_i | \bm{\theta}], \mathbb{E}[X_k | \bm{\theta}]\right)\\
		&= \mathbb{E}\left(\text{Cov}[X_i, X_k \given \bm{\theta}]\right)
			+ \text{Cov}\left(\theta_i, \theta_k\right).
		\label{eq: law of total covariance}
\end{align}
We are not interested in dependence between $X_i$ and $X_k$ arising 
from the dependence of their respective thresholds, i.e., the
$\text{Cov}\left(\theta_i, \theta_k\right)$ term in Equation~\eqref{eq: law of total covariance}. 
A measure of intrinsic association between features $i$ and $k$ 
is captured by $\textnormal{Cov}(V_i, V_k)$
under the threshold model.
The LAMB method detects dependence
between the features of the latent vector $\bm{V}$
from observation of $\bm{X}$ by using latent association.
That is to say, by using latent association 
we are effectively avoiding
estimating the joint distribution $F$ of $\bm{V}$
while still testing for dependence between 
features in $\bm{V}$.

\subsection{Properties of Latent Association}
\label{subsec: prop latent asc}

Latent association shares some of the basic properties of standard correlation.
For example, $|\psi(i,k)| \leq 1$ (see Lemma \ref{lem:uv}) and
$\psi(i,k) \neq 0$ implies dependence between $V_i$ and $V_k$.
Although the latent association of two features in $\bm{X}$ is not necessarily equal to the correlation
of the same two features in the continuous vector $\bm{V}$, there is a monotonic relationship when $\bm{V}$ is 
multivariate normal.

\begin{prop}
\label{prop:rho}
Let $\bm{X} = \mathbb{I}\left[ \bm{V} \leq \bm{F}^{-1}( \bm{\theta}) \right]$  as in Definition~\ref{def:mod}, where
%$\bm{C} \sim \mathcal{N}_d({\bf 0}, \Sigma)$ with $\Sigma_{jj} = \sigma^2$ for all $j$.
$\bm{V} \sim \mathcal{N}_d({\bm{0}}, \mathbf{\Sigma})$ such that $\Sigma_{ii} = 1$ for each $i\in[d]$. 
Then %, for any fixed random vector $\bm{\theta}$, 
$\psi(i,k)$ is a monotone 
nondecreasing function of $\rho(V_i, V_k) = \Sigma_{ik}$.
%\beq
%\label{eq:psipos}
%%\mbox{sgn}(\psi(j,k)) = \mbox{sgn}(\Sigma_{jk}) \, 
%\eeq
%i.e., the sign of the latent coherence between $X_i$ and $X_i$ is equal to the sign of the
%correlation between $V_i$ and $V_j$.
\end{prop}

\begin{remark}
We note that the LAMB method
does not rely on normality assumptions for $\bm{V}$.
However, the above proposition states that whenever $\bm{V}$ has 
a multivariate normal distribution, 
latent association is guaranteed to be
a monotone nondecreasing function 
of the correlation between components 
of $\bm{V}$. %%\footnote{This is useful if
%%there exists a monotone function $\gamma\colon \mathbb{R}^d\to\mathbb{R}^d$
%%that can map a general latent continuous vector
%%$\bm{V}$ (with a given continuous joint distribution function $F$)
%%to a multivariate normal random vector $\gamma(\bm{V})$,
%%because the threshold model is invariant to such a 
%%monotone transformation.}
Note that this matches intuition for the generative model: if the correlation between
$V_i$ and $V_k$ is increasing, then the latent association $\psi(i,k)$
is monotonically nondecreasing.%\footnote{In our
%simulation study we see that the sample estimate of $\psi(i, k)$ increases
%as the true underlying correlation increases, within some small error.}
\end{remark}

\begin{proof}

Without loss of generality assume $d = 2$. 
Let $\bm{\theta}$ be a random threshold
as in Definition~\ref{def:mod}. Let
$\bm{V}:= (V_1, V_2)^T$ and $\bm{W}:= (W_1, W_2)^T$ 
be bivariate normal random vectors
such that $\mathbb{E}(V_i) = \mathbb{E}(W_i) = 0$ and
$\textnormal{Var}(V_i) = \textnormal{Var}(W_i) = 1$, 
for each $i\in \{1, 2\}$.
Further assume that $ r_1 := \rho(V_1, V_2) \in [-1,1]$
and $r_2 := \rho(W_1, W_2) \in [-1,1]$
such that $r_2 > r_1$. 
Then we have
\begin{align*}
	\mathbb{E}(V_1 V_2)
		&= \textnormal{Cov}(V_1, V_2)
			= r_1 < r_2 = \textnormal{Cov}(W_1, W_2)
			= \mathbb{E}(W_1 W_2).
\end{align*}
By Slepian's Lemma \citep{slepian-1962}, for any $a,b\in \mathbb{R}$, 

\begin{align}
	\mathbb{P}\left(V_1\leq a, V_2\leq b\right)
		&\leq \mathbb{P}\left(W_1\leq a, W_2\leq b\right). \label{eq: slep lem}
\end{align}

Define $X_i := \mathbb{I}(V_i \leq \Phi^{-1}(\theta_i))$
and $\tilde{X}_i :=  \mathbb{I}(W_i \leq \Phi^{-1}(\theta_i))$ for each $i\in \{1, 2\}$.
Denote by $\psi(r_1) := \psi_{\bm{X}}(1,2)$ 
and $\psi(r_2) := \psi_{\bm{\tilde{X}}}(1,2)$
the latent associations of the threshold
models for $\bm{X}$ and $\tilde{\bm{X}}$, respectively.
Recall that we assume $\bm{\theta}\indep \bm{V}$
and $\bm{\theta} \indep \bm{W}$.
Fix arbitrary $t_1,t_2\in (0,1)$. Denote $a:= \Phi^{-1}(t_1)$
and $b:= \Phi^{-1}(t_2)$. Then note that,
by inequality \eqref{eq: slep lem}, we have
\begin{align*}
	\mathbb{E} \left[\mathbb{I}(V_1 \leq a)\, \mathbb{I}(V_2\leq b)\right]
		&= \mathbb{P}\left(V_1\leq a, V_2\leq b\right)\\
		&\leq \mathbb{P}\left( W_1\leq a, W_2\leq b\right)\\
		&= \mathbb{E} \left[\mathbb{I}(W_1 \leq a)\, \mathbb{I}(W_2\leq b)\right].
\end{align*}
Therefore, by independence and the above inequality pointwise, we have
\begin{align*}
	\mathbb{E}_{\bm{\tilde{X}}|\bm{\theta}}[ \tilde{X}_1 \tilde{X}_2 
			\big| \bm{\theta} ] 
		- \mathbb{E}_{\bm{X}|\bm{\theta}}[ X_1 X_2 \big| \bm{\theta} ]
		\overset{\textnormal{a.s.}}{\geq} 0.
\end{align*}
Hence, by the linearity of $\mathbb{E}_{\bm{\theta}}$,
\begin{align*}
		\psi(r_2) - \psi(r_1) \geq 0.
\end{align*}
In particular, the latent association $\psi(1,2)$ is a monotonically
nondecreasing function of $r:= \rho(V_1, V_2)$ on $[-1,1]$.
\end{proof}

\begin{remark}
A more general form of monotonicity 
than that of Proposition~\ref{prop:rho} can be shown
through copulas.
This type of monotonicity is present in 
nonparametric association statistics
such as Kendall's tau and Spearman's rho
\citep[Chapter~5]{nelsenCopulas-2010}.
%%First we note that latent association can be expressed as a functional 
%%of the joint and marginal distribution functions of $\bm{V}$. 
%%In particular, this clearly illustrates why
%%$\psi(i,k) = 0$ whenever $V_i\indep V_k$.
%%This functional form of latent 
%%association further shows that 
%%If $C_1$ and $C_2$ are two copulas 
%%such that $C_1 \prec C_2$ (concordance ordering), then
%%the corresponding latent associations
%%$\psi_1(i, k)$ and $\psi_2(i, k)$
%%satisfy $\psi_1(i, k) \leq \psi_2(i, k)$.
\end{remark}

\begin{definition}[Concordance Ordering]
If $C_1$ and $C_2$ are both 2-dimensional copulas,
then we define the \emph{concordance ordering
of copulas} to be $C_1 \prec C_2$
if and only if $C_1(a, b) \leq C_2(a,b)$
for all $a, b \in [0, 1]$.
\end{definition}

\begin{prop}
\label{prop: concordance monotonicity}
	If $C_1$ and $C_2$ are two copulas for the
	threshold model %$(\bm{\theta}, \bm{V})$
	such that $C_1 \prec C_2$ (concordance ordering), then 
	$\psi_1(i, k) \leq \psi_2(i, k)$ for arbitrary $i, k\in [d]$ 
	where $\psi_j$ is the latent association under copula $C_j$ for $j\in \{1, 2\}$.
\end{prop}

\begin{proof}
	Without loss of generality consider the case $d = 2$.
	Recall from Appendix~\ref{app: prop of thresh mod}
	that the copula $C$ of
	$\bm{V}$ is uniquely determined 
	by the continuous joint distribution function $F$:
	\begin{align*}
		C(u_1, u_2)
			&:= \mathbb{P}_{\bm{V}}\left(F_1(V_1) \leq u_1, F_2(V_2) \leq u_2\right).
	\end{align*}
	
	Suppose that $\bm{V}$ 
	has copula $C_1$
	and $\bm{W}$ has copula $C_2$
	such that $C_1 \prec C_2$
	and the marginal distributions 
	$F_1$ and $F_2$ are the same
	for $\bm{V}$ and $\bm{W}$.
	In particular, the only difference
	between the distributions
	of $\bm{V}$ and $\bm{W}$ 
	is that, for all $a, b\in [0,1]$,
	\begin{align*}
		\mathbb{P}_{\bm{V}}\left(F_1(V_1) \leq a, F_2(V_2) \leq b\right)
			&\overset{\textnormal{def}}{=} 	C_1(a, b)	\leq C_2(a, b)
				\overset{\textnormal{def}}{=}
				\mathbb{P}_{\bm{W}}\left(F_1(W_1) \leq a, F_2(W_2) \leq b\right)\,.
	\end{align*}
	Since we assume that $\bm{V} \indep \bm{\theta}$
	and $\bm{W} \indep \bm{\theta}$, the above inequality implies that
	\begin{align*}
		\mathbb{P}_{\bm{V}}\left(F_1(V_1) \leq \theta_1, F_2(V_2) \leq \theta_2 \given \bm{\theta} \right)
			&\overset{\textnormal{a.s.}}{\leq}
				\mathbb{P}_{\bm{W}}\left(F_1(W_1) \leq \theta_1, F_2(W_2) \leq \theta_2 \given \bm{\theta} \right)\,.
	\end{align*}
	From here the proof proceeds just as in the proof of
	Proposition~\ref{prop:rho}, i.e., we have
	\begin{align*}
		\psi_1(1, 2)
			&\leq \psi_2(1, 2)
	\end{align*}
	where $\psi_1$ is the latent association
	under the threshold model for $(\bm{\theta}, \bm{V})$
	and $\psi_2$ is the latent association under 
	the threshold model for $(\bm{\theta}, \bm{W})$.
\end{proof}

\subsection{Random Thresholds are Identifiable}
\begin{prop}
\label{prop: identifiable}

Setting $\alpha_1 := 1$ makes the model for 
$\mathbf{\Theta} = [\bm{\theta}_{\cdot 1}, \ldots, \bm{\theta}_{\cdot n}]$ identifiable.
In particular, $(\bm{\alpha}, \bm{\tau}^0) \equiv (\bm{\beta}, \bm{\nu}^0)$
if and only if $\mathbf{\Theta}(\bm{\alpha}, \bm{\tau}^0) \equiv \mathbf{\Theta}(\bm{\beta}, \bm{\nu}^0)$.
\end{prop}

\begin{proof}
	This follows from carefully using that $\exp(x)$ is an injective function. 
%%	Suppose that $(\alpha_1,\ldots, \alpha_d, \tau_1^0,\ldots, \tau_n^0)$
%%	and $(\beta_1, \ldots, \beta_d, \nu_1^0, \ldots, \nu_n^0)$
%%	are two vectors of parameters to be estimated, such that
%%	for every $i\in[d]$ and $j\in [n]$  we have
%%	\begin{align*}
%%		\theta_{ij}(\bm{\alpha}, \bm{\tau}^0)
%%			&\overset{\textnormal{def}}{=} 1 - \exp[-\alpha_i \tau_j^0]
%%				= 1 - \exp[-\beta_i \nu_j^0] \overset{\textnormal{def}}{=} \theta_{ij}(\bm{\beta}, \bm{\nu}^0).
%%	\end{align*} 
%%	Recall that $\alpha_i, \beta_i, \tau_j^0, \nu_j^0 > 0$ in our model 
%%	for each $i\in[d]$ and $j\in[n]$.
%%	
%%	Since we must have $\alpha_1 = 1$ and $\beta_1 = 1$
%%	by the assumed constraint, this implies that for every $j\in [n]$
%%	\begin{align*}
%%		\theta_{1j}(\bm{\alpha}, \bm{\tau}^0)
%%			&= 1 - \exp[- \tau_j^0]
%%				= 1 - \exp[- \nu_j^0] = \theta_{1j}(\bm{\beta}, \bm{\nu}^0).
%%	\end{align*}
%%	Hence, $\tau_j^0 = \nu_j^0$ for each $j\in [n]$ because $\exp(-x)$ is injective.
%%	Therefore, for arbitrarily fixed $j\in [n]$ and $i\in [d]$, 
%%	\begin{align*}
%%		\theta_{ij}(\bm{\alpha}, \bm{\tau}^0)
%%			&= 1 - \exp[-\alpha_i \tau_j^0]
%%				= 1 - \exp[-\beta_i \tau_j^0] = \theta_{ij}(\bm{\beta}, \bm{\tau}^0)
%%	\end{align*} 
%%	further implies that $\alpha_i = \beta_i$, since 
%%	$\exp(-c\,x)$ is injective for $c> 0$. Thus,
%%	we must have $(\bm{\alpha}, \bm{\tau}^0) \equiv (\bm{\beta}, \bm{\nu}^0)$
%%	if and only if 
%%	$\mathbf{\Theta}(\bm{\alpha}, \bm{\tau}^0) \equiv \mathbf{\Theta}(\bm{\beta}, \bm{\nu}^0)$.
\end{proof}

\newpage

\section{Central Limit Theorem for Latent Association}
\label{sec: CLT}

As discussed in Section~\ref{sec:test}, 
the set update step in the LAMB method's search procedure relies on approximate p-values
for estimates of the latent association. %$\psi(i, A_{-i})$.
The approximate p-values are derived from a central limit theorem that we now present in full detail.

In what follows, let the $d$-dimensional random vectors $\bm{V}$, $\bm{\theta}$, and
$\bm{X} = \mathbb{I}\left[ \bm{V} \leq \bm{F}^{-1}(\bm{\theta}) \right]$
be as in threshold model of Definition~\ref{def:mod}.  
Recall that $X_i \given \bm{\theta} \sim \textnormal{Bern}(\theta_i)$.
For each $i \in [d]$, let
\begin{align}
	\label{eq:u}
	U_{i}  
		&:=  \frac{X_{i} - \theta_{i}}{ \sqrt{ \theta_{i} (1- \theta_{i}) } }
\end{align}
be the conditionally standardized $X_{i}$.  
For any subset $A \subset [d]$, denote
the average of $U_{i}$ over the features $i \in A$ by
\begin{align}
	\label{eq:v}
	U_{A}  
		&:=  \frac{1}{|A|} \sum_{i \in A} U_{i} \, ,
\end{align}
and define
\begin{align}
\label{eq:barpsi}
	\Psi(A)
		&:= \frac{1}{|A|^2} \sum_{i,k \in A} \psi(i,k)\,.
\end{align}
%be the average pairwise latent coherence between the variables in $A$.  
The next two lemmas establish some basic
properties of the quantities defined above in 
relation to latent association
(see Equations~\eqref{eq:lat_cor} and \eqref{eq:avg_coh}).

\begin{lemma}
\label{lem:uv}
For each $i,k \in [d]$ and $A \subseteq [d]$:
%%\blue{such that $j\notin A$},
\begin{enumerate}

\item $\mathbb{E}(U_i) = 0$ and $\mathbb{E}( U_i^2) = 1$;

\item $\psi(i,k) = \mathbb{E}(U_i U_k)$, $|\psi(i,k)| \leq 1$,
	and $|\Psi(A)| \leq 1$;

\item $\mathbb{E}(U_i \, U_A) = \psi(i,A)$ and $\mathbb{E}(U_A^2) = \Psi(A)$;

\item If $V_i$ is independent of $\{ V_k : k \in A \}$, then
	$\mathbb{E}(U_i \, U_A) = 0$ and 
	$\mathbb{E}( U_i^2 \, U_A^2) = \Psi(A)$. %= \mathbb{E} (U_A^2)$.
\end{enumerate}
\end{lemma}

\begin{proof}
Fix arbitrary $i, k\in [d]$ and $A\subset [d]$.
The definition of $U_i$ in \eqref{eq:u} 
ensures that both $\mathbb{E}(U_i \given \bm{\theta}) = 0$ and
$\mathbb{E}(U_i^2 \given \bm{\theta}) = 1$ almost surely.
It is clear from the definitions that $\psi(i, k) = \mathbb{E}(U_i \, U_k)$. 
Consequently, the bound
$|\psi(i, k)| \leq 1$ follows from the Cauchy-Schwarz inequality.
Note that the bound $|\Psi(A)| \leq 1$
trivially follows by the definition of $\Psi(A)$
and the triangle inequality.
By the linearity of expectation, we have
\begin{align*}
	\mathbb{E}(U_i \,U_A)
		&= \frac{1}{|A|} \sum_{k \in A} \mathbb{E}( U_i \, U_k)
			= \frac{1}{|A|} \sum_{k \in A} \psi(i, k)
			\overset{\text{def}}{=}
			\psi(i, A)\quad\text{and} \\ 
	\mathbb{E}(U_A^2)
		&= \frac{1}{|A|^2} \sum_{k,l \in A} \mathbb{E}( U_k \, U_l) 
			= \frac{1}{|A|^2} \sum_{k,l \in A} \psi(k, l)\,
			\overset{\text{def}}{=}
			\Psi(A).
\end{align*}

Now suppose that $V_i$ is independent of $\{ V_k \colon k \in A \}$.
Then $U_i$ is conditionally independent of
$U_A$ given $\bm{\theta}$,
since $\bm{V} \indep \bm{\theta}$. 
%Because, for each $\bm{t}\in(0,1)^d$, $U_j | \bm{\theta} = \bm{t}$
%is $\sigma(V_j)$-measurable,
%$U_A | \bm{\theta} = \bm{t}$ is 
%$\sigma( V_k ; k\in A)$-measurable, 
%and $\sigma(\bm{V})$ is independent 
%of $\sigma(\bm{\theta})$, so that for each $B, D\in \mathcal{B}(\mathbb{R})$ we have
%\begin{align*}
%	\mathbb{P}\left( U_j\in B, U_A\in D | \bm{\theta} = \bm{t}\right)
%		&= \mathbb{P}\left( U_j\in B | \bm{\theta} = \bm{t}\right) \mathbb{P}\left( U_A\in D | \bm{\theta} = \bm{t}\right).
%\end{align*}
Therefore, since $\mathbb{E}(U_i \given \bm{\theta}) = 0$ almost surely,
\begin{align*}
	\mathbb{E}(U_i \, U_A)
		&= \mathbb{E} \left(\mathbb{E} [U_i \, U_A \given \bm{\theta}] \right) 
			= \mathbb{E} \left( \mathbb{E}[ U_i \given \bm{\theta}] \, \mathbb{E}[ U_A \given \bm{\theta}]  \right) 
			=  0 \,.
\end{align*}
%A similar conditioning argument shows that $\Ex (U_j^2 \, U_A^2) = \Ex (U_A^2)$
Similarly, $\mathbb{E}(U_i^2 \given \bm{\theta}) = 1$ almost surely and so
\begin{align*}
	\mathbb{E}\left( U_i^2 \, U_A^2\right)
		&= \mathbb{E}\left( \mathbb{E}[ U_i^2 \, U_A^2 \given \bm{\theta}]\right)
			= \mathbb{E}\left( \mathbb{E}[ U_i^2 \given \bm{\theta}] \, \mathbb{E}[ U_A^2 \given \bm{\theta}]\right)
			= \mathbb{E}\left(U_A^2\right)
			= \Psi(A)\,.
\end{align*}
\end{proof}

\begin{lemma}
\label{lem:v4}
%Let $U_i$ and $U_A$ be defined as in \eqref{eq:u} and \eqref{eq:v}. Then
For each $i \in [d]$ and any $A \subset [d]$:
\begin{enumerate}
\item $\mathbb{E} (U_i^4) \leq  \mathbb{E} \left(\frac{1}{\theta_i (1 - \theta_i)}\right)$
	and $\mathbb{E}(U_A^4 )
		\leq  \frac{1}{|A|} \sum_{k \in A} \mathbb{E}\left( \frac{1}{\theta_k (1 - \theta_k)}\right).$

\item If $V_i$ is independent of $\{ V_k : k \in A \}$, then 
	\begin{align*}
		\mathbb{E}(U_i^4 \, U_{A}^4) 
			&\leq  \frac{1}{|A|}  \sum_{k \in A} \mathbb{E} \left(\frac{1}{\theta_i (1 - \theta_i) \theta_k (1 - \theta_k)}\right)\,.
	\end{align*}
\end{enumerate}
\end{lemma}

\begin{proof}
Fix arbitrary $i \in [d]$ and $A\subset [d]$. 
Note that, because $X_i$ is \{0,1\}-valued, we can express
\begin{align*}
	U_i ^4
		&= X_i\, \left( \frac{ 1 - \theta_i }{ \sqrt{ \theta_i (1 - \theta_i) } } \right)^4
			+ (1 - X_i)\, \left( \frac{ - \theta_i }{ \sqrt{ \theta_i (1 - \theta_i) } } \right)^4 \, .
\end{align*}
Using $\mathbb{E}(X_i \given \bm{\theta}) = \theta_i$ almost surely, 
we have
\begin{align*}
	\mathbb{E}( U_i^4 \given \bm{\theta})
		&= \frac{ (1-\theta_i)^2 }{ \theta_i }  +  \frac{ \theta_i^2 }{ (1-\theta_i) }
			\leq \frac{ 1}{ \theta_i }  +  \frac{1}{ (1-\theta_i) }
			= \frac{ 1 }{ \theta_i (1 - \theta_i) },
\end{align*}
because $\theta_i$ is $(0,1)$-valued. 
Therefore,
\begin{align*}
	\mathbb{E}(U_i^4)
		&\leq \mathbb{E}\left( \frac{1}{\theta_i (1 - \theta_i)} \right).
\end{align*}
Moreover, it follows from Jensen's inequality that pointwise we have
\begin{align*}
	U_{A}^4 
		&= \left( \frac{1}{|A|}  \sum_{k \in A} U_k \right)^4 \leq \frac{1}{|A|} \sum_{k \in A} U_k^4\,,
\end{align*}
and so by the above work we have that
\begin{align*}
	\mathbb{E}(U_A^4 \given \bm{\theta})
		&\leq \frac{1}{|A|} \sum_{k \in A} \frac{ 1 }{ \theta_k (1 - \theta_k) }\,.
\end{align*}
Hence, by the monotonicity of expectation,
\begin{align*}
	\mathbb{E}(U_A^4)
		&\leq \frac{1}{|A|} \sum_{k \in A} \mathbb{E}\left(\frac{ 1 }{ \theta_k (1 - \theta_k) } \right)\,.
\end{align*}

Now suppose that $V_i$ is independent of $\{ V_k \colon k \in A \}$.
Then $U_i$ is conditionally independent of
$U_A$ given $\bm{\theta}$,
since $\bm{V} \indep \bm{\theta}$. 
So conditioning on $\bm{\theta}$ yields
\begin{align*}
	\mathbb{E}( U_i^4 \,  U_k^4 )
		&= \mathbb{E}\left( \mathbb{E}[ U_i^4 \,U_k^4 \given \bm{\theta}] \right)
		 	= \mathbb{E}\left( \mathbb{E}[ U_i^4 \given \bm{\theta}] \, 
				\mathbb{E} [ U_k^4 \given \bm{\theta}] \right)\\
		&\leq \mathbb{E} \left(\frac{1}{\theta_i (1 - \theta_i) \theta_k (1 - \theta_k)}\right)\,
\end{align*}
for each $k\in A$.
Finally, by the pointwise Jensen inequality above, we have
\begin{align*} 
	\mathbb{E}(U_i^4 U_{A}^4)
		&\leq \frac{1}{|A|} \, \sum_{k \in A}  \mathbb{E}( U_i^4 \, U_k^4 )
			\leq \frac{1}{|A|} \, 
			\sum_{k \in A} \mathbb{E}\left(\frac{ 1 }{\theta_i(1- \theta_i) \theta_k (1 - \theta_k) } \right)\,.
\end{align*}  
\end{proof}

With Lemmas~\ref{lem:uv} and \ref{lem:v4}, we can now establish a 
triangular array central limit theorem 
for latent association. First 
we must set up the asymptotic setting for the
threshold model of Definition~\ref{def:mod}.
For each $n \in\mathbb{N}$, let 
\begin{enumerate}
	\item[(i)] $d_n \in \mathbb{N}$,
	
	\item[(ii)] $\bm{V}_{\cdot 1}, \ldots, \bm{V}_{\cdot n}$ 
		be an iid sample of $\mathbb{R}^{d_n}$-valued random 
		vectors with distribution $\mu_n$ and 
		continuous joint distribution function $F_n$,

	\item[(iii)] $\bm{\theta}_{\cdot 1}, \ldots, \bm{\theta}_{\cdot n}$ 
		be an iid sample of $(0,1)^{d_n}$-valued 
		random vectors with distribution $\nu_n$, 
		
	\item[(iv)] and $\bm{V}_{\cdot j}$ 
		be independent of $\bm{\theta}_{\cdot j}$ for each $j\in [n]$. 
\end{enumerate}
Note that the distributions $\mu_n$ and $\nu_n$
as well as the dimension $d_n$
may depend on $n$.
For each $n\in\mathbb{N}$ and $j \in [n]$, let 
$\bm{X}_{\cdot j} = \mathbb{I}[ \bm{V}_{\cdot j} \leq \bm{F}_n^{-1}(\bm{\theta}_{\cdot j} )]$
as in the shorthand for the threshold model
of Definition~\ref{def:mod},
with $\bm{X}_{\cdot j} = (X_{1j}, \ldots, X_{d_n j})^T$.

For now fix arbitrary $i, k \in [d]$ and 
$A\subset [d]$.
The sample latent association $\widehat{\psi}_n(i, k)$
between features $i$ and $k$ is
defined to be
\begin{align*}
	\widehat{\psi}_n(i,k)
		&:=  \frac{1}{n} \sum_{j = 1}^n U_{ij} U_{kj}\,, \quad\text{where}\quad 
			U_{ij} := \frac{X_{ij} - \theta_{ij} }{ \sqrt{ \theta_{ij} (1-\theta_{ij})  } } \, .
\end{align*}
Let $U_{Aj}$ be defined as in  \eqref{eq:v} for each $j\in [n]$.  
Then a natural estimate of $\psi(i, A) = \mathbb{E}(U_i \, U_A)$ 
(see Lemma~\ref{lem:uv}) is the sample average
\begin{align}
\label{eq:psihat}
	\widehat{\psi}_n (i, A)
		&:= \frac{1}{|A|} \sum_{k\in A} \widehat{\psi}_n(i,k)
			= \frac{1}{n} \sum_{j \in [n] } U_{ij} \, U_{Aj}\,.
\end{align}
In particular, it follows from Lemma~\ref{lem:uv} that 
$\mathbb{E}\left(\widehat{\psi}_n (i, A) \right)= \psi(i, A)$.

Now suppose that $V_i$ is independent of $\{ V_k : k \in A \}$.
Then, by Lemma~\ref{lem:uv}, 
\begin{align*}
	\textnormal{Var}\left(\sqrt{n} \, \widehat{\psi}_n (i, A) \right) 
		&= \frac{1}{n} \sum_{j\in [n]}\textnormal{Var}( U_{ij} \, U_{Aj}) 
			= \mathbb{E} ( U_{i1}^2 \, U_{A1}^2 ) = \Psi(A) \,.
\end{align*}
Hence, %when $V_i$ is independent of $\{V_k\in k\in {A}\}$ we have
%%\beq
%%\label{eq:sample_var}
%%\hat{\sigma}^2_n (i, A)  := \frac{1}{n} \sum_{j \in [n]}  U_{ij}^2 U_{Aj}^2 
%%\eeq
\begin{align}
	\widehat{\sigma}^2_n (i, A)  
		&:= \frac{1}{n}\sum_{j\in [n]} U_{Aj}^2 \label{eq:sample_var}
\end{align}
is an unbiased estimator of the variance of $\sqrt{n} \, \widehat{\psi}_n (i, A)$.
We note that %, if
%$V_i$ is independent of $\{ V_k : k \in A \}$, then 
another unbiased estimator of the 
variance of $\sqrt{n}\,\widehat{\psi}_n(i, A)$ is
\begin{align*}
	\widehat{s}^2_n(i, A) 
		&:= \frac{1}{n}\sum_{j\in [n]} U^2_{ij}U^2_{Aj}\,.
\end{align*} 
Both $\widehat{\sigma}^2_n (i, A)$ and 
$\widehat{s}^2_n(i, A)$
converge in probability to $\Psi(A)$, 
under the same assumptions used below
in Lemma~\ref{lem:sigma}.
However, $\widehat{\sigma}^2_n(i, A)$ is 
more efficient computationally\footnote{This is 
because $\widehat{\sigma}^2_n(i, A)$ can be calculated and stored outside of 
the search procedure. The key 
difference is the lack of dependence on $i$ in the statistic
$\widehat{\sigma}^2_n(i, A)$, which makes calculations more efficient
as the set $A$ is updated during the search procedure 
(see Section~\ref{sec:lamb}).} 
than $\widehat{s}^2_n(i, A)$,
so it is the estimator used in the LAMB method
when calculating approximate p-values.

We are now ready to formally state the central limit theorem for latent association.
This establishes the approximate p-values used in the
LAMB method's iterative testing based search procedure, 
as described in Sections~\ref{sec:test} and~\ref{sec:lamb}.
%%The proof follows from two lemmas 
%%that are proved in Appendix~\ref{app:theory}.

\begin{theorem}{\bf(CLT for Latent Association)}
\label{thm:csm_clt}
Fix a feature index $i\in \mathbb{N}$ and, for each $n\in \mathbb{N}$, 
let $A_n \subset [d_n]\backslash \{i\}$ be a set of index features, %with cardinality $m_n :=|A_n|$.
let $\widehat{\psi}_n (i, A_n)$ be the sample latent association of $i$ and $A_n$ under 
the threshold model for $(\nu_n, \mu_n)$ 
(expressed as $(\bm{\theta}, \bm{V})$ in Section~\ref{sec: model}),
and let $\widehat{\sigma}^2_n (i, A_n)$ be defined as in \eqref{eq:sample_var}.
Assume:
\begin{enumerate}
	\item For each $n\in \mathbb{N}$, $V_i$ is independent of $\{ V_k : k \in A_n \}$ under $\mu_n$;
	\item $\frac{1}{\Psi(A_n)^2 |A_n|} \sum_{k \in A_n} \mathbb{E} \left(\frac{1}{\theta_{i1}(1-\theta_{i1}) \theta_{k1}(1-\theta_{k1})}\right)
	  		= o(n)$.
\end{enumerate}
Then,
\begin{align}
\label{eq:clt2}
 	\frac{\sqrt{n}\, \widehat{\psi}_n (i, A_n)} {\widehat{\sigma}_n(i, A_n)}
		&\overset{d}{\longrightarrow}  \mathcal{N}(0,1) \quad\textnormal{as  $n \to \infty$} .
\end{align}
\end{theorem}

\begin{proof}
The proof follows
from Slutsky's lemma %\citep[Chapter~2]{vaartAsymptoticStatistics-1998} 
with Lemmas~\ref{lem:csm_clt} and \ref{lem:sigma}.
\end{proof}

\begin{lemma}
\label{lem:csm_clt}
Under the notation and assumptions of Theorem~\ref{thm:csm_clt},
\begin{align}
	\label{eq:clt}
	\frac{\sqrt{n} \, \widehat{\psi}_n (i, A_n)}{\sqrt{\Psi(A_n)}}
		&\overset{d}{\longrightarrow} \mathcal{N}(0,1) \quad\textnormal{as  $n \to \infty$.}
\end{align}
\end{lemma}

\begin{proof}
Note that, under the assumptions on $V_i$ and $\{V_k\colon k\in A_n\}$, 
$\widehat{\psi}_n(i, A_n)$  is the %\, = \, n^{-1} \sum_{j = 1}^n U_{ij} \, U_{A_n j}$ is the
average of %i.i.d.\ random variables distributed as $U_j  U_{A_n}$, where $U_j$ and$U_{A_n}$ are defined in (\ref{eq:u}) and (\ref{eq:v}), respectively.
independent random variables with mean zero (see Lemma~\ref{lem:uv}).
It suffices to verify the Lindeberg condition
for the Lindeberg Central Limit Theorem. %\citep[Chapter~3]{durrettProbability-2019}.
The Lindeberg condition requires that, for all $\epsilon > 0$,
\begin{align*}
	\lim_{n \to \infty} \frac{1}{ n } \sum_{j \in [n]} 
		 \mathbb{E}\left( \frac{ U_{ij}^2 \, U_{A_n j}^2 }{ \Psi(A_n) } \, \mathbb{I}\left[ |U_{ij} \, U_{A_n j}| > \epsilon \, \sqrt{n \, \Psi(A_n)} \right] \right)=  0  \, .
\end{align*}
After applying the Cauchy-Schwarz inequality, it is sufficient to show that
\begin{align}
	\label{eq:cs}
	\lim_{n \to \infty} \frac{ \Ex (U_{i1}^4  \, U_{A_n 1}^4) }{ \Psi(A_n)^2 }
		\, \mathbb{P}\left( |U_{i 1} \, U_{A_n 1}| > \epsilon \, \sqrt{n \, \Psi(A_n)} \right) 
		&= 0 \quad\textnormal{for all $\epsilon > 0$.}
\end{align}
By Markov's inequality and Lemma \ref{lem:uv}, we have
\begin{align}
	\mathbb{P}\left( |U_{i1} \, U_{A_n 1}| > \epsilon \, \sqrt{n \, \Psi(A_n)}\right)
		&\leq \frac{ \mathbb{E}(U_{i1}^2 \, U_{A_n 1}^2) }{n \, \epsilon^2 \, \Psi(A_n)}
			=  \frac{\Psi(A_n)}{n \, \epsilon^{2} \Psi(A_n)}
			=  \frac{1}{n \, \epsilon^{2}} \,.
	\label{app eq: key markov}
\end{align}
Lemma~\ref{lem:v4}, Inequality~\eqref{app eq: key markov}, 
and the second assumption 
of Theorem~\ref{thm:csm_clt} ensure that 
Equation~\eqref{eq:cs} holds,
and therefore the Lindeberg condition holds.
\end{proof}

\begin{lemma}
\label{lem:sigma}
Under the notation and assumptions of Theorem~\ref{thm:csm_clt},
\begin{align}
	\label{eq:var_consis}
		\left| \frac{\widehat{\sigma}_n^2(i, A_n) }{\Psi(A_n)} - 1 \right| \overset{p}{\longrightarrow} 0 \quad\textnormal{as $n\to \infty$}.
\end{align}
\end{lemma}

\begin{remark}
A similar result holds for the estimator $\widehat{s}_n^2(i, A_n)$.
\end{remark}

\begin{proof}
To reduce notation, let us denote $\widehat{\sigma}_n^2 := \widehat{\sigma}_n^2(i, A_n)$.
%%Note that $\Ex( \widehat{\sigma}_n^2) = \Ex (U_{i1}^2 \, U_{A_n 1}^2) = \Psi(A_n)$ by Lemma \ref{lem:uv},
%%so (\ref{eq:var_consis}) is equivalent to%% old version
Note that $\mathbb{E}( \widehat{\sigma}_n^2) = \mathbb{E} (U_{A_n 1}^2) = \Psi(A_n)$ 
by Lemma~\ref{lem:uv}.
So \eqref{eq:var_consis} is equivalent to
\begin{align*}
	\left| \frac{\widehat{\sigma}_n^2  -  \mathbb{E}( \widehat{\sigma}_n^2)}{\Psi(A_n)} \right| 
		\overset{p}{\longrightarrow} 0 
		\quad\textnormal{as $n\to\infty$}.
\end{align*}
By Chebyshev's inequality, it suffices to show that
$\frac{\textnormal{Var}(\widehat{\sigma}^2_n )}{\Psi(A_n)^{2}} = o(1)$.  
Using independence and 
Lemmas~\ref{lem:uv} and \ref{lem:v4}, it is clear that
%%\[
%%\textnormal{Var}( \widehat{\sigma}^2_n )
%%\, = \,
%%\frac{1}{n}  \textnormal{Var} (U_{i1}^2 \, U_{A_n 1}^2)
%%\, = \, \frac{1}{n} \left( \, \Ex (U_{i1}^4 \, U_{A_n 1}^4 ) - \Psi(A_n)^2 \right) ,
%%\]
\begin{align*}
	\textnormal{Var}( \widehat{\sigma}^2_n )
		&= \frac{1}{n} \textnormal{Var}(U_{A_n1}^2)
%			\leq \frac{1}{n} \left[ \frac{1}{|A_n|} \sum_{k\in A_n} \mathbb{E}(U_{k1}^4) - \Psi(A_n)^2\right]
			\leq \frac{1}{n} \left( \frac{1}{|A_n|} \sum_{k\in A_n} \mathbb{E}\left(\frac{1}{\theta_{k1} (1- \theta_{k1})} \right) - \Psi(A_n)^2\right)\,.
\end{align*}
%%where we use %Jensen's inequality on $\mathbb{E}(U^4_{A1})$ for the first inequality and 
%%Lemma \ref{lem:v4} for the %second 
%%inequality.
Therefore, it is enough to show that 
\begin{align*}
	\frac{1}{|A_n| \, \Psi(A_n)^2} \sum_{k\in A_n} \mathbb{E}\left(\frac{1}{\theta_{k1} (1- \theta_{k1})} \right) = o(n)\,,
\end{align*}
which follows from the second assumption 
of Theorem~\ref{thm:csm_clt}
combined with the monotonicity of expectation.
%%, because
%%$\theta_{i1}(1 - \theta_{i1}) \in (0,1)$ almost surely implies that 
%%$1 \leq \theta_{i1}^{-1}(1 - \theta_{i1})^{-1}$ almost surely.
\end{proof}

Finally, note that the sample quantities $\widehat{\psi}_n(i,k)$ 
and  $\widehat{\psi}_n(i,A)$ are not guaranteed 
to fall between -1 and 1.
Under mild conditions, however, their 
values will converge to the interval $[-1,1]$ 
as $n$ tends to infinity. %We now state this formally.

\begin{prop}
\label{prop:u_consist}
If $\max_{i \in [d_n]} \mathbb{E} \left(\frac{1}{\theta_{i1} (1 - \theta_{i1})}\right) = o(n)$
then, for any $\epsilon > 0$,
\begin{align*}
	\max_{i \in [d_n]} \max_{A \subset [d_n]} \mathbb{P}\left( | \widehat{\psi}_n(i, A) | > 1 + \epsilon \right)  
		&\longrightarrow 0 \qquad \textnormal{ as $n \to \infty$.}
\end{align*}
\end{prop}

\begin{proof}
Fix arbitrary $n\in \mathbb{N}$, $i\in [d_n]$, and $A\subset[d_n]$.
Since $\mathbb{E} \left(\widehat{\psi}_n(i, A)\right) = \psi(i, A)$ 
and $|\psi(i, A)| \leq 1$, a routine argument implies that
\begin{align*}
	\mathbb{P}\left( | \widehat{\psi}_n(i, A) | > 1 + \epsilon \right) 
		&\leq \mathbb{P}\left( \left| \widehat{\psi}_n(i, A)  - \mathbb{E} \left(\widehat{\psi}_n(i, A)\right) \right|  > \epsilon \right).
\end{align*}
By Chebyshev's inequality, it suffices to show that $\textnormal{Var}(\widehat{\psi}_n(i, A)) = o(1)$.  
By Jensen's inequality,
$U_{A1}^2 \leq \frac{1}{|A|} \sum_{k \in A} U_{k1}^2$ pointwise, and it follows that
\begin{align*}
	\textnormal{Var}(\widehat{\psi}_n(i, A))
		&= \frac{1}{n} \textnormal{Var}(U_{i1} U_{A1}) 
			\leq \frac{1}{n} \mathbb{E} (U_{i1}^2 U_{A1}^2)
			\leq \frac{1}{|A|} \sum_{k \in A} \frac{ \mathbb{E} (U_{i1}^2 U_{k1}^2) }{n}\,.
\end{align*}
Hence, it is enough to show that $\max_{i,k \in [d_n]} \Ex (U_{i1}^2 U_{k1}^2) = o(n)$. 
By the Cauchy-Schwarz inequality and Lemma~\ref{lem:v4},
\begin{align*}
 \mathbb{E} (U_{i1}^2 U_{k1}^2) 
 	&\leq \sqrt{ \mathbb{E}(U_{i1}^4)\, \mathbb{E}(U_{k1}^4) } 
		\leq \sqrt{\mathbb{E}\left(\frac{1}{\theta_{i1} (1 - \theta_{i1})}\right)\,  \mathbb{E}\left(\frac{1}{\theta_{k1} (1 - \theta_{k1})}\right)}.
\end{align*}
The condition $\max_{i \in [d_n]} \mathbb{E}\left(\frac{1}{\theta_{i1}(1 - \theta_{i1})}\right) = o(n)$ completes the proof.
\end{proof}

\newpage

\section{Theory for Estimating the Random Thresholds}
\label{sec: est thresholds}

The hypothesis testing approach outlined in 
Section~\ref{sec:test} and Appendix~\ref{sec: CLT}
requires that the random thresholds 
$\mathbf{\Theta} = [\bm{\theta}_{\cdot 1}, \ldots, \bm{\theta}_{\cdot n}]$ 
be known alongside the observed $\mathbb{X} = [X_{ij}]
\in \{0, 1\}^{d \times n}$. %= [\bm{X}_{\cdot 1}, \ldots, \bm{X}_{\cdot n}]$.  
In the threshold model of Definition~\ref{def:mod}, however,
$\bm{\theta}_{\cdot 1}, \ldots, \bm{\theta}_{\cdot n}$ are latent vectors.
In this appendix we derive consistent estimators of %$\mathbf{\Theta}$ 
$\bm{\theta}_{\cdot 1}, \ldots, \bm{\theta}_{\cdot n}$
from the data matrix $\mathbb{X}$. 
This implies that the inference problem
outlined in Section~\ref{sec:tildes} is well-posed, and 
so the approximate solution of the corresponding optimization procedure
is meaningful.  
As discussed throughout Section~\ref{sec: search procedure}, 
we plug in the estimates $\widehat{\bm{\theta}}_{\cdot 1}, \ldots, \widehat{\bm{\theta}}_{\cdot n}$
directly into the sample latent association defined in
Equation~\eqref{eq:sam_coh}.  
The successful results obtained from 
applying the LAMB method 
to both the artificial and real datasets discussed in this paper, 
as well as the asymptotic consistency of 
$\widehat{\bm{\theta}}_{\cdot 1}, \ldots, \widehat{\bm{\theta}}_{\cdot n}$
discussed below, 
suggests that this is a reasonable procedure.

Recall that, under the truncated Poisson factorization model (TPFM)
for the random threshold, 
$\bm{\theta}_{\cdot 1}, \ldots, \bm{\theta}_{\cdot n}$ 
depend on a random sample $\tau_1, \ldots, \tau_n$ with density $\pi$
and deterministic parameters $1, \alpha_2, \cdots, \alpha_d$
(see Equation~\ref{eq:theta} in Definition~\ref{def:mod}
and Proposition~\ref{prop: identifiable}).\footnote{The 
TPFM resembles a stochastic ``rank-one approximation'' of 
$\mathbf{\Theta}$.}
Therefore, the number of quantities to estimate is reduced from 
$d \cdot n$ realized values of $\{ \theta_{ij} \}$ to $d + n - 1$ parameters 
$(1, \alpha_2, \ldots, \alpha_d, \tau_1^0, \ldots, \tau^0_n)$,
where $\tau_j^0$ is a realized value of $\tau_j\sim \pi$ for $j\in [n]$. 
Reducing the effective dimension of $\mathbf{\Theta}$
from $d\cdot n$ to $d + n - 1$ %parameters
allows us to efficiently estimate $\mathbf{\Theta}$ from $\mathbb{X}\in \{0, 1\}^{d \times n}$
under mild conditions.

%U, described below,
%we can efficiently estimate $\mathbf{\Theta}$ from $\mathbb{X}$. %,
%which we consider in Section~\ref{subsec: parameter consistency}. %\ref{sec:params}. 
%%In particular, since in many practical cases \blue{most} $\theta_{ij}$'s are small, 
%%we focus on the case when their expected values tend towards 0.

%As discussed earlier, for a given $j\in [n]$ and all $i\in [d]$,
%the TPFM defines $\theta_{ij}$ marginally as a function of 
%a positive-valued random variable $\tau_j$
%and a fixed positive-valued rate parameter $\alpha_i$.
%The parameter $\tau_j$ captures  a notion of resource or budget
%for sample $j$, while the parameter 
%$\alpha_i$ captures an intrinsic popularity or utility  of feature $i$
%for all samples in the population.
%This assumption for the random thresholds helps us model 
%heterogeneity in samples. %, i.e.,
%drop the assumption of identically distributed samples.
%%In other words, we assume that the dependence structure and 
%%randomness of $\bm{\theta}_{\cdot j}$ derives from a single shared random variable $\tau_j$.  
%%Differences between the marginal distributions of $\theta_{1j}, \ldots, \theta_{dj}$ 
%%are then entirely captured by the fixed parameters $\bm{\alpha} = (\alpha_1, \ldots, \alpha_d)$.

\subsection{Asymptotic Consistency}
\label{subsec: parameter consistency}

Throughout this appendix we consider the asymptotic 
setting as in Appendix~\ref{sec: CLT} for 
the threshold model of Definition~\ref{def:mod} 
and Equation~\eqref{eq:theta}, i.e., 
$\bm{\theta}_{\cdot 1}, \ldots, \bm{\theta}_{\cdot n}\overset{\textnormal{iid}}{\sim} \nu_n$.
Denote by $\pi$ the density of $\tau_1, \ldots, \tau_n$ inherited
from the distribution $\nu_n$ for all $n\in \mathbb{N}$.
First, we consider the consistency in estimating the expected value.
Note that, for each $j\in [n]$,
\begin{align*}
	\mathbb{E}(X_{ij}) 
		&= \mathbb{E}(\mathbb{E}[X_{ij} \given \bm{\theta}_{\cdot j}])
			= \mathbb{E}(\theta_{ij})
			= \mathbb{E}(\theta_{i1})\,.
\end{align*} 
Therefore, unconditionally for each $i\in[d_n]$, 
$X_{i1}, \ldots, X_{in} \overset{\textnormal{iid}}{\sim} \textnormal{Bern}(\mathbb{E}(\theta_{i 1}))$,
and a natural estimate of $\mathbb{E}(\theta_{i1})$ is  $ \overline{X}_{i} := \frac{1}{n} \sum_{j=1}^n X_{ij}$. 
We now present the following proposition on the consistency of $\overline{X}_{i}$ when 
the expected values $\{\mathbb{E}(\theta_{i1})\}_{i\in[d_n]}$ are not too small.

\begin{prop}
\label{prop:theta_mean}
Suppose we have the following condition on $\{\mathbb{E}(\theta_{i1})\}_{i\in[d_n]}$
for $n\in \mathbb{N}$:
\begin{align*}
	\varlimsup_{n \rightarrow \infty} \frac{1}{n} \max_{i \in [d_n]} \frac{1}{\mathbb{E}(\theta_{i1})}
		&= 0\,.
\end{align*} 
Then we have the following convergence:
 \begin{align*}
  	\max_{i\in [d_n]} \left| \frac{\overline{X}_{i} }{ \mathbb{E}(\theta_{i1}) } - 1 \right|  
		\stackrel{p}{\longrightarrow} 0
			\quad\textnormal{as $n\to\infty$.}
\end{align*}
\end{prop}

\begin{remark}
The condition in Proposition~\ref{prop:theta_mean}
trivially holds when $\{\mathbb{E}(\theta_{i1})\}_{i\in[d_n]}$ is uniformly bounded from below.
\end{remark}

\begin{proof}
Fix arbitrary $n\in\mathbb{N}$ and $\epsilon > 0$.
Note that $\mathbb{E}(X_{ij}^2 \given \bm{\theta}_{\cdot j}) = \theta_{ij}$,
and so, unconditionally, $\mathbb{E}(X_{ij}^2) = \mathbb{E}(\theta_{ij})$
for each $j\in [n]$.
Then, by Chebyshev's inequality and the condition on $\{\mathbb{E}(\theta_{i1})\}_{i\in[d_n]}$, 
we have
\begin{align*}
	\max_{ i \in [d_n]} 
	\mathbb{P}\left(\left| \frac{\overline{X}_{i} }{ \mathbb{E}(\theta_{i1}) } - 1 \right|  \geq \epsilon \right) 
		&\leq \frac{1}{n \, \epsilon^2} \max_{ i \in [d_n]}  \left[\frac{1}{\mathbb{E}(\theta_{i1})} -1\right]
			= o(1) \,.
\end{align*}
\end{proof}

Suppose that the condition on $\{\mathbb{E}(\theta_{i1})\}_{i\in[d_n]}$ 
in Proposition~\ref{prop:theta_mean} is satisfied.
An immediate consequence of Proposition~\ref{prop:theta_mean} 
is the estimation of $\alpha_{i}$
for  $i\in [d_n]$. From Equation~\eqref{eq:theta} we see that 
\begin{align*}
	\mathbb{E}(\theta_{i1}) 
		&= h(\alpha_i)  :=   \int_\mathcal{T} \left( 1 - e^{- t \alpha_{i}} \right)  \pi(t) \,\textnormal{d}t,
\end{align*} 
where $\mathcal{T}\subset (0, \infty)$ is the support corresponding to the density $\pi$.
If the density $\pi$ leads to a function $h$ that is %continuous and 
invertible and $\mathbb{E}(\theta_{i1})\in C_{h^{-1}}:= \{x: h^{-1} \textnormal{ is continuous at } x\}$, 
then the Continuous Mapping Theorem %\citep[Chapter~3]{durrettProbability-2019} 
implies that $h^{-1} (\overline{X}_{i})$ is 
consistent for $\alpha_i$. %$h^{-1} (\mathbb{E}(\theta_i))=\alpha_{i}$

To completely estimate 
$\mathbf{\Theta} = [\bm{\theta}_{\cdot 1}, \ldots, \bm{\theta}_{\cdot n}]$ 
we must also estimate 
the realized values of the random variables 
$\tau_1, \ldots, \tau_n\overset{\textnormal{iid}}{\sim} \pi$,
which we denote by $\tau_1^0, \ldots, \tau_n^0$.  
Consider the posterior density of 
$\tau\sim \pi$ given $\bm{X}_{\cdot j}$ %= \bm{x}_{\cdot j}:=(x_{1j}, \ldots, x_{dj})^T$ 
and $\bm{\alpha}_n := (\alpha_1^n,\ldots, \alpha_{d_n}^n)^T$, 
which we denote by $\pi(\cdot \given \bm{X}_{\cdot j}, \bm{\alpha}_n)$.  
A natural estimator for $\tau_j^0$ is the posterior mean
\begin{align}
	\mathbb{E}( \tau_j \given \bm{X}_{\cdot j}, \bm{\alpha}_n )
		&=  \int_{\mathcal{T}}  t \,\pi(t \given \bm{X}_{\cdot j}, \bm{\alpha}_n) \, \textnormal{d}t  \, .
\end{align}
The following result guarantees consistency of the posterior mean.  We appeal directly to Theorem~4.1 of \citet{choiPosteriorConsistency-2008}, which requires the following condition on the prior $\pi$ for $\tau$.

\begin{condition}   
\label{cond: tau prior}
For all $\delta>0$ there exist sets $\{S_k\}_{k\in\mathbb{N}}$ 
such that the diameter of each set is less than 
$\delta$, $\cup_{k \in \mathbb{N}} S_k = \mathcal{T}\subseteq (0, \infty)$, 
and $\sum_{k \in\mathbb{N}}{\sqrt{\mathbb{P}_{\pi}(S_k)}} < \infty$.
\end{condition}

\begin{remark} 
In essence, Condition~\ref{cond: tau prior} is a 
concentration condition, guaranteeing 
that the measure $\mathbb{P}_{\pi}$ is not too 
spread out over the support $\mathcal{T}$ of $\tau$.
\end{remark}

\begin{theorem}{\bf (\citet{choiPosteriorConsistency-2008})}
\label{thm:taus}
    Suppose that Condition~\ref{cond: tau prior} holds and that  
    $\mathrm{\pi}(\cdot \given \bm{X}_{\cdot j}, \bm{\alpha}_n)$ is bounded. Then, for every $\epsilon > 0$,
    \begin{align*}
    	\mathbb{P}\left( \left| \mathbb{E}[ \tau_j \given \bm{X}_{\cdot j}, \bm{\alpha}_n ] - \tau_j^0 \right| \geq \epsilon \right) \to 0
    \end{align*}
    as $n \to \infty$.
%%    The probability is taken over 
%%    the distribution of $(\bm{X}_n, \bm{\theta}_n)$ derived
%%    from $(\bm{\theta}_n, \bm{V}_n)$.%%should this be $(\bm{X}_{\infty}, \bm{\theta}_{\infty}$ derived from $\nu_{\infty} \otimes\mu_{\infty}$? -CM
\end{theorem}

In practice $\bm{\alpha}_n$ is not known, so we instead estimate 
$\tau_j^0$ by plugging in consistent estimates 
$(\widehat{\alpha}^n_{1}, \ldots, \widehat{\alpha}^n_{d_n} )$, i.e.,
\begin{align}
	\label{eq:tau}
	\widehat{\tau}_j^n
		&:= \mathbb{E}[ \tau_j \given \bm{X}_{\cdot j}, \widehat{\bm{\alpha}}_n ]
			=  \int_{\mathcal{T}}  t\, \pi(t \given \bm{X}_{\cdot j}, \widehat{\bm{\alpha}}_n)\, \textnormal{d}t  \, .
\end{align}
Then, for all $n\in\mathbb{N}$, $j \in [n]$, and $i \in [d_n]$, 
$\theta_{ij}$ is estimated 
by $\widehat{\theta}_{ij}(n) := 1-\exp{(-\widehat{\alpha}_{i}^n \,\widehat{\tau}_j^n )}$.

The following proposition provides a general family 
of models which satisfy the conditions in 
Theorem~\ref{thm:csm_clt} and Proposition~\ref{prop:theta_mean}.
\begin{prop}
\label{prop:prior}
Consider the prior distribution Gamma($\zeta$, $\beta$) for $\tau$, with density
\begin{align*}
	\pi(t) %:= f_\tau(t) 
		&:= \frac{\beta^\zeta}{\Gamma(\zeta)} t^{\zeta -1} e^{-\beta \,t}.
\end{align*}
As in Theorem~\ref{thm:csm_clt},
fix $i \in [d_n]$ and for each $n\in \mathbb{N}$ let $A_n \subset [d_n]\backslash \{i\}$ 
be an index set with cardinality $m_n := |A_n|$. Assume the following:
\begin{enumerate}
  \item $\min_{i \in [d_n]} n\,\alpha_i \rightarrow \infty$ as $n\to \infty$.
  \item $\max_{i \in [d_n]} \alpha_i \leq M\in (0, \infty)$ for all $n\in \mathbb{N}$.
  \item If we denote $c_n := \min_{i \in [d_n]}\alpha_i$ 
  	and $\rho_n := \lambda_{\min} \left(\Sigma(A_n)\right)$,
	the minimal eigenvalue of the matrix
	$\Sigma(A_n) := [\psi(l,k)]_{l,k \in A_n}$,
	then $\frac{m_n^2}{\rho_n^2 \, c_n^2 } = o(n)$.
\end{enumerate}
Then the Gamma($\zeta$, $\beta$) prior with $\beta>6M$ and $\zeta > 3$ satisfies:
\begin{enumerate}

	  \item $\varlimsup_{n \rightarrow \infty}\frac{1}{n} \max_{i \in [d_n]} \frac{1}{\mathbb{E}(\theta_{i1})} =0$ 
  		as in Proposition~\ref{prop:theta_mean}.
		
 	 \item $\frac{1}{\Psi(A_n)^2 |A_n|} \sum_{k \in A_n}\mathbb{E}\left(\frac{1}{\theta_{i1}  (1-\theta_{i1}) \theta_{k1}(1-\theta_{k1})}\right) = o(n)$ as in Theorem~\ref{thm:csm_clt}.
  
\end{enumerate}
\end{prop}

%%\begin{remark}  
%%\red{need to double check this because we made a change to assumption 3:}
%%The three assumptions in Proposition~\ref{prop:prior} are mild for practical use. 
%%For example, if $\rho_n =O(1)$ and $c_n>> n^{-\frac{1}{4}}$, then the assumptions
%%hold when $m_n << n^{\frac{1}{4}}$.
%%\end{remark}

\begin{proof}
Fix arbitrary $n\in \mathbb{N}$ and $k\in [d_n]$.
First recall that $\theta_{k1} = 1-\exp{\left(- \alpha_k\, \tau_1\right)}$. 
Then, since $\alpha_k > 0$, the corresponding moment-generating
function implies that
\begin{align*}
	\mathbb{E}(\theta_{k1}) 
		&= 1-\left(\frac{\beta}{\beta+\alpha_k}\right)^\zeta\geq 
			1-\left(\frac{\beta}{\beta+\alpha_k}\right) = \frac{\alpha_k}{\beta+\alpha_k}\,.
\end{align*}
Therefore, we have the following inequality:
\begin{align*}
	\frac{1}{n} \max_{i \in [d_n]} \frac{1}{\mathbb{E}(\theta_{i1}) }
		&\leq \frac{1}{n}+\frac{\beta}{\min_{i \in [d_n] }n\, \alpha_j} \,.
\end{align*}
Hence, $\varlimsup_{n \rightarrow \infty}\frac{1}{n} \max_{i \in [d_n]} \frac{1}{\mathbb{E}(\theta_{i1})} =0$
because we assumed that $\min_{i \in [d_n]} n\,\alpha_i \rightarrow \infty$ as $n\to \infty$.

Now notice that
$\Psi(A_n) = \frac{1}{m_n^2} \mathbf{1}^T\Sigma(A_n)\mathbf{1}$. 
Therefore, by our notation and the min-max theorem for symmetric matrices, we have
\begin{align}
	\label{app eq: Psi eigenvalue inequality}
	\Psi(A_n) 
		&= \frac{1}{m_n} \left(\frac{\mathbf{1}}{\sqrt{m_n}}\right)^T\Sigma(A_n)\left(\frac{\mathbf{1}}{\sqrt{m_n}}\right) \geq \frac{\rho_n}{m_n} \,.
\end{align}
Fix an arbitrary $k\in A_n$.
Holder's inequality implies that
\begin{align}
	\label{app eq: theta holders ineq}
	\mathbb{E}\left(\frac{1}{\theta_{i1}\, \theta_{k1}\, (1-\theta_{i1})\, (1-\theta_{k1})}\right)
		&\leq \left[\mathbb{E}\left(\frac{1}{\theta_{i1}^3}\right)
				\, \mathbb{E}\left(\frac{1}{\theta_{k1}^3}\right)
				\, \mathbb{E}\left(\frac{1}{(1-\theta_{i1})^3 (1-\theta_{k1})^3}\right) \right]^{\frac{1}{3}} \,.
\end{align}
The moment generating function of the Gamma($\zeta$, $\beta$)
distribution and the assumption that $\beta > 6M$
further implies that
\begin{align}
 	\mathbb{E}\left(\frac{1}{(1-\theta_{i1})^3 (1-\theta_{k1})^3}\right)
		&= \mathbb{E}\left(e^{3(\alpha_i+\alpha_k)\tau_1}\right) \nonumber \\
		&= \left(1-\frac{3(\alpha_i+\alpha_k)}{\beta}\right)^{-\zeta} 
			\leq \left(1-\frac{6M}{\beta}\right)^{-\zeta} \,. \label{app eq: mgf theta cube}
\end{align}
Now note that $\frac{1}{1-e^{-t}}$
is a decreasing function of $t$  and $1- e^{-t} \geq t - \frac{t^2}{2}$ when $t > 0$.
Therefore,
\begin{align}
	\mathbb{E}\left(\frac{1}{\theta_{i1}^3}\right) 
		&= \int_{t>0} \left(\frac{1}{1-\exp(-\alpha_i t)}\right)^3 \,\pi(t) \,\textnormal{d}t \nonumber\\
		&\leq \left(\frac{1}{1- e^{-\frac{1}{2}}}\right)^3 + 
			\int_0^{\frac{1}{2\,\alpha_i}} \left(\frac{1}{1-\exp(-\alpha_i t)}\right)^3 \,\pi(t) \,\textnormal{d}t \nonumber\\
		&\leq \left(\frac{1}{1- e^{-\frac{1}{2}}}\right)^3 + 
			\int_0^{\frac{1}{2\,\alpha_i}} \left(\frac{1}{\alpha_i\,t - \frac{\alpha_i^2\,t^2}{2}}\right)^3 \,\pi(t) \,\textnormal{d}t\,. \label{app eq: rhs sum upper bnd}
\end{align}
For the second term in Equation~\eqref{app eq: rhs sum upper bnd},
there exists a constant $C(\beta,\zeta) \in (0, \infty)$ such that
\begin{align}
	\int_0^{\frac{1}{2\,\alpha_i}} \left(\frac{1}{\alpha_i\,t - \frac{\alpha_i^2\,t^2}{2}}\right)^3 \,\pi(t) \,\textnormal{d}t
		&= \int_0^{\frac{1}{2\,\alpha_i}} \left(\frac{1}{\alpha_i\,t\left[1 - \frac{\alpha_i\,t}{2}\right]}\right)^3 
			\,\pi(t) \,\textnormal{d}t \nonumber \\
		&\leq \frac{64}{27} \int_0^{\frac{1}{2\,\alpha_i}} \left(\frac{1}{\alpha_i\,t}\right)^3 
			\,\pi(t) \,\textnormal{d}t \nonumber \\
		&= \frac{64}{27\, \alpha_i^3} \frac{\beta^\zeta}{\Gamma(\zeta)} 
			\int_0^{\frac{1}{2\,\alpha_i}} t^{\zeta - 4}\, e^{-\beta t}\,\textnormal{d}t \nonumber \\
		&\leq \frac{C(\beta,\zeta)}{c_n^3}\,, \label{app eq: rhs sum second term}
\end{align}
where we use the monotonicity of integration in the first 
inequality and the assumption that $\zeta > 3$ in the second inequality.

Mutatis mutandis, Equations~\eqref{app eq: theta holders ineq}, 
\eqref{app eq: mgf theta cube}, 
\eqref{app eq: rhs sum upper bnd}, and
\eqref{app eq: rhs sum second term} imply that there exists a 
constant $\bar{C}(\beta, \zeta) \in (0, \infty)$ that is independent 
of $i$ and $k$ such that
\begin{align}
	\mathbb{E}\left(\frac{1}{\theta_{i1}\, \theta_{k1}\, (1-\theta_{i1})\, (1-\theta_{k1})}\right)
		&\leq \frac{\bar{C}(\beta, \zeta)}{c_n^2} \,. \label{app eq: key gamma upper bnd}
\end{align}
Combining Equations~\eqref{app eq: Psi eigenvalue inequality}
and \eqref{app eq: key gamma upper bnd} yields
\begin{align*}
	\frac{1}{\Psi(A_n)^2 \, |A_n|} \sum_{k \in A_n} 
		\mathbb{E}\left(\frac{1}{\theta_{i1}\, \theta_{k1}\, (1-\theta_{i1})\, (1-\theta_{k1})}\right)
		&\leq  \frac{m_n^2}{\rho_n^2 \, c_n^2 } \, \bar{C}(\beta, \zeta)\,.
\end{align*}
Finally, using the assumption $\frac{m_n^2}{\rho_n^2 \, c_n^2 } = o(n)$,
we have shown that 
\begin{align*}
	\frac{1}{\Psi(A_n)^2 \, |A_n|} \sum_{k \in A_n} 
		\mathbb{E}\left(\frac{1}{\theta_{i1}\, \theta_{k1}\, (1-\theta_{i1})\, (1-\theta_{k1})}\right)
		&= o(n)\,.
\end{align*}
%\qed
\end{proof}

\newpage

\section{Additional LAMB Method Details}
\label{app: additional lamb method details}

The LAMB method is presented in
Section~\ref{sec: search procedure}.
In this appendix we will discuss additional 
details relevant to using the LAMB method in practice.

\subsection{Full LAMB Algorithm}
\label{sec: full LAMB algorithm}

%%We can now write out the full LAMB method
%%using pseudocode. As noted above, 
%%in Appendix~\ref{app: additional lamb method details}
%%we give additional details for using the LAMB method.
%%The details in Appendix~\ref{app: additional lamb method details}
%%complement the analysis done on 
%%artificial and real datasets in Sections~\ref{sec: sim study}
%%and \ref{sec: applications}.
%%%%Post-processing estimated coherent sets is 
%%%%also discussed in Appendix~\ref{app: search procedure details}.

 \begin{framed}
 \begin{enumerate}
 	\item {\sc Given:} Binary data matrix $\mathbb{X}\in \{0, 1\}^{d\times n}$ and 
		parameters $\delta \in (0,1)$ and $N\geq 2$.
 	\item {\sc Estimation (Algorithm~\ref{alg: est rand thresh} in Section \ref{sec:tildes}):} 
			Compute $\widehat{\mathbf{\Theta}}$, 
			a matrix of estimates of the random thresholds
			$\mathbf{\Theta} = [\theta_{ij}]$,
			from $\mathbb{X}$.
	\item {\sc Initialization:} Initialize the set  $A^0 := \{ i \}$ for some $i \in [d]$.
	\item {\sc Search Procedure (Algorithm~\ref{alg: lamb search proc} in Section~\ref{sec:lamb}):}
		\begin{itemize}
		\renewcommand\labelitemi{$\rhd$}
			\item Given $A^t$ and notation $A^t_{-k}:= A^t\setminus\{k\}$, 
			compute $\widehat{\psi}(k, A^t_{-k})$ and 
			$\hat{\sigma}(k, A^t_{-k})$ from $\mathbb{X}$ and $\widehat{\mathbf{\Theta}}$, 
			for each $k \in [d]$,
			as in Section~\ref{sec:test} and Appendix~\ref{sec: CLT}.
			\item Compute approximate p-values 
				$\{ \textnormal{pv}(1, A^t), \ldots, \textnormal{pv}(d, A^t) \}$ 
				as in Equation~\eqref{eq:pvs} in Section~\ref{sec:test}.
			\item Simultaneously test the hypotheses 
				\begin{align*}
					\textnormal{H}_0(k, A^t): \psi(k, A^t_{-k}) \leq 0 
						\quad\textnormal{vs}\quad 
						\textnormal{H}_1(k, A^t): \psi(k, A^t_{-k}) > 0
						\quad \textnormal{for each $k\in [d]$,}
			\end{align*}
			by applying the Benjamini-Yekutieli 
			multiple testing procedure at level $\delta$ to 
			the set of approximate p-values.
		\end{itemize}
	\item {\sc Update: }  Define $A^{t + 1} := \{  k \in[d] \colon \textnormal{H}_0(k, A^t) \, \textnormal{was rejected} \}$.
	\item {\sc Iteration: }  Repeat steps 4 and 5 until $A^{t} = A^{t'}$ for some $t' < t$
					or a maximum number of iterations is reached.
	\item {\sc Output:} Output $A^* := A^{t}$ if $|A^*| \geq N$ and $t = t'+1$. 
					In this case $A^*$ is a nonempty fixed point and an estimated coherent set.
	\item {\sc Repetition: }  Repeat 3-7 as many times as desired, for a given 
		subset of indices $\mathcal{I}\subset [d]$, or for every $i \in [d]$,
		or for a given class of features $\mathcal{C} \subset 2^{[d]}$.
	 \end{enumerate}
 \end{framed}
 
\subsection{Embarrassingly Parallel Search Procedure}
\label{app: lamb parallel search}

The overall time expense of the LAMB method 
comes from the iterative testing based search procedure
(Algorithm~\ref{alg: lamb search proc} in Section~\ref{sec:lamb}).
Unless explicitly specified otherwise, 
the LAMB method's search procedure is run $d$ times, 
for initial sets $\{1\}$, \ldots, $\{d\}$,
and within each search is a for $k\in [d]$ loop. 
This search procedure, however, is embarrassingly parallel
from a computational perspective. 

Before the search procedure is run, a $d\times d$ dimensional
matrix is computed and temporarily stored 
so that the statistics $\widehat{\psi}_n(k, A^t_{-k})$ 
and $\widehat{\sigma}_n^2(k, A^t_{-k})$
can be easily calculated %,
%and consequently the approximate p-values can 
%be easily calculated, 
for an arbitrary set $A^t\subset [d]$ without 
dependence on the step $t$ or initialized set 
$A^0$ (see Section~\ref{sec:lamb}). 
%%by simply using 
%%appropriate submatrices that depend
%%only on the elements of the set $A$.
The variance estimator $\widehat{\sigma}_n^2(k, A^t_{-k})$
is actually independent of the feature $k\in [d]$.
Consequently, it is feasible to 
have multiple searches in parallel
starting from different initialized sets, 
and to calculate the Benjamini-Yekutieli adjusted approximate
p-values $\textnormal{BY}(k, A^t)$ in parallel across the for $k\in [d]$ loop.
%%A future update to  the LAMB method's  code
%%will utilize this parallel search algorithm,
%%and this update will make a dramatic difference in the total computational time.

\subsection{Multiple Testing Procedures}
\label{app: multiple testing procedures}

The set update process in Algorithm~\ref{alg: lamb search proc} 
in Section~\ref{sec:lamb} uses a multiple testing procedure to 
update and refine a set $A^t\subset [d]$ of features. 
In principle, any multiple testing procedure can be applied in this step
of the algorithm.
A Bonferroni adjustment would guarantee 
family wise error control at each step, but would, in many cases, greatly reduce the
sensitivity of the algorithm.  The default implementation of the LAMB 
method uses the FDR-controlling procedure of \citet{yekutieli-2001},
which controls the expected false discovery rate even when the p-values of the hypotheses %$H_0 (j, A^t)$ 
are correlated. The FDR-controlling procedure of
\citet{hochberg-1995} can be easily used instead.
This set update step only controls the expected false discovery rate
\emph{per iteration}.
Currently, we do not have theoretical guarantees 
for controlling the expected false discovery rate
over the entire search procedure.

\subsection{Fixed Points and Cycles}
\label{app: search procedure details}

The LAMB method's search procedure
stops when $A^{t} = A^{t'}$ for some $t>t'$.
If $t = t'+1$, then $A^t$ is a fixed point
of the search procedure and further set updates will not change $A^t$.
As mentioned in Section~\ref{sec:lamb},
nonempty fixed points of the search procedure 
satisfy Definition~\ref{def:coherent_set} 
up to a level of statistical significance,
and they are considered estimated coherent sets.
Nonempty fixed points, however, are not the only 
result from the LAMB method's search 
procedure; nonempty cycles occur in both real and 
artificial data settings.

If $A^{t} = A^{t'}$ for $t' +1 < t$, then the LAMB method's search procedure
has reached a terminating cycle $A^{t'}, \ldots, A^{t}$ of three or more sets.
When the algorithm cycles through three or more sets, the final set 
$A^* := A^t$ is included in the output,
but labeled differently than the fixed points, 
since cycles are not considered estimated coherent sets.  In our experience, 
cycling is rare in artificial datasets, but not in real datasets. 
For example, cycles occur
when the LAMB method is applied to the 
text dataset in Section~\ref{sec: text}, and these cycles 
typically differ from similar fixed points only by a few 
different features,
out of hundreds or thousands of features.

\subsection{Iterative Testing Based Search Procedure is Greedy}
\label{app: lamb greedy search proc}

Proposition~\ref{prop: coh set greedy} below
demonstrates that the set update step 
in the iterative testing based search procedure
is Greedy. 
We only test $\psi(k, A_{-k}^t)$ 
for some $A^t\subset [d]$ and $k \in [d]$ (see Section~\ref{sec:lamb}).
In particular, 
%%we do not test the result of adding 
%%or removing more than one element from 
%%the set of features $A^t$.
by only testing $\psi(k, A^t_{-k})$ 
for a fixed set $A^t\subset [d]$ and 
each $k \in [d]$ we are only considering 
a local change of one element to the set, instead
of a global change of two or more elements.
However, because of the multiple testing procedure
used in the set update step,
it is still possible for two or more features
to be added to $A^t$ or removed from $A^t$
in the updated set $A^{t+1}$.

\begin{prop}
	\label{prop: coh set greedy}
	Let $A\subset [d]$ be a coherent set for latent association.
	Denote the elements $A = \{k_l\colon l\in [q]\}$ for some $q \geq 2$.
	Suppose there exists $\tilde{k}_1, \tilde{k}_2 \in [d]\setminus A$
	such that 
	\begin{align*}
		\max_{j\in [2]}\max_{i\in [d]\setminus\{\tilde{k}_1, \tilde{k}_2\}} &\psi(\tilde{k}_j, i) 
			\leq 0\,,
			\qquad \psi(\tilde{k}_1, \tilde{k}_2)
				> \max_{j\in [2]} \left\{ - \sum_{l\in [q]} \psi(\tilde{k}_j, k_l) \right\}\,,\\
			&\textnormal{and}\quad \min_{\tilde{l}\in [q]} 
				\left\{\sum_{l\in [q]\setminus\{\tilde{l}\}} \psi(k_{\tilde{l}}, k_l) 
					+ \sum_{j\in [2]} \psi(k_{\tilde{l}}, \tilde{k}_j) \right\} > 0\,.
	\end{align*}
	Then neither $A\cup\{\tilde{k}_1\}$ nor $A\cup\{\tilde{k}_2\}$
	are coherent sets for latent association. However,
	$A\cup\{\tilde{k}_1, \tilde{k}_2\}$ is a coherent set for latent association.
\end{prop}

\begin{remark}
	Note that the assumptions in Proposition~\ref{prop: coh set greedy} 
	are trivially satisfied if $\{V_{\tilde{k}_1}, V_{\tilde{k}_2}\}$
	are positively associated with each other 
	and independent of $\{V_i\colon i\in [d]\setminus\{\tilde{k}_1, \tilde{k}_2\}\}$.
	In particular, $\psi(\tilde{k}_1, \tilde{k}_2) > 0$ 
	and $\psi(i, \tilde{k}_j) = 0$ for all $j\in [2]$ and $i\in [d]\setminus\{\tilde{k}_1, \tilde{k}_2\}$.
\end{remark}

\begin{remark}
	The intuition is that two features not in the 
	coherent set $A$ are negatively associated with, or are independent of, all other features,
	but they have a strong positive association with each other.
\end{remark}

\begin{proof}
	Denote $B:= A\cup\{\tilde{k}_1, \tilde{k}_2\}$.
	By Definition~\ref{def:coherent_set} of coherent sets, we immediately know that $\psi(\tilde{k}_j, A) \leq 0$
	for each $j\in [2]$ since
	\begin{align*}
		\max_{j\in [2]}\max_{i\in [d]\setminus\{\tilde{k}_1, \tilde{k}_2\}} \psi(\tilde{k}_j, i) 
			&\leq 0\,.
	\end{align*}
	Hence, $A\cup \{\tilde{k}_j\}$ is not a coherent set 
	for latent association for any $j\in [2]$. Now note that 
	\begin{align*}
		\psi(\tilde{k}_1, B\setminus \{\tilde{k}_1\})
			&= \frac{1}{|A| + 1} \left[\sum_{l\in [q]} \psi(\tilde{k}_1, k_l)
				+ \psi(\tilde{k}_1, \tilde{k}_2) \right]
				> 0\,,
	\end{align*}
	because
	\begin{align*}
		\psi(\tilde{k}_1, \tilde{k}_2)
				> \max_{j\in [2]} \left\{ - \sum_{l\in [q]} \psi(\tilde{k}_j, k_l) \right\}\,.
	\end{align*}
	Mutatis mutandis, $\psi(\tilde{k}_2, B\setminus \{\tilde{k}_2\}) > 0$.
	Similarly, for arbitrary $p\in [q]$, we have
	\begin{align*}
		\psi(k_p, B\setminus\{k_p \})
			&= \frac{1}{|A| + 1}\, \left[ \sum_{l\in[q]\setminus\{ p \}} \psi(k_p, k_l) 
				+ \sum_{j\in [2]} \psi(k_p, \tilde{k}_j)\right]\\
			&\geq \frac{1}{|A| + 1}\, \min_{\tilde{l}\in [q]} 
				\left\{\sum_{l\in [q]\setminus\{\tilde{l}\}} \psi(k_{\tilde{l}}, k_l) 
					+ \sum_{j\in [2]} \psi(k_{\tilde{l}}, \tilde{k}_j) \right\}
				> 0\,.
	\end{align*}
	Finally, fix arbitrary $i\notin B$ and note that $\psi(i, A)\leq 0$
	implies that
	\begin{align*}
		\psi(i, B)
			&= \frac{1}{|A| + 2} \left[|A|\psi(i, A) + \sum_{j\in [2]} \psi(i, \tilde{k}_j)\right] 
				\leq 0
	\end{align*}
	by again using
	\begin{align*}
		\max_{j\in [2]}\max_{i\in [d]\setminus\{\tilde{k}_1, \tilde{k}_2\}} \psi(\tilde{k}_j, i) 
			&\leq 0\,.
	\end{align*} 
	Thus, $B$ is a coherent set for latent association by definition.
\end{proof}

\subsection{Effective Number of Sets}
\label{app: eff num sets}

In practice, the LAMB method and other
association mining methods can produce
many estimated sets with high overlap
in constituent features.\footnote{Minimal
coherent sets are ideally discovered, but that 
is not guaranteed by the LAMB method's
search procedure. Indeed, because of the Greedy
set update step, estimated coherent sets often 
exhibit either substantial overlap 
of features or very little overlap.} 
It is important to have a measure
of distinct sets, to allow for post-processing
of the output into a more straightforward result.
This motivates
a measure as in~\citet{nobelLAS-2009} that
is adapted for the LAMB method and coherent sets.

\begin{definition}[\textbf{Effective Number of Sets}]
\label{def: eff num}
Given nonempty sets $A_1, \ldots, A_K \subset [d]$ for some $K\in \mathbb{N}$,
we define the observed count of feature $i\in [d]$ in the sets
$A_1, \ldots, A_K$ as
\begin{align*}
	N_i(A_1, \ldots, A_K)
		&:= \sum_{j = 1}^K \mathbb{I}\left( i \in A_j \right).
\end{align*}
We then define the \emph{effective number} of the sets
$A_1, \ldots, A_K$ to be
\begin{align*}
	\textbf{eff num}(A_1, \ldots, A_K)
		&:= \sum_{j=1}^K \frac{1}{|A_j|} \sum_{i \in A_j} \frac{1}{N_i(A_1, \ldots, A_K)}\,.
\end{align*}
\end{definition}

The measure $N_i(A_1, \ldots, A_K)$
increases for at least some $i\in [d]$
as the nonempty sets $A_1, \ldots, A_K \subset [d]$ increase in their overlap
of elements.\footnote{Alternatively, as the Jaccard index 
\citep{jaccard-1901} between some pairs of
sets in $A_1, \ldots, A_K$ tends to 1, $N_i(A_1, \ldots, A_K)$
monotonically increases for at least some $i\in [d]$.}
For the special case where exactly $r\in [K]$ of the sets $A_1, \ldots, A_K$
are distinct we have $\textbf{eff num}(A_1, \ldots, A_K) = r$.
The effective number of the estimated coherent sets
discovered by the LAMB method
gives a simple measure of how many different
sets of associated features have actually been discovered. 
See Appendix~\ref{app: post process} for details about
post-processing the LAMB method's estimated coherent sets.

\subsection{Post-processing LAMB Method Output}
\label{app: post process}

The LAMB method may discover many 
estimated coherent sets, not all of which are very distinct.
Often there is either substantial overlap of elements in the
estimated coherent sets, or there is very little, if any, overlap 
in the estimated coherent sets. For this reason,
we find that it is helpful to have an automated post-processing step
that yields effectively distinct estimated coherent sets in a
simple manner. The following procedure
was adapted from the work in~\citet{miheerBimodules-2020}.

Consider estimated coherent sets $A_1, \ldots, A_K \subset [d]$
for some $K\in \mathbb{N}$.
Here we have $A_i \neq A_j$ for all $i, j \in [K]$.
However, there might be some sets that are very 
similar (in terms of Jaccard index \citep{jaccard-1901}).
First we apply hierarchical clustering to the sets 
$A_1, \ldots, A_K$ based on Jaccard distance  between the sets. 
The corresponding dendrogram is cut such that 
$\lceil \textbf{eff num}(A_1, \ldots, A_K) \rceil$
clusters of sets are formed (see Definition~\ref{def: eff num} in Appendix~\ref{app: eff num sets}).
Finally, representative sets of each cluster
are selected based on a centrality score defined by
\begin{align*}
	\textbf{score}(A_j | A_1, \ldots, A_K)
		&:= \frac{1}{|A_j|} \sum_{i \in A_j} N_i(A_1, \ldots, A_K)\,,
\end{align*}
where $N_i(A_1, \ldots, A_K)$ is defined as in Definition~\ref{def: eff num}.
Note that it is possible to use
$1 \vee \lfloor \textbf{eff num}(A_1, \ldots, A_K) \rfloor$
or $\lceil \textbf{eff num}(A_1, \ldots, A_K) \rceil + c$
for some $c < K$ to cut the dendrogram.
The use of $\lceil \textbf{eff num}(A_1, \ldots, A_K) \rceil$
is simple and conservative in real data applications.

\subsection{Convexity for Optimization in Section~\ref{sec:tildes}}
\label{app: conv opt theory}

Recall the optimization procedure from Section~\ref{sec:tildes}
requires optimizing two seperate equations coordinate-wise.
In this appendix we show that the optimization procedure 
for estimating $\alpha_i$ given $\bm{\tau}^0$
can be turned into minimizing
a convex function over an interval.
We restate the two optimization problems below:
\begin{framed}
\begin{enumerate}
		
	\item Given $\bm{\alpha}$, for each $j\in [n]$, 
	solve the coordinate-wise optimization problem 
	\begin{align}
		\hat{\tau}_j^0 | \bm{\alpha}
			&:= \argmin_{t\in (0, 8)} \sum_{i\in [d]} \big[\alpha_i t - X_{ij}\big( \alpha_i t + \ln[1 - \exp\{-\alpha_i t\} ]\big) \big] \,.
				\label{app eq: opt eq ell}
	\end{align}

	\item Given $\bm{\tau}^0$, for each $i\in [d]\setminus\{1\}$, 
	solve the coordinate-wise optimization problem
	\begin{align}
		\hat{\alpha}_i | \bm{\tau}^0
			&\in \argmin_{a\in (0, 8)} \left(1 - \overline{X}_i - \frac{1}{n} \sum_{j\in [n]} \exp\{- a \tau_j^0\} \right)^2\,. 
			\label{app eq: opt eq mom}%\\
	%%		&\textnormal{subject to} \quad \frac{1}{n}\sum_{i\in [n]} \exp(-a \tau_i^0)
	%%			\geq 1 - \overline{X}_j - s(a, \bm{\tau}^0) 
	%%			\label{eq: opt eq convexity}\\
	%%			&\textnormal{where} \quad s(a, \bm{t}) := \frac{\left[\frac{1}{n} \sum_{i\in[n]} t_i e^{-a t_i} \right]^2}{\frac{1}{n} \sum_{i\in [n]} t_i^2 e^{-a t_i}}. 	\label{eq: opt eq slack var}
	\end{align}
	
\end{enumerate}
\end{framed}

\begin{prop}
	The parameterized  family of functions $\ell_{a, x}(t) := at - x(at +\ln[1 - e^{-at}])$ 
	for $a \in (0, \infty)$ and $x\in \{0,1\}$ is convex on $t\in (0,\infty)$.
\end{prop}

\begin{remark}
Note that this proposition is sufficient for showing that \eqref{app eq: opt eq ell}
is a convex optimization problem because the sum of these parameterized 
functions is also convex.
\end{remark}

\begin{prop}
	Let $\bm{t}\in (0, \infty)^n$ and $b\in (0, 1)$ be given.
	There exists an interval $(0, S(\bm{t}, b)]\subset (0, \infty)$
	that satisfies
	\begin{align*}
		\frac{1}{n} \sum_{j\in [n]} e^{-a\, t_j}
			&\geq 1 - b\quad \textnormal{for all}\, a \in (0, S(\bm{t}, b)]\,.
	\end{align*}
	Furthermore, 
	the parameterized family of functions 
	$m_{\bm{t}, b}(a) := \left(b - 1 +  \frac{1}{n} \sum_{j\in [n]}  e^{-a t_j} \right)^2$
	for fixed $\bm{t}\in (0,\infty)^n$ and $b\in (0,1)$ is convex on
	$(0, S(\bm{t}, b)]$.
\end{prop}

\begin{remark}
Note that this proposition is sufficient for showing that \eqref{app eq: opt eq mom}
is a convex optimization problem when $a_i$ is 
restricted to the feasible set $(0, S(\bm{t}, b)]$.
\end{remark}

\begin{proof}
	Fix arbitrary $b\in (0,1)$ and $\bm{t} \in (0,\infty)^n$.
	Let $M:= \max_{j\in[n]} t_j$ and note that $M > 0$.
	For each $j\in[n]$ we have
	\begin{align*}
		e^{-a t_j}
			&\geq e^{-a M} \geq 1 - a M 
	\end{align*}
	using the inequality $1 + x \leq e^x$.
	Define $S(\bm{t}, b) := \frac{b}{M}$.
	Then if $a\in (0, S(\bm{t}, b)]$ we have
	\begin{align*}
		\frac{1}{n} \sum_{j\in [n]} e^{-a\, t_j}
			&\geq 1 - a M \geq 1 - b\,.
	\end{align*}
	Now note that 
	the second derivative of $m_{\bm{t}, b}$ with respect to $a$ is
	\begin{align*}
		m_{\bm{t}, b}''(a)
			&:= 2\left\{ \left[b - 1 + \frac{1}{n}\sum_{j\in[n]} e^{-a t_j} \right] \left( \frac{1}{n}\sum_{j\in [n]} t_j^2 e^{-a t_j}\right)
				+ \left( \frac{1}{n}\sum_{j\in [n]} t_j e^{-a t_j}\right)^2
				\right\}.
	\end{align*}
	Since $m_{\bm{t}, b}''(a) \geq 0$ for all $a\in (0, S(\bm{t}, b)]$,
	the function is convex on this interval.
\end{proof}

\begin{remark}
In practice, we found that
the interval $(0, S(\bm{t}, b)]$ is not
large enough for robust estimation of $\widehat{\bm{\alpha}}$ and 
$\widehat{\bm{\tau}}^0$.
%%Without using this convex optimization constraint, we are 
%%still able to compute $\widehat{\bm{\alpha}}$ and 
%%$\widehat{\bm{\tau}}^0$ to within arbitrary tolerance in both the
%%artificial and real data settings that are discussed in this paper.
\end{remark}

\newpage

\section{Additional Results}
\label{app: add details and results}

\subsection{Toy Market Basket Dataset}
\label{app: toy data results}

\begin{figure}[h]
\begin{subfigure}[b]{0.49\textwidth}
\centering
\includegraphics[width = .9\textwidth]{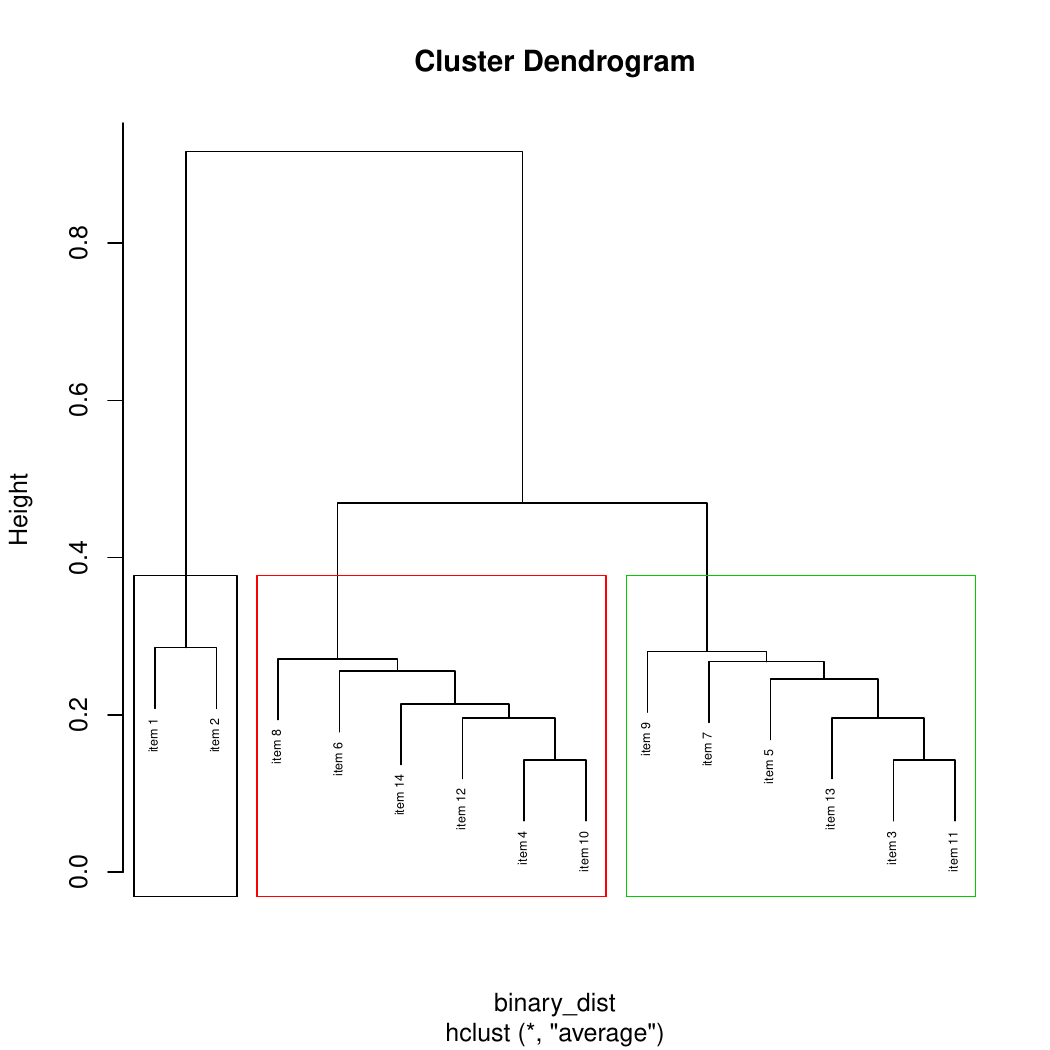}\\
\hspace{1em}
\caption{Toy market basket dataset 
with binary distance and average linkage.}
\end{subfigure}
~
\begin{subfigure}[b]{0.49\textwidth}
\centering
\includegraphics[width = .9\textwidth]{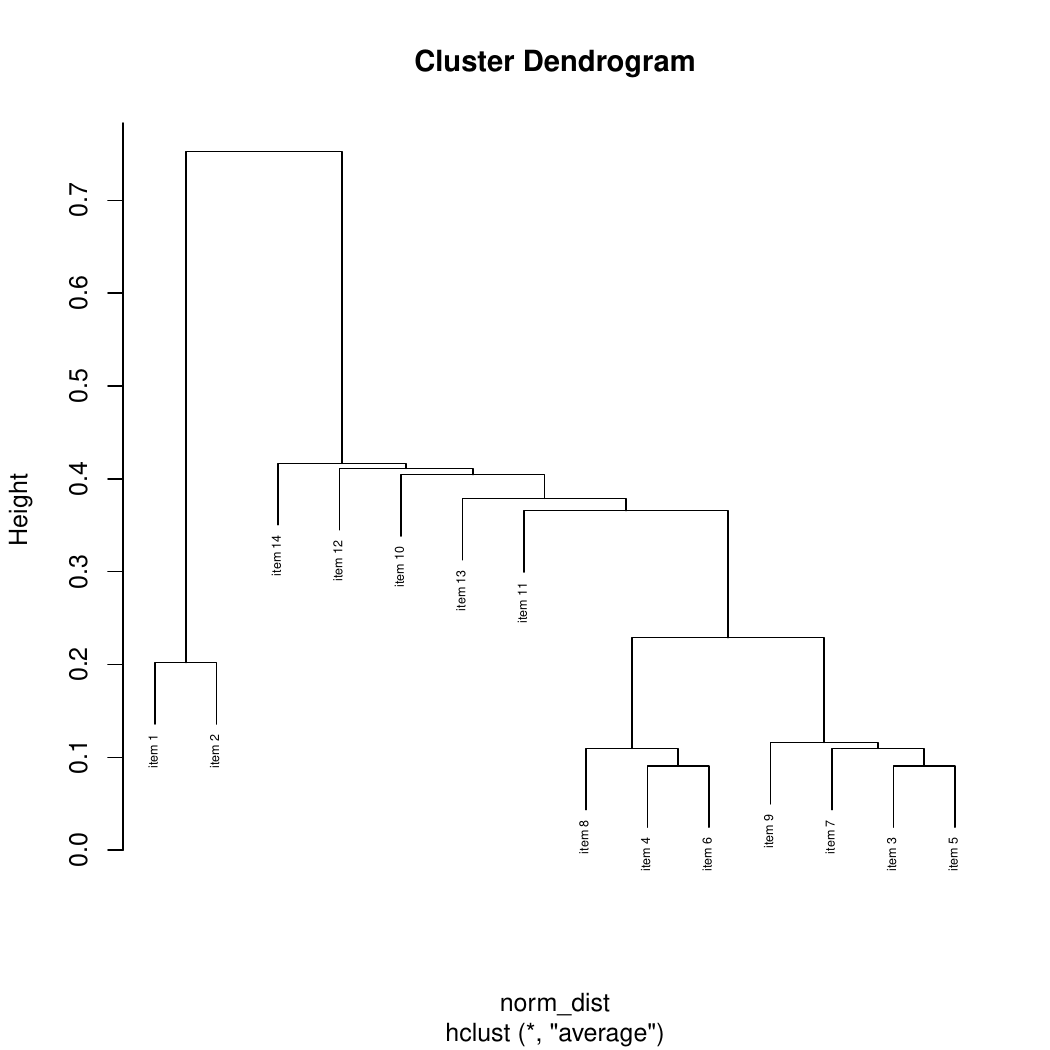}\\
\hspace{1em}
\caption{``Buyer normalized'' dataset 
with Euclidean distance and average linkage.}
\end{subfigure}

\caption{Dendrograms from hierarchical clustering.}
\label{fig: toy hclust}
\end{figure}

A number of conventional 
statistical methods could be applied to the toy market basket dataset
in Section~\ref{subsec: motivating example}.
Hierarchical clustering using average linkage
and binary distance discovers the item sets
$\{1, 2 \}$, $\{3, 5, 7, 9, 11, 13\}$,
and $\{4, 6, 8, 10, 12, 14\}$ (see Figure~\ref{fig: toy hclust}(a)).
The alternating pattern in the medium volume buyers (i.e., buyers 6 and 7) 
drives the latter two item sets.

Recall the sign change in the correlations
among items 10--14 between the 
heat maps in Figures~\ref{fig: app toy coh}(b)
and \ref{fig: app toy coh}(c). The 
negative correlations among items 10--14
arise from the buyer normalization
giving more weight to purchases made by low volume buyers
(i.e., buyers 8--12). Using Euclidean distance 
on the buyer normalized dataset with average linkage
discovers the item sets $\{1, 2\}$ and $\{3, 4, 5, 6, 7, 8, 9\}$
(see Figure~\ref{fig: toy hclust}(b)).
Buyer normalization does not
appropriately account for the sample heterogeneity
in this example. In particular, when considering
the entire toy dataset, there is no reason to consider
associations between items 3--9 
to be any stronger than the associations
between items 10--14 or 
other subsets of items 3--14.
 
Nonnegative Matrix Factorization (NMF) was
also applied to this toy dataset. 
We considered matrix factorizations of rank 2, rank 3, and rank 4.
Similar item sets as the ones discovered by hierarchical 
clustering were discovered by the NMF method. 

Conventional Itemset Mining (IM) methods
(described in Section~\ref{subsec: rel work})
could be applied to the toy dataset.
However, because conventional IM methods do not scale
well to high-dimensional datasets, we do not consider
them in the simulation study (see Section~\ref{sec: sim study})
or the real data applications (see Section~\ref{sec: applications}).
For this reason we do not apply any conventional 
IM methods to the toy dataset.

\subsection{Simulation Study}
\label{app: sim study add results}

\subsubsection{Generating Artificial Datasets}
\label{subsec: artificial datasets}

Recall the setting of Section~\ref{subsec: high level artificial datasets}.
In particular, we are generating an
artificial binary data matrix $\mathbb{X}\in \{0, 1\}^{d\times n}$
in accordance with the threshold
model of Definition~\ref{def:mod}.
For the simulation study $n = 200$ and $d = 2,000$.
The problem discussed in this appendix
is to systematically generate a matrix of 
latent continuous vectors 
$\mathbb{V} := [\bm{V}_{\cdot 1}, \ldots, \bm{V}_{\cdot n}]
\in \mathbb{R}^{d \times n}$
and a matrix of random thresholds
$\mathbf{\Theta} := [\bm{\theta}_{\cdot 1}, \ldots, \bm{\theta}_{\cdot n}]
\in (0, 1)^{d \times n}$.

Continuous random vectors 
$\bm{V}_{\cdot 1}, \ldots, \bm{V}_{\cdot n}$ 
%in the threshold model of Definition~\ref{def:mod} 
were sampled iid from a multivariate normal distribution
$\mathcal{N}_d(\bm{\mu}, \mathbf{\Sigma})$ to generate a 
matrix $\mathbb{V} := [\bm{V}_{\cdot 1}, \ldots, \bm{V}_{\cdot n}]
\in \mathbb{R}^{d \times n}$.
We set $\bm{\mu} := \bm{0}$ and the diagonal entries of $\mathbf{\Sigma}$
to be 1. Hence, when generating different matrices, say $\mathbb{V}^{(1)}$
and $\mathbb{V}^{(2)}$, the only population quantity that we change
is the off diagonal entries of $\mathbf{\Sigma}$.
Recall that the threshold model of Definition~\ref{def:mod}
is invariant under component-wise
monotone transformations (see Section~\ref{subsec: threshold model}).
Therefore, these two constraints on the values of $\bm{\mu}$ and $\mathbf{\Sigma}$ 
do not effect the robustness of the results obtained from the LAMB method
to samples $\bm{V}_{\cdot 1}, \ldots, \bm{V}_{\cdot n}$  
from a more general multivariate normal distribution.

Five associated sets of features $A_1, \ldots, A_5 \subset [d]$ were embedded
into the artificial binary datasets by using five disjoint blocks of nontrivial correlation
in $\mathbf{\Sigma}$.
In particular, $\mathbf{\Sigma}$ 
was constrained to be a block diagonal matrix with five nontrivial blocks 
$\mathbf{B}_1, \ldots, \mathbf{B}_5$ and one block the identity matrix $\mathbf{I}$;
for each sample of $\bm{V}_{\cdot 1}, \ldots, \bm{V}_{\cdot n}$ 
from $\mathcal{N}_d(\bm{0}, \mathbf{\Sigma})$
we used $\mathbf{\Sigma} := \textnormal{diag}[\mathbf{B}_1, 
\ldots, \mathbf{B}_5, \mathbf{I}]$.\footnote{Here $\textnormal{diag}$
refers to a diagonal block matrix, so that
entries in $\mathbf{\Sigma}$ other than the listed block matrices are zero. This is the easiest way
to create a nontrivial correlation matrix while maintaining
the positive definiteness of $\mathbf{\Sigma}$.}
The size of each embedded associated set
is 200, i.e., each block $\mathbf{B}_k$ is a $200\times 200$
dimensional matrix (of correlations) for $k\in [5]$.
A nontrivial block $\mathbf{B}_k$ generates a truly 
associated set of features $A_k\subset [d]$
in the artificial dataset, for each $k\in [5]$. 
Features corresponding to the identity matrix
block $\mathbf{I}$ in $\mathbf{\Sigma}$ are independent.
Note that the block diagonal structure of $\mathbf{\Sigma}$
means that any two features from different blocks
are independent.

Let $\rho_k\in [0, 1]$ be the off-diagonal value 
of the block $\mathbf{B}_k$, for each $k\in [5]$.
The off-diagonal values $\rho_1, \ldots, \rho_5$
of the correlation blocks $\mathbf{B}_1, \ldots, \mathbf{B}_5$
were varied to study the sensitivity of the LAMB method 
to the strength of the association in the embedded 
sets $A_1, \ldots, A_5$. To accomplish this, 
if $\mathcal{R} := \{0, 0.02, 0.04, \ldots, 0.96, 0.98, 1\}$,
then 
\begin{align*}
	\rho_1, \ldots, \rho_5 \overset{\textnormal{iid}}{\sim} \textnormal{Unif}(\mathcal{R}),
\end{align*}
i.e., we sample the off-diagonal (correlation) values 
$\rho_1, \ldots, \rho_5$ from $\mathcal{R}$
independently with replacement. 
This sampling was independently done 50 times, resulting in 50 
independent and nontrivial (correlation) matrices
$\mathbf{\Sigma}^{(1)}, \ldots, \mathbf{\Sigma}^{(50)}$.
As discussed above, sampling 
$\bm{V}_{\cdot 1}^{(s)}, \ldots, \bm{V}_{\cdot n}^{(s)}$ 
from $\mathcal{N}_d\left(\bm{0}, \mathbf{\Sigma}^{(s)}\right)$
for each $s\in [50]$
generated matrices $\mathbb{V}^{(1)}, \ldots, \mathbb{V}^{(50)}$.

Random thresholds $\bm{\theta}_{\cdot 1}, \ldots, \bm{\theta}_{\cdot n}$
were sampled independently of $\bm{V}_{\cdot 1}, \ldots, \bm{V}_{\cdot n}$
to generate a matrix $\mathbf{\Theta} := [\bm{\theta}_{\cdot 1}, \ldots, \bm{\theta}_{\cdot n}]$.
For each matrix $\mathbf{\Theta}$, $\bm{\alpha} := (\alpha_1, \ldots, \alpha_d)^T$
was sampled once while $\tau\sim \pi$ was sampled iid $n$ times to generate 
the vector of realized values $\bm{\tau}^0 := (\tau_1^0, \ldots, \tau_n^0)^T$
(see Equation~\eqref{eq:theta} in Definition~\ref{def:mod}).
The parameters $\bm{\alpha}$ were generated from a scaled beta distribution, i.e.,
\begin{align*}
	\alpha_2, \ldots, \alpha_d 
		&\overset{\textnormal{iid}}{\sim} 2\cdot \textnormal{Beta}(a , b)
\end{align*}
for shape parameters $a$ and $b$ and $\alpha_1 := 1$ for identifiability
(see Proposition~\ref{prop: identifiable}).
The values for $\bm{\alpha}$ are scaled by a factor of 2 to make the 
generating distributions of $\bm{\theta}_{\cdot 1}, \ldots, \bm{\theta}_{\cdot n}$ 
more heterogeneous.\footnote{Recall that $\bm{\alpha}$ is not random 
in the threshold model 
of Definition~\ref{def:mod}. However, because
these datasets are artificial, we must generate $\bm{\alpha}$. 
A scaled Beta distribution is a simple way to
generate heterogeneous values for $\bm{\alpha}$.} 
The parameters $\bm{\tau}^0$ were generated from a gamma distribution, i.e.,
\begin{align*}
	\tau_1^0, \ldots, \tau_n^0
		&\overset{\textnormal{iid}}{\sim} \textnormal{Gamma}(s, r)
\end{align*}
for shape parameter $s$ and rate parameter $r$.

\begin{figure}[H]
\centering
\includegraphics[width = \textwidth]{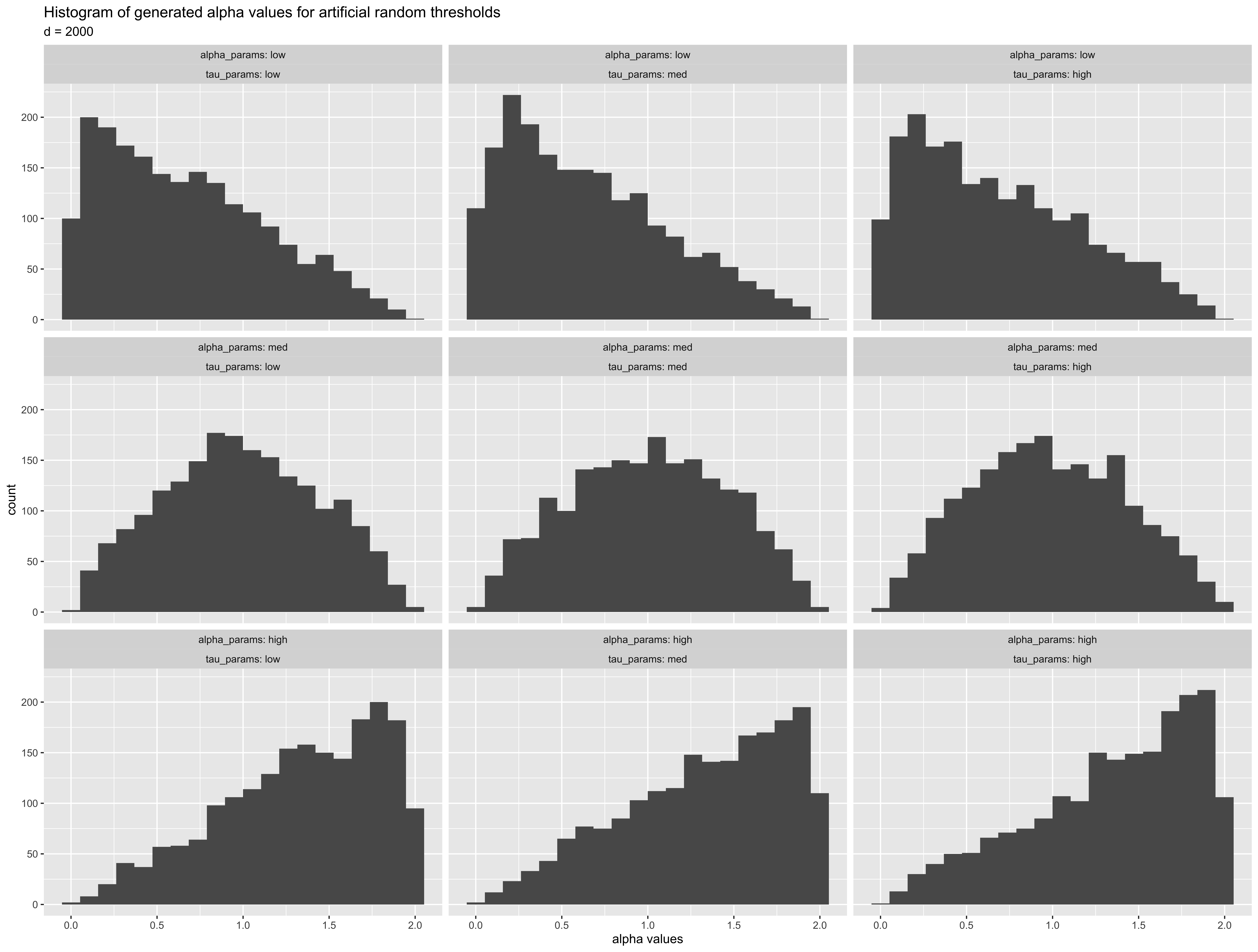}%\\
%\hspace{1em}

\caption{Histograms for the generated $\bm{\alpha}$ parameters in the 
artificial datasets of Section~\ref{sec: sim study}.}
\label{app fig: alphas histogram}
\end{figure}

\begin{figure}[H]
\centering
\includegraphics[width = \textwidth]{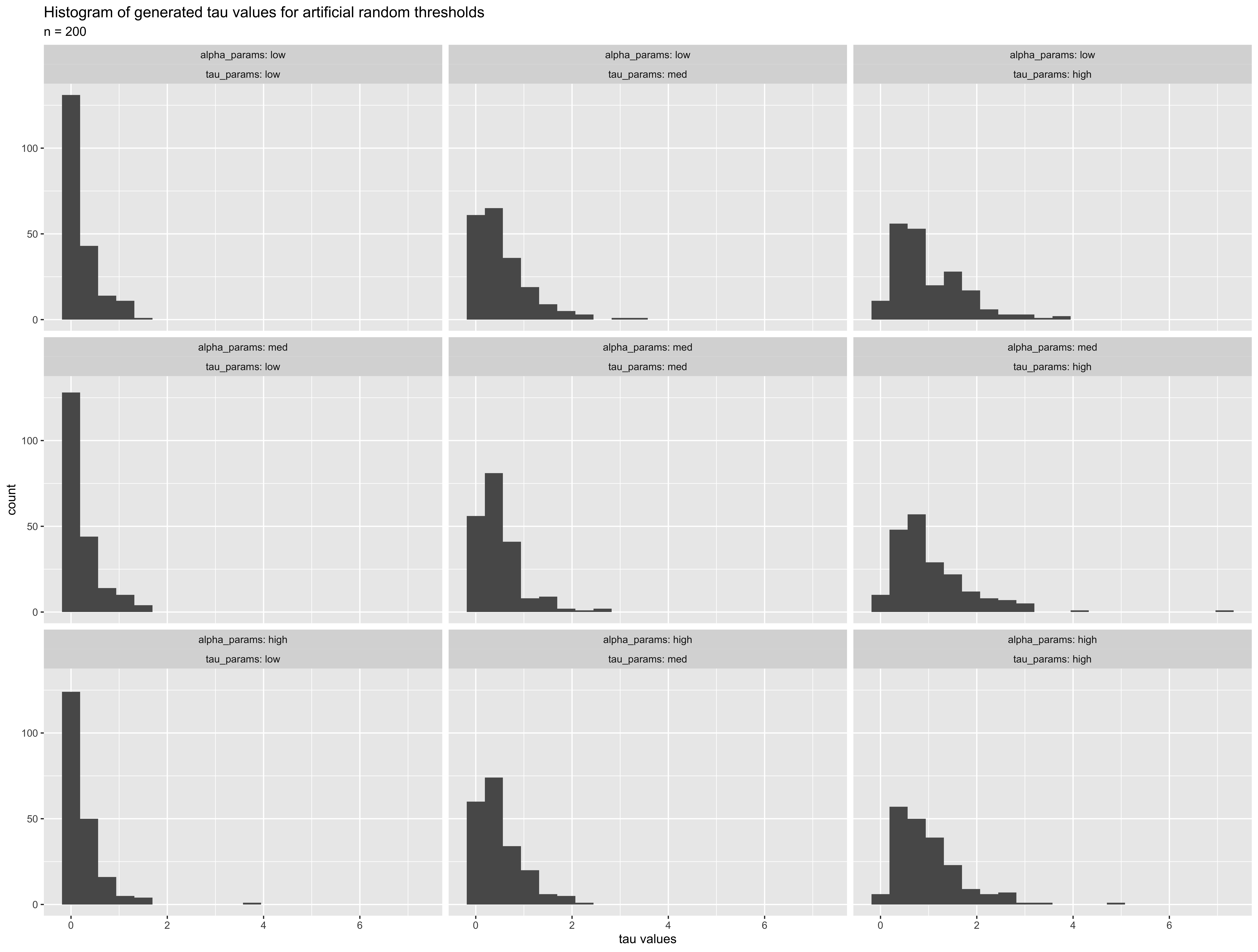}%\\
%\hspace{1em}

\caption{Histograms for the generated $\bm{\tau}^0$ parameters in the 
artificial datasets of Section~\ref{sec: sim study}.}
\label{app fig: taus histogram}
\end{figure}

Recall that the LAMB method is interested in discovering the 
intrinsic association encoded in $\mathbb{V}$
and that estimating the random thresholds $\mathbf{\Theta}$ 
is considered a nuisance parameter.
Different random threshold matrices $\mathbf{\Theta}$ were generated
by independently varying the distribution parameters $a,b$ for $\bm{\alpha}$
and $s, r$ for $\bm{\tau}^0$. This allows for a more robust
study of the results obtained from the LAMB method.
We are able to study the sensitivity of the LAMB
method to the values of the random thresholds
by leaving the matrix $\mathbb{V}$ the same
and changing the matrix $\mathbf{\Theta}$ to 
generate a different binary data matrix $\mathbb{X}$.
Three different pairs of parameter
values were used for $\bm{\alpha}$ and $\bm{\tau}^0$: 
\begin{enumerate}
	\item For ``low'' $\bm{\alpha}$ parameter values, $(a, b) = (1, 2)$.
	
	\item For ``medium'' $\bm{\alpha}$ parameter values, $(a, b) = (2, 2)$.
	
	\item For ``high'' $\bm{\alpha}$ parameter values, $(a, b) = (2, 1)$.
	
	\item For ``low'' $\bm{\tau}$ parameter values, $(s, r) = \left(\frac{1}{2}, 2\right)$.
	
	\item For ``medium'' $\bm{\tau}$ parameter values, $(s, r) = (1, 2)$.
	
	\item For ``high'' $\bm{\tau}$ parameter values, $(s, r) = (2, 2)$.
\end{enumerate}
The labels ``low'', ``medium'', and ``high'' simply refer
to the skew of the densities for the corresponding distributions
based on these parameters. 
See Figures~\ref{app fig: alphas histogram} and \ref{app fig: taus histogram}
for histograms of the corresponding values of $\bm{\alpha}$
and $\bm{\tau}^0$, respectively. 

Note that, because
of the functional form of the random thresholds, i.e.,
$\theta_{ij} := 1- \exp{(-\alpha_i \tau_j^0)}$, larger values of 
$\alpha_i$ or $\tau_j^0$ correspond to values of $\theta_{ij}$
exponentially close to 1.\footnote{Larger values
of $\theta_{ij}$ makes it more likely that $X_{ij}$ equals 1
in the artificial binary data matrix $\mathbb{X}$. So changing 
the parameters for $\bm{\alpha}$
and $\bm{\tau}^0$ effects the sparsity of 
the data matrix $\mathbb{X}$.}
In total, 9 different random threshold matrices 
$\mathbf{\Theta}^{(1)}, \ldots, \mathbf{\Theta}^{(9)}$
were generated; there is one random threshold matrix 
$\mathbf{\Theta}^{(r)}$, where $r\in [9]$, for each combination of $\bm{\alpha}$
and $\bm{\tau}$ parameter values.

The matrices $\mathbb{V}$ and $\mathbf{\Theta}$ were combined
as in the threshold model of Definition~\ref{def:mod} to generate
an artificial binary data matrix $\mathbb{X}$ with nontrivial association structure
and sample heterogeneity.
This resulted in a total of $450$ different artificial binary datasets,
since 50 different matrices $\mathbb{V}^{(1)}, \ldots, \mathbb{V}^{(50)}$
and 9 different matrices $\mathbf{\Theta}^{(1)}, \ldots, \mathbf{\Theta}^{(9)}$ 
were generated in total. 
Many values of all of the parameters discussed above were considered
for this simulation study. We expect that the results
from the LAMB method in Section~\ref{subsec: sim study analysis}
will be robust to changes of any of the values of the parameters used above,
as long as $n \geq 200$.

\begin{figure}[H]
\centering
\includegraphics[width = \textwidth]{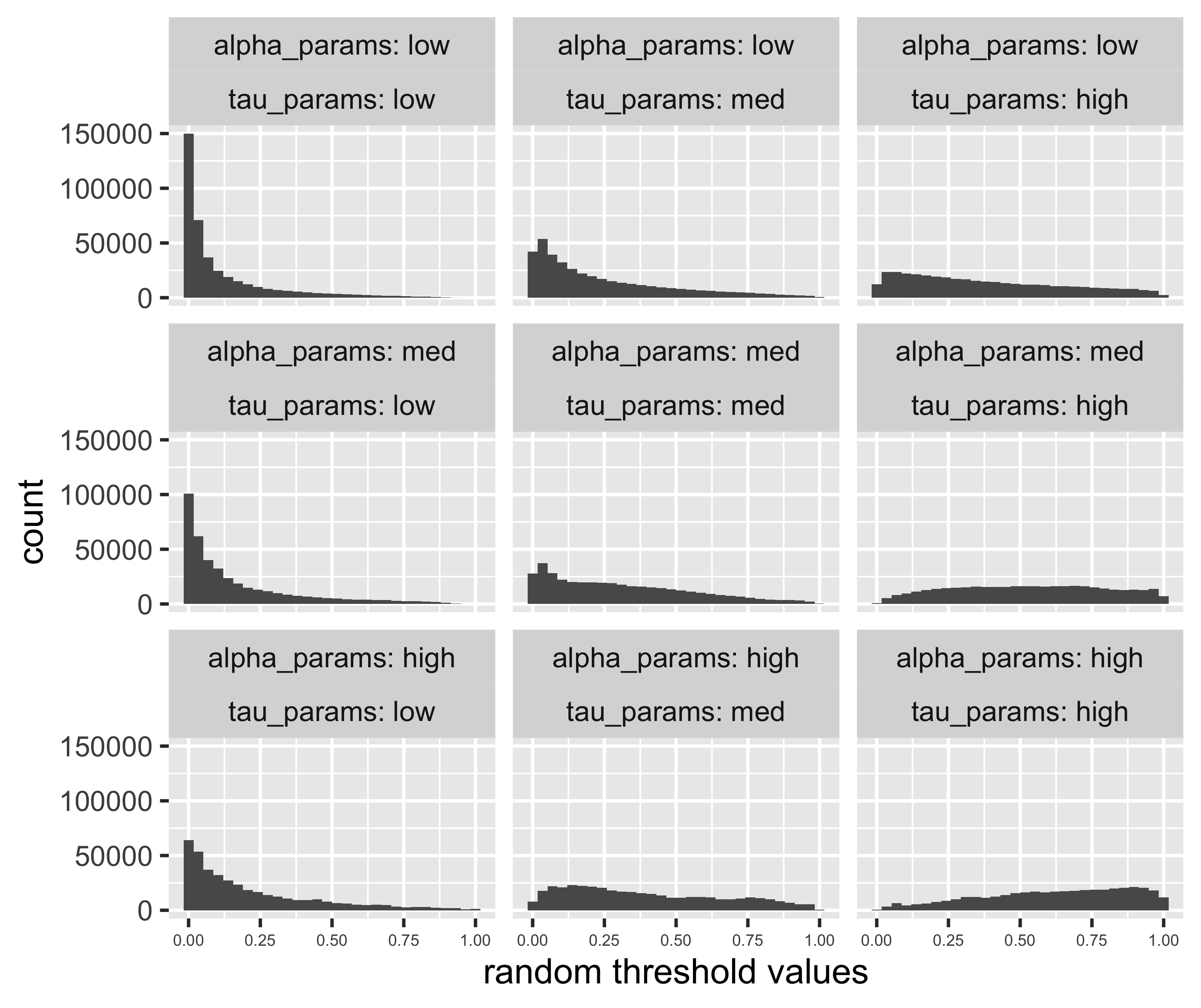}%\\
%\hspace{1em}

\caption{Histograms for the generated random thresholds in the 
artificial datasets of Section~\ref{sec: sim study}.}
\label{app fig: random threshold histogram}
\end{figure}

\subsubsection{Additional Figures}
\label{subsec: additional sim study figures}

\begin{figure}[H]
\centering
\includegraphics[width = \textwidth]{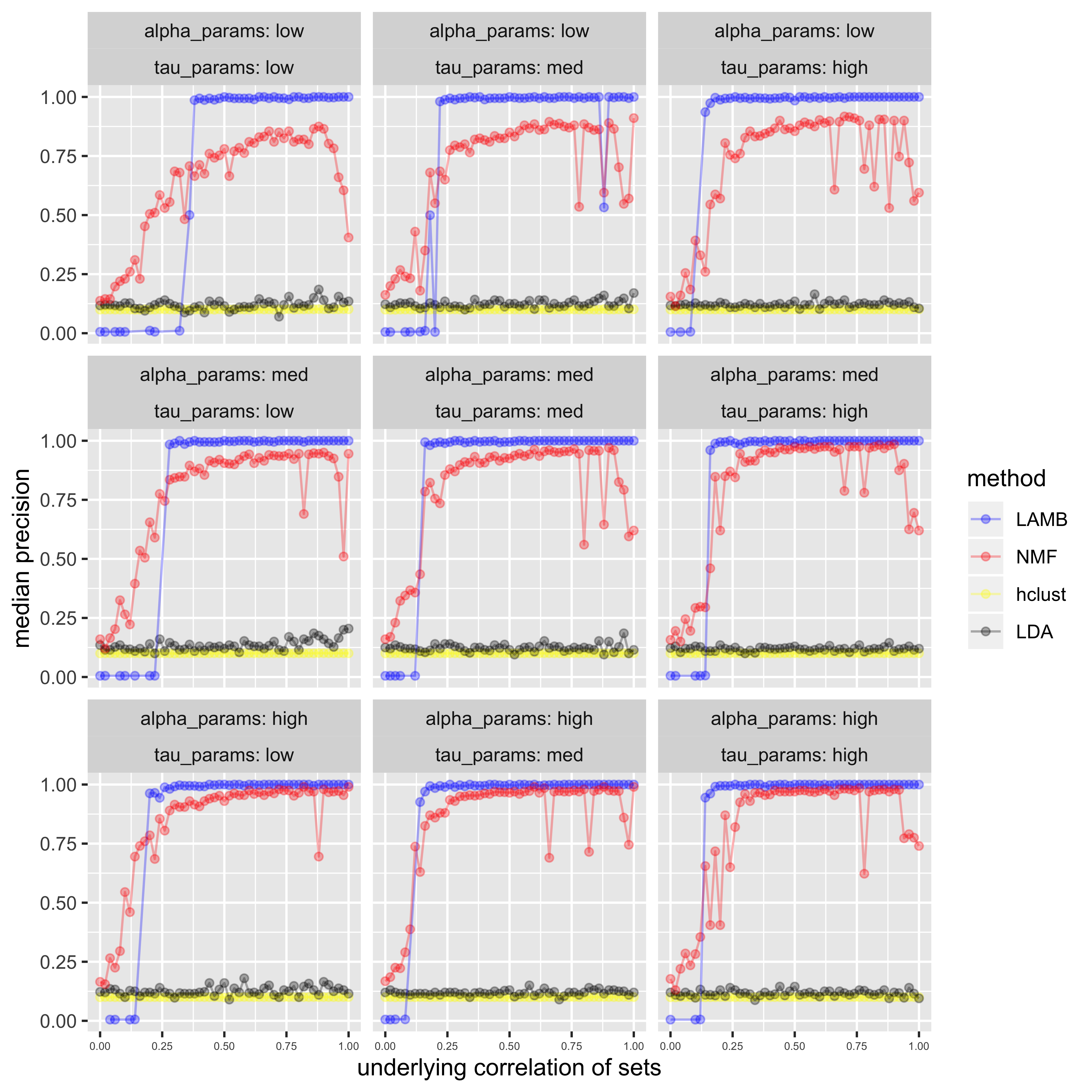}%\\
%\hspace{1em}

\caption{Comparison of the LAMB, NMF, and LDA methods on the artificial datasets in Section~\ref{sec: sim study} using the (median) precision of their estimated associated sets. 
See Section~\ref{subsec: sim study analysis} for a detailed analysis.
The LAMB method produces estimated coherent sets while
the NMF and LDA methods produce soft clusters  
of all the features per topic. Only the ``top 200'' features
in the soft clusters of the NMF and LDA methods are used
since the embedded sets all contained 200 features.
Figures~\ref{app fig: alphas histogram}, \ref{app fig: taus histogram},
and \ref{app fig: random threshold histogram}
in Appendix~\ref{subsec: artificial datasets}
show the histograms of the corresponding values of $\bm{\alpha}$,
$\bm{\tau}^0$, and $\bm{\theta}_{\cdot 1}, \ldots, \bm{\theta}_{\cdot n}$, respectively.
For this figure we used $\delta = .05$ for the LAMB method
and fit 6 latent topics for the NMF and LDA methods.
%%See Figures~\ref{app fig: sim study lamb precision}, 
%%\ref{app fig: sim study lamb recall}, 
%%\ref{app fig: sim study nmf recall}, 
%%and \ref{app fig: sim study lda recall}
%%in Appendix~\ref{app: add details and results}
%%for a comparison of different values of these parameters.
For reference we include results obtained from hierarchical clustering 
with binary distance and average linkage 
using 7 clusters.}
\label{fig: sim study comp precision}
\end{figure}

\begin{figure}[H]
\centering
\includegraphics[width = \textwidth]{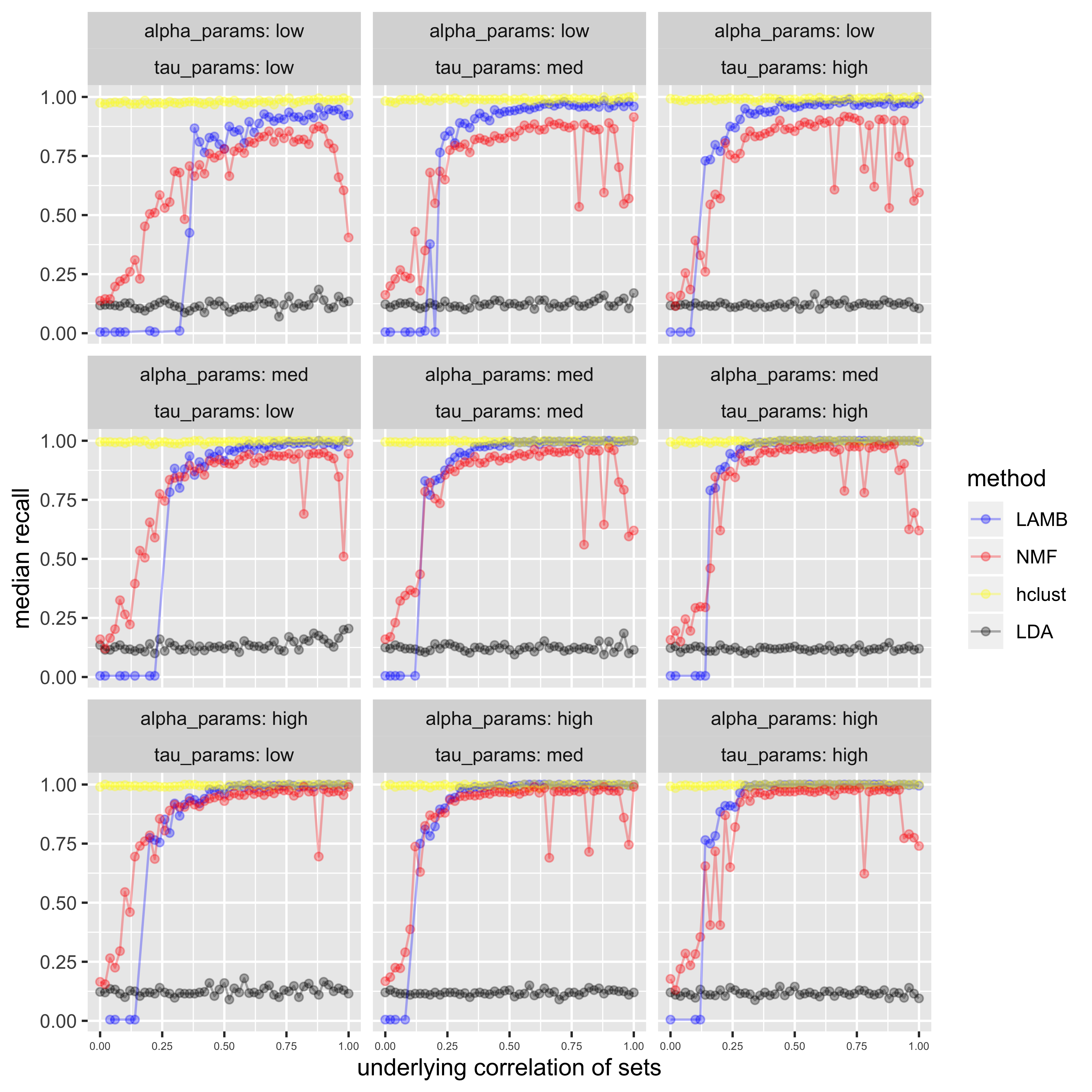}%\\
%\hspace{1em}

\caption{Comparison of the LAMB, NMF, and LDA methods on the artificial datasets of Section~\ref{sec: sim study} using the (median) recall of their estimated associated sets. 
See Section~\ref{subsec: sim study analysis} for a detailed analysis.
The LAMB method produces estimated coherent sets while
the NMF and LDA methods produce soft clusters  
of all the features per topic. Only the ``top 200'' features
in the soft clusters of the NMF and LDA methods are used
since the embedded sets all contained 200 features.
Figures~\ref{app fig: alphas histogram}, \ref{app fig: taus histogram},
and \ref{app fig: random threshold histogram}
in Appendix~\ref{subsec: artificial datasets}
show the histograms of the corresponding values of $\bm{\alpha}$,
$\bm{\tau}^0$, and $\bm{\theta}_{\cdot 1}, \ldots, \bm{\theta}_{\cdot n}$, respectively.
%%For this figure we used $\delta = .05$ for the LAMB method
%%and fit 6 latent topics for the NMF and LDA methods.
%%See Figures~\ref{app fig: sim study lamb precision}, 
%%\ref{app fig: sim study lamb recall}, 
%%\ref{app fig: sim study nmf recall}, 
%%and \ref{app fig: sim study lda recall}
%%in Appendix~\ref{app: add details and results}
%%for a comparison of different values of these parameters.
For reference we include results obtained from hierarchical clustering 
with binary distance and average linkage 
using 7 clusters.}
\label{fig: sim study comp recall}
\end{figure}

Recall that the LAMB method relies on estimating the 
random thresholds $\mathbf{\Theta}$ to discover associated sets of features. 
It is therefore important that the results obtained from the LAMB method
in Figures~\ref{fig: sim study comp precision}
and \ref{fig: sim study comp recall} are robust to 
changes in the values of $\mathbf{\Theta}$.
The lone exception is in the upper left plots of 
Figures~\ref{fig: sim study comp precision}
and \ref{fig: sim study comp recall}, 
where the corresponding artificial binary datasets are exceptionally sparse.\footnote{These
artificial binary datasets are sparse because both $\bm{\alpha}$ and 
$\bm{\tau}^0$ are generated from distributions that are skewed towards low values.
Low values of both $\bm{\alpha}$ and $\bm{\tau}^0$ generate low 
values of $\theta_{ij}$ in $\mathbf{\Theta}$, which significantly decreases the chances of
generating 1s in the threshold model of Definition~\ref{def:mod}
independent of the values in $\bm{V}_{\cdot 1}, \ldots, \bm{V}_{\cdot n}$.}

\begin{figure}[H]
\centering
\includegraphics[width = \textwidth]{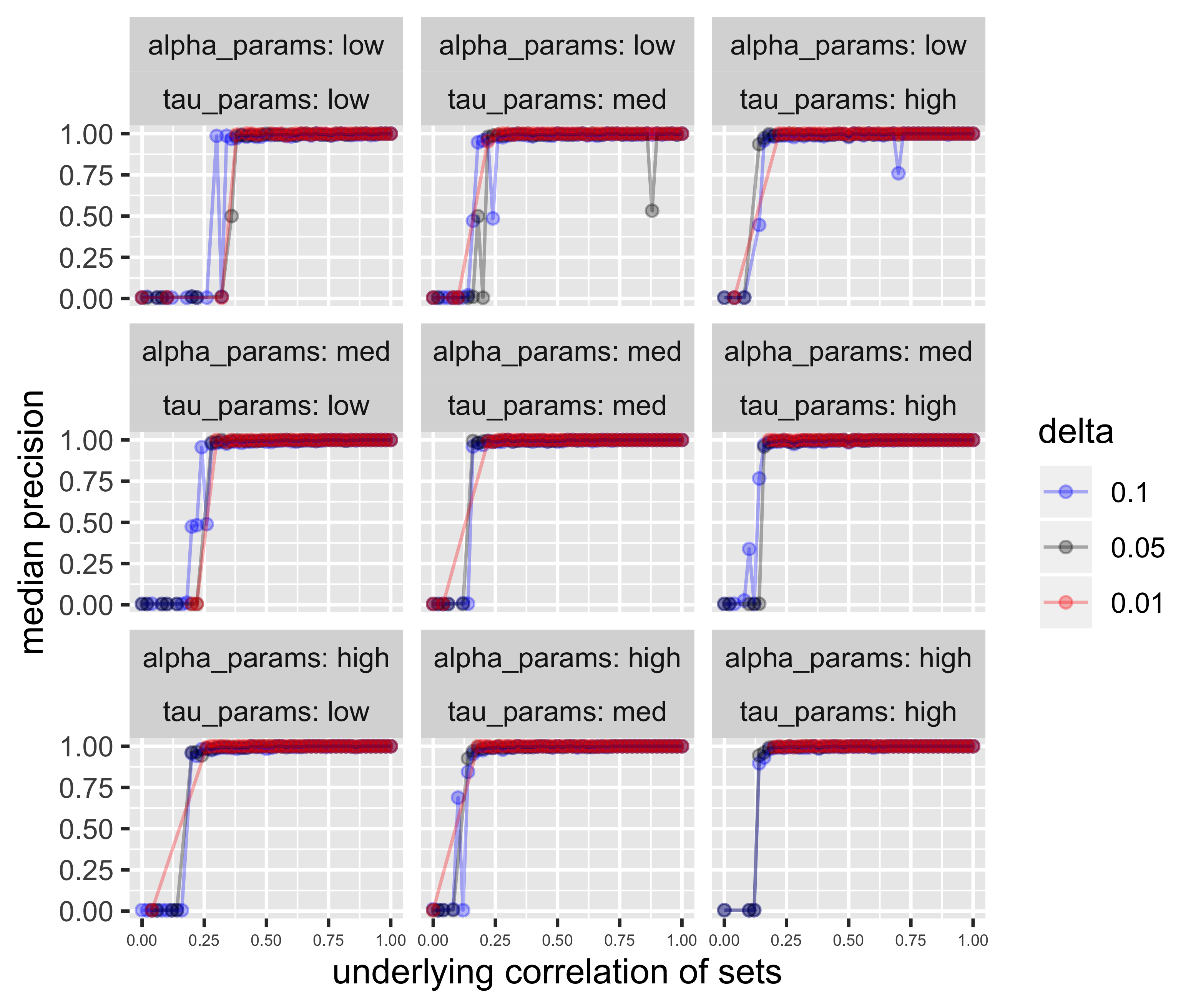}%\\
%\hspace{1em}

\caption{Analysis of the LAMB method based on precision of estimated sets. 
The value of  ``delta'' in this figure
is used during the multiple testing procedure of the LAMB method's search. 
See Algorithm~\ref{alg: lamb search proc} in Section~\ref{sec:lamb}, where it is denoted $\delta$.
Note that the LAMB method is robust to changes in the value of delta.}
\label{app fig: sim study lamb precision}
\end{figure}

\begin{figure}[H]
\centering
\includegraphics[width = \textwidth]{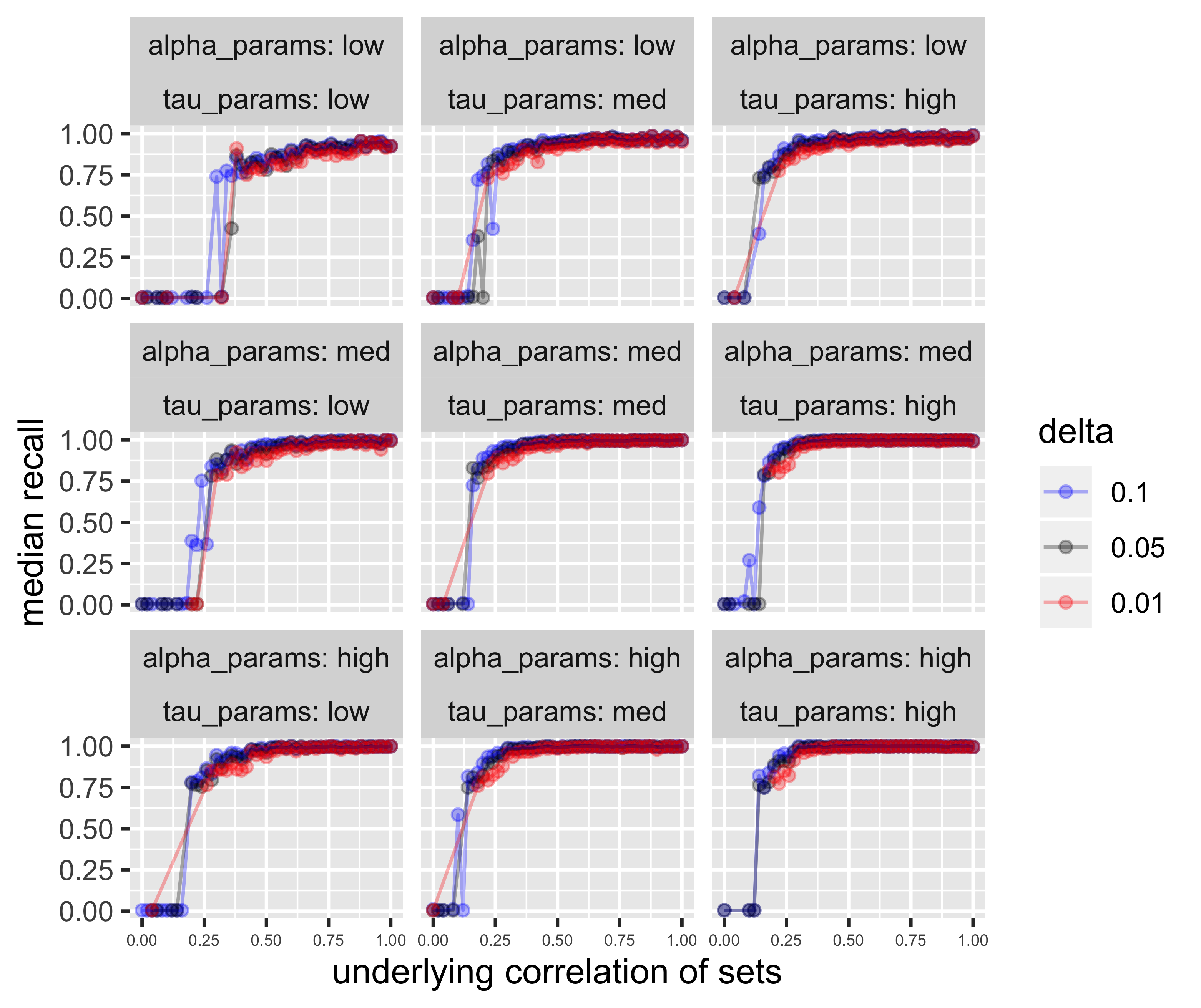}%\\
%\hspace{1em}

\caption{Analysis of the LAMB method based on recall of estimated sets. 
The value of  ``delta'' in this figure
is used during the multiple testing procedure of the LAMB method's search. 
See Algorithm~\ref{alg: lamb search proc} in Section~\ref{sec:lamb}, where it is denoted $\delta$.
Note that the LAMB method is robust to changes in the value of delta.}
\label{app fig: sim study lamb recall}
\end{figure}

\begin{figure}[H]
\centering
\includegraphics[width = \textwidth]{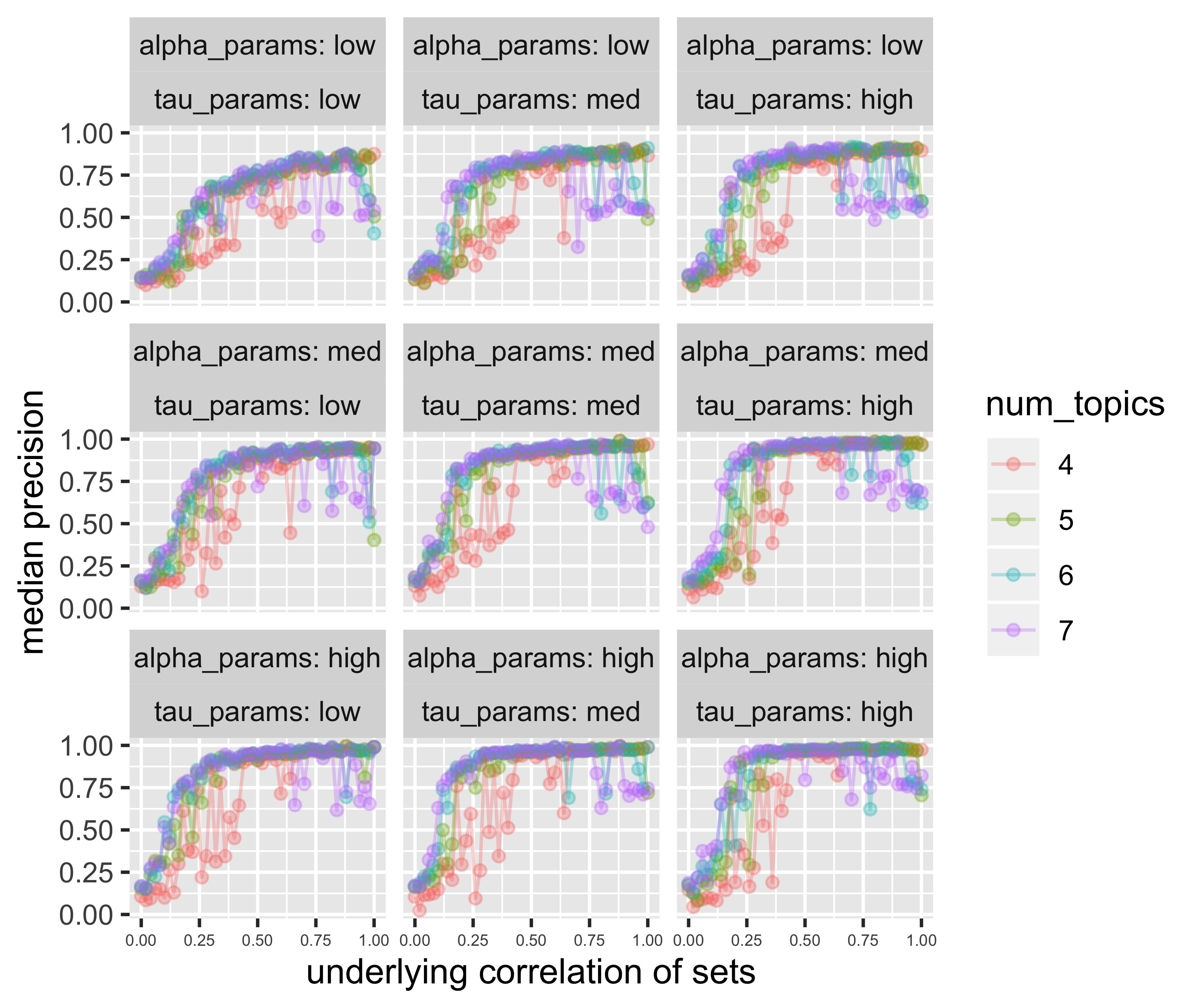}%\\
%\hspace{1em}

\caption{Analysis of the NMF method based on precision of estimated sets. 
Recall that the precision and recall statistics are the same for the NMF
method (see Section~\ref{sec: sim study}). We considered
different numbers of latent topics. Note that the 5 topic NMF method has better results for large
values of underlying correlation in the embedded sets. On the other hand, 
the 6 topic NMF method has better results for small values of underlying
correlation in the embedded sets. When the underlying correlation is strong
enough, the NMF method does not benefit as much from an extra topic.
However, when the underlying correlation is not very strong, the NMF method benefits
from an extra topic that clusters the ``noisy'' features.}
\label{app fig: sim study nmf recall}
\end{figure}

\begin{figure}[H]
\centering
\includegraphics[width = \textwidth]{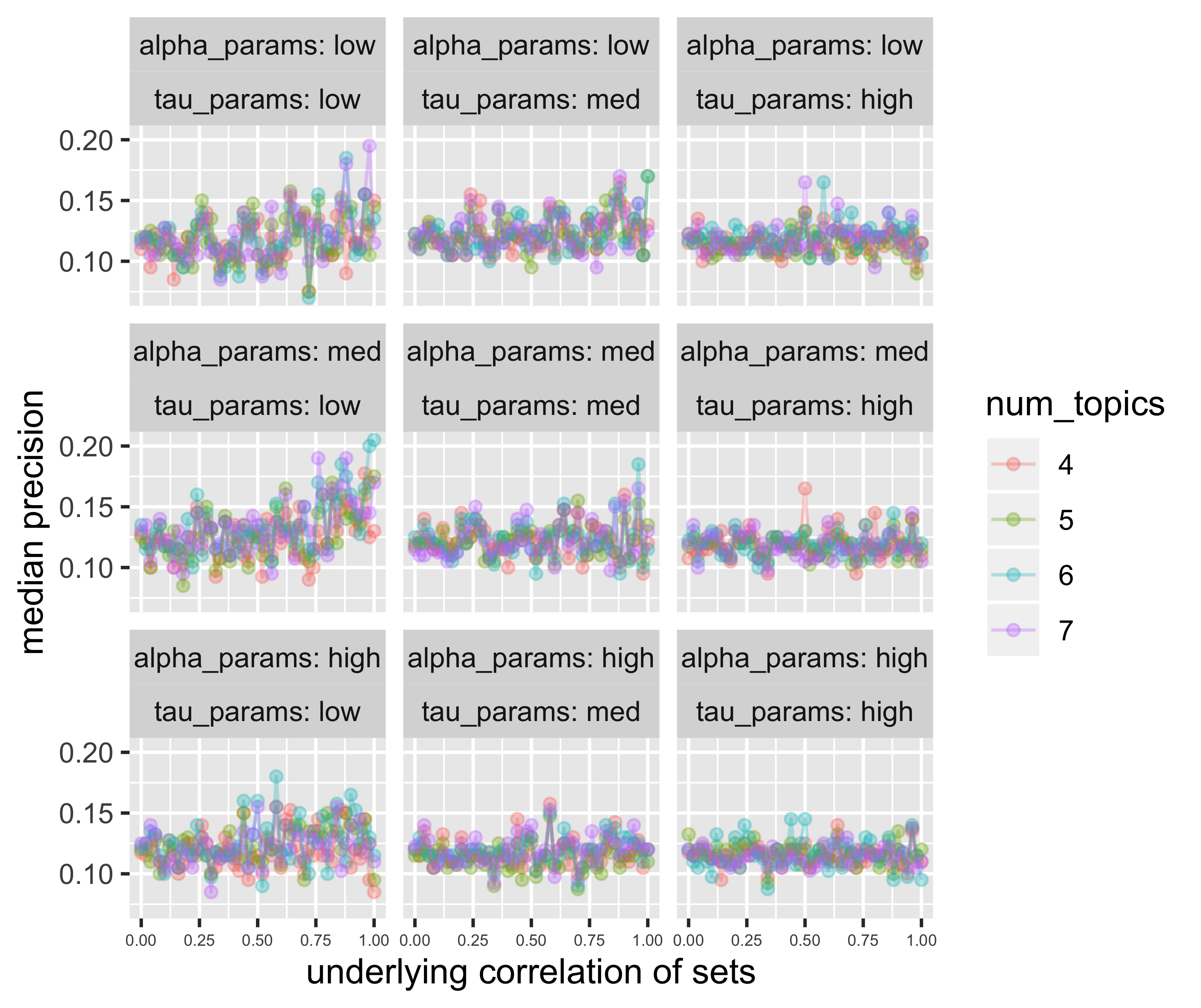}%\\
%\hspace{1em}

\caption{Analysis of the LDA method based on precision of estimated sets. 
Recall that the precision and recall statistics are the same for the LDA
method (see Section~\ref{sec: sim study}). We considered
different numbers of latent topics.
As mentioned in Section~\ref{subsec: sim study analysis},
it appears that the binary-valued artificial datasets do not provide enough information for 
the inference problem used in the LDA method. 
These results are not an indictment
of the LDA method. Rather, the results
of this simulation study indicate the difficulty 
that binary-valued data can present for inference
and association mining problems.
}
\label{app fig: sim study lda recall}
\end{figure}

\begin{figure}[H]
\centering
\includegraphics[width = \textwidth]{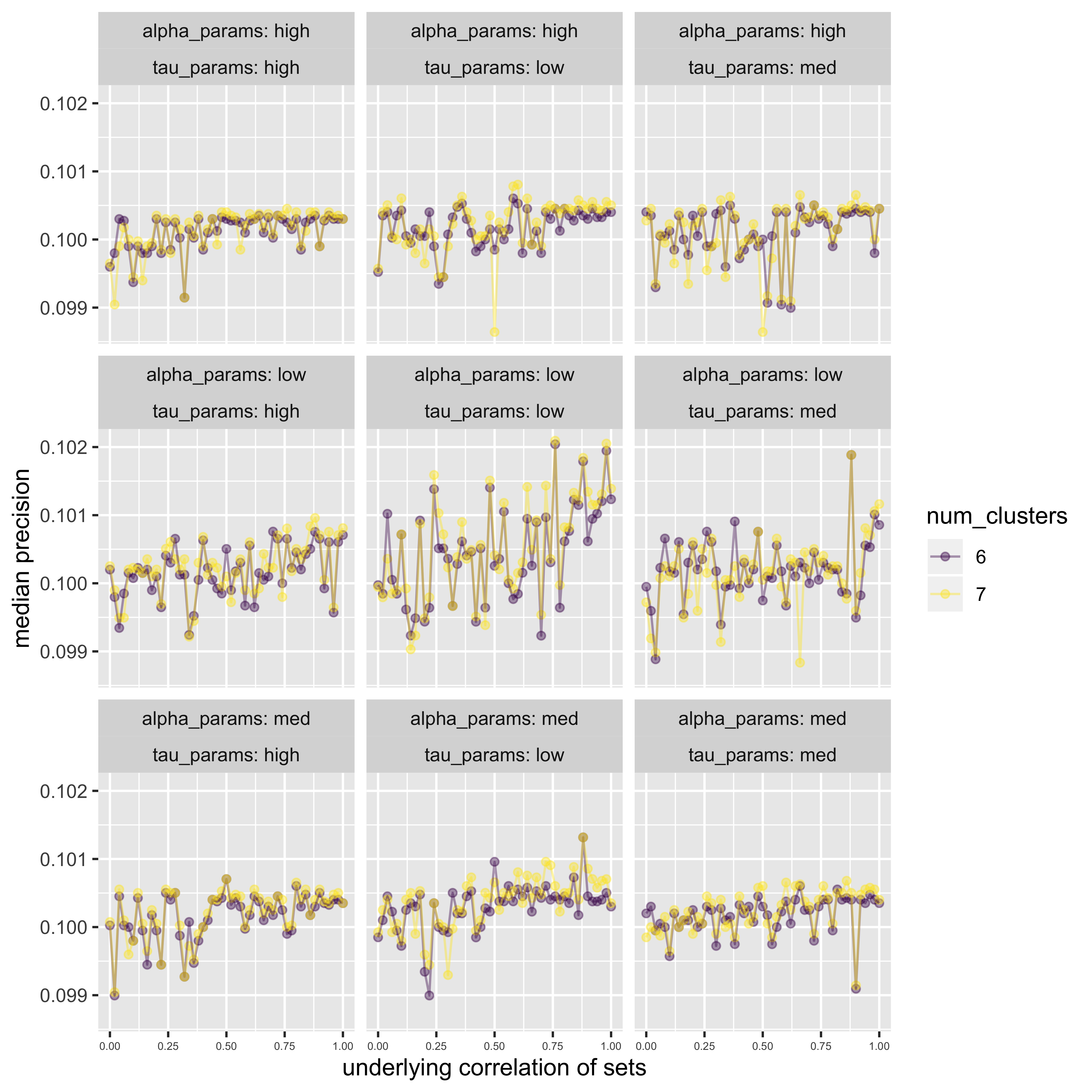}%\\
%\hspace{1em}

\caption{Analysis of hierarchical clustering based on precision of estimated sets. 
This is using binary distance on the artificial datasets with average linkage.
We considered different numbers of clusters.}
\label{app fig: sim study binary hclust recall}
\end{figure}

\begin{figure}[H]
\centering
\includegraphics[width = \textwidth]{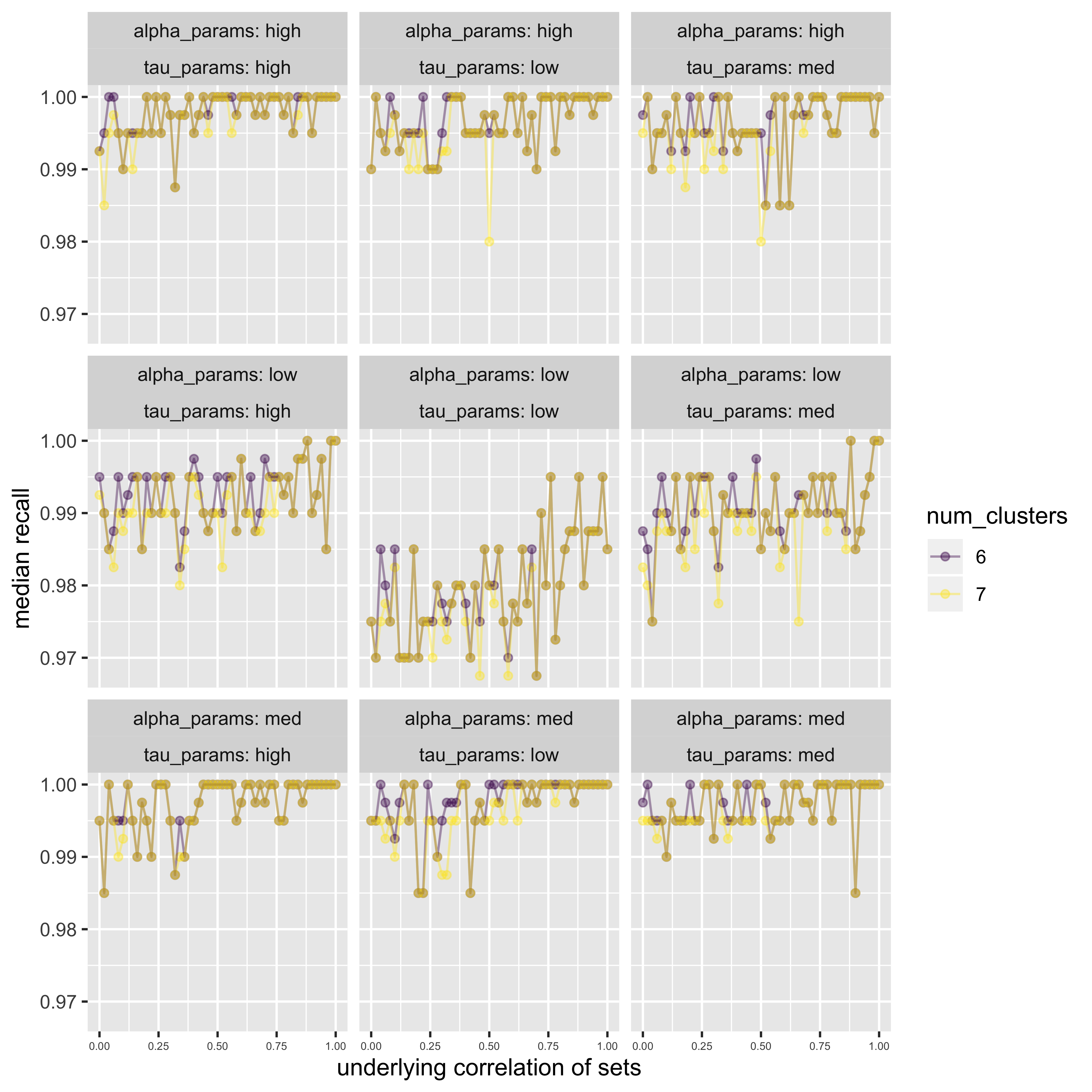}%\\
%\hspace{1em}

\caption{Analysis of hierarchical clustering based on recall of estimated sets. 
This is using binary distance on the artificial datasets with average linkage.
We considered different numbers of clusters.}
\label{app fig: sim study binary hclust recall}
\end{figure}

\begin{figure}[H]
\centering
\includegraphics[width = \textwidth]{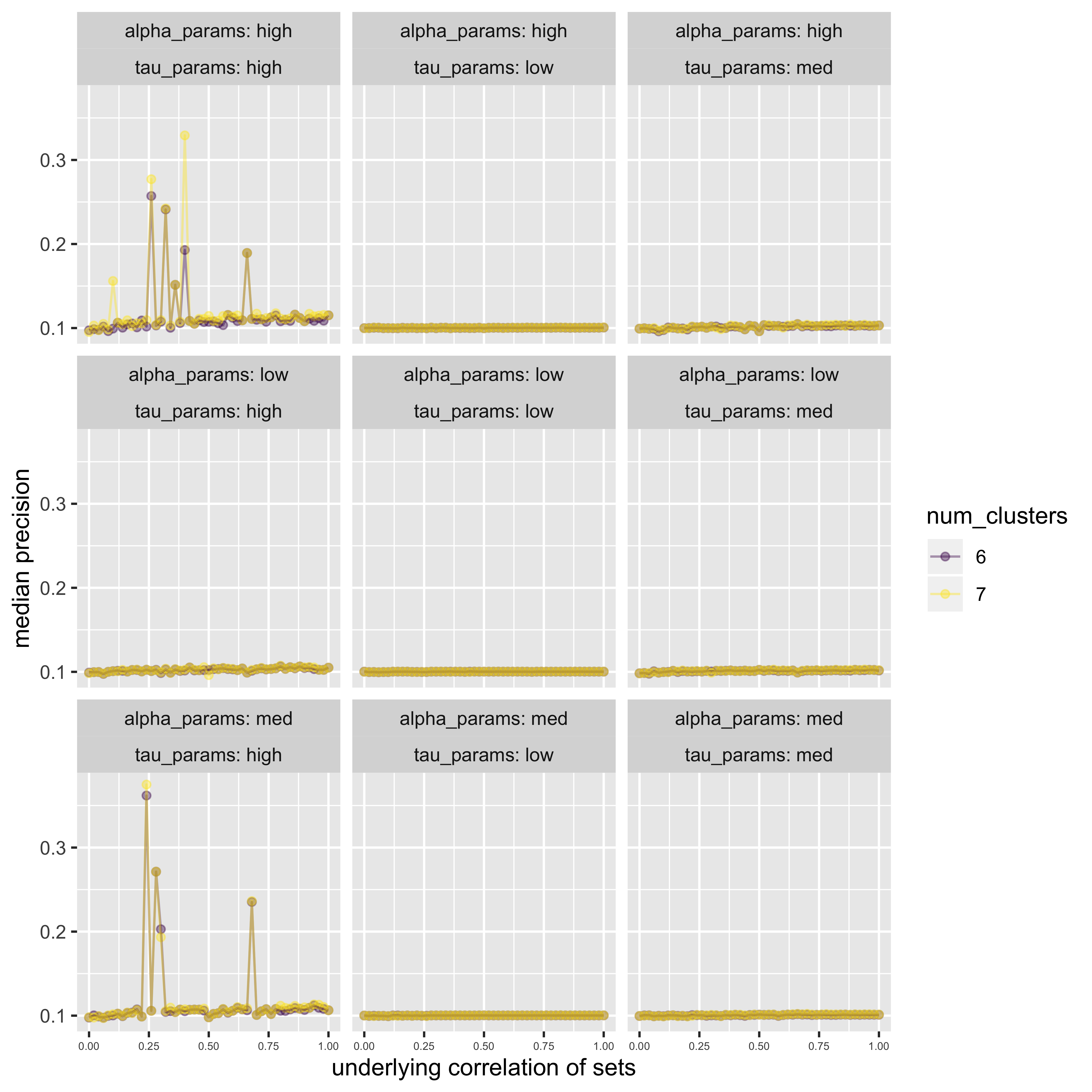}%\\
%\hspace{1em}

\caption{Analysis of hierarchical clustering based on precision of estimated sets. 
This is using Euclidean distance on the ``buyer normalized'' artificial datasets with average linkage.
We considered different numbers of clusters.}
\label{app fig: sim study norm hclust recall}
\end{figure}

\begin{figure}[H]
\centering
\includegraphics[width = \textwidth]{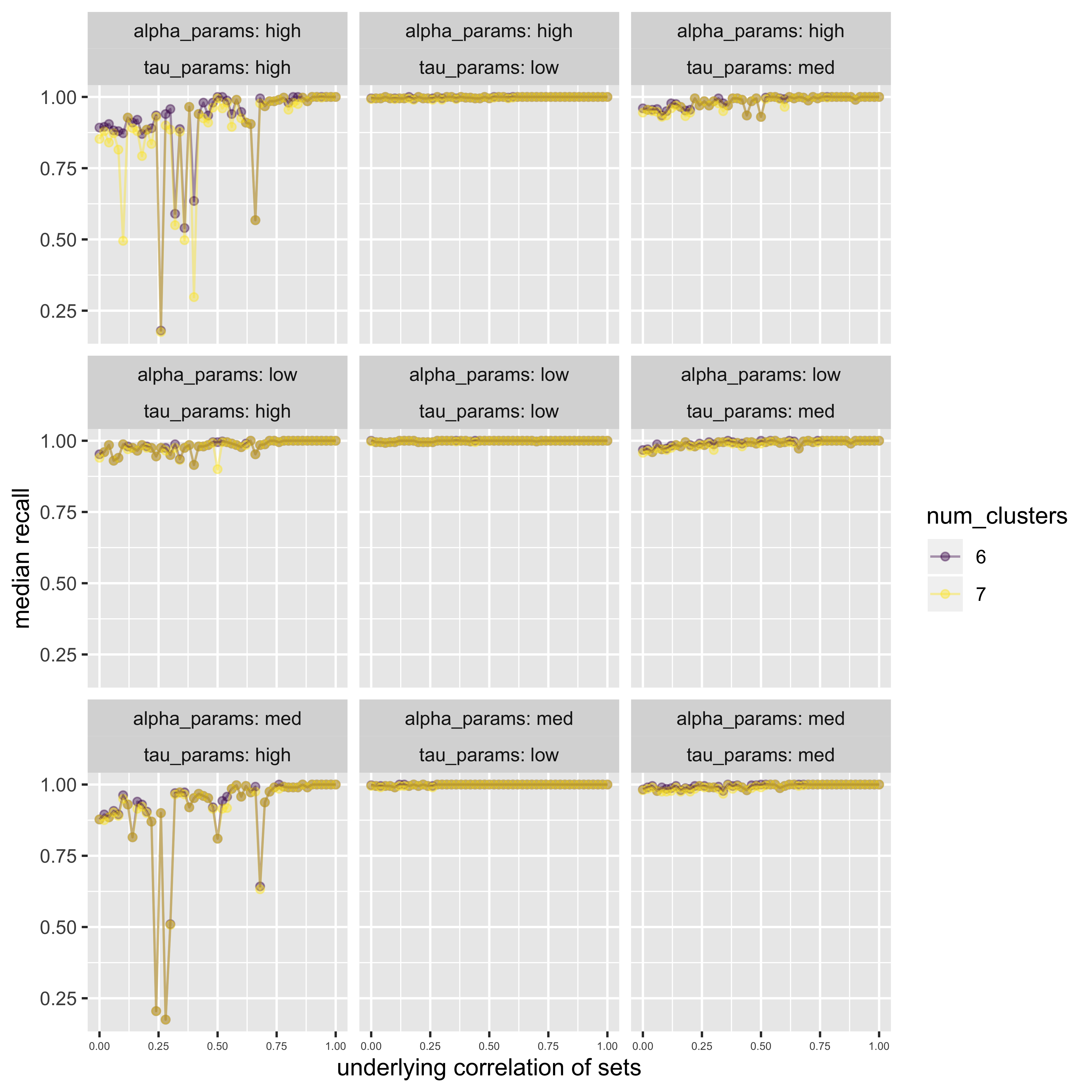}%\\
%\hspace{1em}

\caption{Analysis of hierarchical clustering based on recall of estimated sets. 
This is using Euclidean distance on the ``buyer normalized'' artificial datasets with average linkage.
We considered different numbers of clusters.}
\label{app fig: sim study binary hclust recall}
\end{figure}

\subsection{Additional Text Data Analysis}
\label{app: text add results}

\begin{table}[H]
\centering
\begin{tabular}{ || c || c || c || } 
 \hline
 Moby Dick & The Count of Monte Cristo & Pride and Prejudice \\          
 \hline
 Adventures of Huckleberry Finn & The Picture of Dorian Gray &  Great Expectations \\  
 \hline
 The Adventures of Tom Sawyer & Alice's Adventures in Wonderland & Little Women \\
 \hline
 The Call of the Wild & Through the Looking-Glass &   \\
 \hline
\end{tabular}
\caption{List of non-Shakespeare text documents used in Section~\ref{sec: text}.}
\label{tab: west lit non-shake docs}
\end{table}

\begin{table}[H]
\centering
\begin{tabular}{ || c || c || c || } 
 \hline
 The Taming of the Shrew & Romeo and Juliet &  A Midsummer Night's Dream \\          
 \hline 
 Julius Caesar & As You Like It & Hamlet, Prince of Denmark \\
 \hline
 Twelfth Night & All's Well That Ends Well & Othello \\
 \hline
 King Lear & Macbeth & Antony and Cleopatra \\
 \hline
 The Tempest & Shakespeare's Sonnets & \\
 \hline
\end{tabular}
\caption{List of Shakespeare text documents used in Section~\ref{sec: text}.}
\label{tab: west lit shake docs}
\end{table}

In this appendix we provide additional results
from the LAMB, NMF, and LDA methods
applied to the text dataset discussed in 
Section~\ref{sec: text}. 
See Tables~\ref{tab: west lit non-shake docs}
and \ref{tab: west lit shake docs} for
the different works used in this text dataset.
Note that the works in both 
Tables~\ref{tab: west lit non-shake docs}
and \ref{tab: west lit shake docs}
are all contained in a single text dataset; 
the works are separated into two tables 
to better demonstrate the taxonomy of text sources 
contained in this dataset.

The term sets discovered by the LAMB method,
which uses only binary co-occurence data,
include characters, themes, and settings.
In contrast, the NMF and LDA methods
are applied to count-valued term-document
matrices.
Table~\ref{tab: example lamb text coh sets} includes a sample
of important terms for some of the LAMB method's estimated
coherent sets.
%%Table~\ref{tab: lamb text coh sets group 2}
%%contains more information on the 
%%text sources or themes for the different estimated coherent
%%sets produced by the LAMB method.

Tables~\ref{tab: nmf text effective numbers}, 
\ref{tab: lda probs text effective numbers},
and \ref{tab: lda term scores text effective numbers} 
demonstrate that the sets of associated terms 
discovered by the LAMB method
are reasonably discriminative relative to 
the results obtained from the NMF and LDA methods. 
The top 50 terms in the soft clusters from the NMF and LDA methods
are the most discriminative terms between the different latent topics.
In particular,
the effective number of distinct term sets for the NMF and LDA methods
is largest when only the top 50 terms are used instead of the top 100, 500, or 1000 terms.
%%(See Tables~\ref{tab: nmf text effective numbers}, \ref{tab: lda probs text effective numbers},
%%and \ref{tab: lda term scores text effective numbers}.)
Note that the LAMB method produces a similar effective number
of distinct term sets as the best results (in the context of effective number) from 
the NMF and LDA methods. In contrast to the best
results obtained from the NMF and LDA methods, the term sets produced
by the LAMB method have
100 to 1,000 terms as opposed to only 50 terms.
Neither the number nor the size of the estimated coherent sets is 
a parameter of the LAMB method.\footnote{A minimum set size for
fixed points can be set in the LAMB method's code. However, 
in practice we use a low threshold for this type of set filtering.}

\begin{table}[H]
\centering
\begin{tabular}{ || c || c | c | c || } 
 \hline
 \textbf{Method} & \textbf{Parameter} & \textbf{Set Filtering} & \textbf{Effective Number} \\          
 \hline \hline
 LAMB & $\delta = .10$ & $\geq 25$ terms & 6.895 \\
 \hline
 LAMB & $\delta = .05$ & $\geq 25$ terms & 7.510 \\
 \hline
 LAMB & $\delta = .01$ & $\geq 25$ terms & 10.004 \\
 \hline
 LAMB & $\delta = .001$ & $\geq 25$ terms & 19.416 \\
 \hline
 NMF & 20 topics & top 50 terms & 10.30 \\
 \hline
 NMF & 20 topics & top 1000 terms & 6.33 \\
 \hline
 NMF & 25 topics & top 50 terms & 12.42 \\
 \hline
 NMF & 25 topics & top 1000 terms & 7.11 \\
 \hline
  NMF & 30 topics & top 50 terms & 13.84 \\
 \hline
 NMF & 30 topics & top 1000 terms & 7.91 \\
 \hline
 LDA & 20 topics & top 50 terms (probabilities) & 9.88 \\
 \hline
 LDA & 20 topics & top 1000 terms (probabilities) & 5.73 \\
 \hline
 LDA & 25 topics & top 50 terms (probabilities) & 11.12 \\
 \hline
 LDA & 25 topics & top 1000 terms (probabilities) & 6.40 \\
 \hline
 LDA & 30 topics & top 50 terms (probabilities) & 12.70 \\
 \hline
 LDA & 30 topics & top 1000 terms (probabilities) & 7.06 \\
 \hline
\end{tabular}
\caption{Some effective numbers of distinct term sets obtained
from applying the LAMB, NMF, and LDA methods
to the text dataset in Section~\ref{sec: text}. The LAMB method outputs estimated coherent sets
while the NMF and LDA methods can produce soft clusters of terms.}
\label{tab: full lamb text effective numbers}
\end{table}

\begin{table}[H]
\centering
\begin{tabular}{ || c || c | c | c | c || } 
 \hline
 \textbf{Method} & \textbf{Parameter} & \textbf{Set} & \textbf{Set Size} & \textbf{Sample of Representative Terms} \\          
 \hline \hline
 LAMB & $\delta = .05$ & 1 & 975 & elizabeth, bennet, darcy, bingley, \\
  & & & & longbourn, wickham, netherfield, \\
  & & & & hertfordshire, behaviour, feelings \\
 \hline
 LAMB & $\delta = .05$ & 2 & 1485 & love, faith, gods, judgement, power, \\
  & & & & fools, wrongs, mortal, death, \\
  & & & & fortune, antony, cleopatra, macbeth, \\
  & & & & othello, ophelia, hamlet, sebastian \\
 \hline                   
 LAMB & $\delta = .05$ & 3 & 1215 & whale, captain, ahab, ship, deck, pequod,\\
  & & & & starbuck, crew, queequeg, moby, dick, leviathan, \\
  & & & & sail, pacific, indian, ocean, water, blubber \\
 \hline
 LAMB & $\delta = .05$ & 4 & 918 & jim, huck, finn, tom, sawyer, \\
  & & & & river, runaway, woods, cussing, fooling \\
 \hline
 LAMB & $\delta = .05$ & 5 & 872 & beth, jo, amy, meg, march, \\
  & & & & motherly, sisters, family, laurie, lessons, prim, \\
  & & & & feeling, longed, fun, nice, afraid, shy, scolded,  \\
  & & & & romance, busy, cozy, pretty, piano, sang \\
 \hline
 LAMB & $\delta = .05$ & 6 & 122 &  alice, dinah, kitten, queen, rank, moving, \\
  & & & & poetry, walrus, carpenter, oysters, \\
  & & & & dreaming, woke, memories, \\
  & & & &  sicken, welfare, ills, purge, drugs, medicine \\
 \hline
 LAMB & $\delta = .05$ & 7 & 131 & artic, chilled, flame, smoke, warming, \\
  & & & & darkness, blankets, jackets, sharing,  \\ 
  & & & & household, quiet, enjoy, pleasant, alive,   \\
  & & & & awake, consciousness, sleeping, skeleton \\
 \hline
 LAMB & $\delta = .05$ & 8 & 290 & vindictive, vengeful, destructive, tearful, \\
  & & & & gratification, entitled, festive, bliss,  \\
  & & & & schoolmaster, pupils, schoolhouse,  \\
  & & & & tyranny, banishment, mayor, political \\
 \hline
 \end{tabular}
\caption{A small sample of important terms for some of the LAMB method's estimated
coherent sets when applied to the text dataset in Section~\ref{sec: text}.}
\label{tab: example lamb text coh sets}
\end{table}

\begin{figure}
\centering
\includegraphics[width = \textwidth]{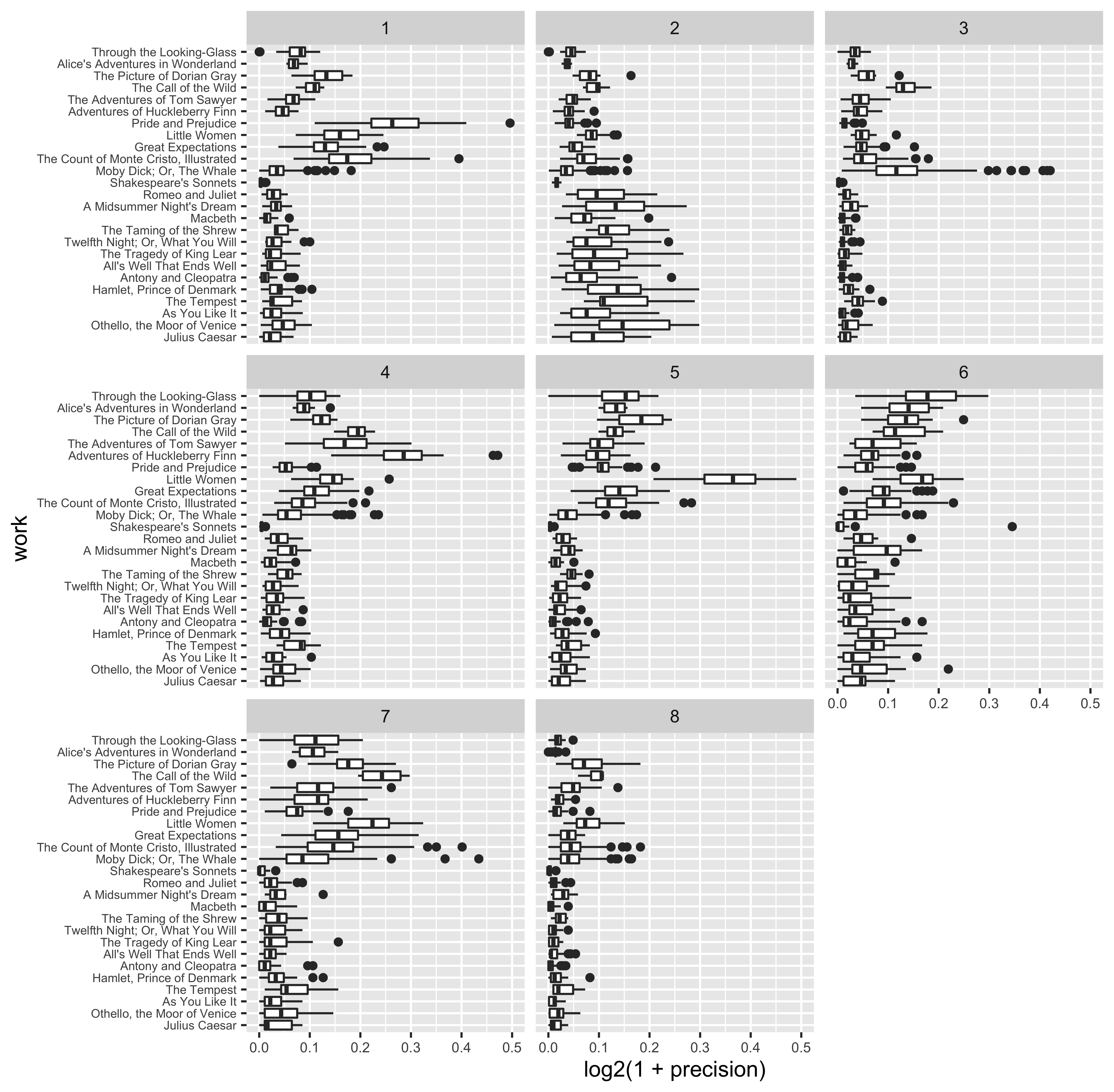}

\caption{Box plots for each estimated coherent set obtained from the LAMB method
with $\delta = .05$ on the text dataset in Section~\ref{sec: text}.
Each data point within a box plot row corresponds to a chapter or scene in a given
work. Higher values of precision correspond to the estimated
coherent sets containing more terms in the work.}
\label{fig: lamb coh set box plots alpha 05}
\end{figure}

\begin{table}[H]
\centering
\begin{tabular}{ || c || c | c | c ||} % c || } 
 \hline
 \textbf{Method} & \textbf{Parameter} & \textbf{Set} & \textbf{Set Size}  \\ %& \textbf{Related Documents or Themes} \\          
 \hline \hline
 LAMB & $\delta = .10$ & 1 & 1602 \\ %&  Shakespeare Works \\
 \hline
 LAMB & $\delta = .10$ & 2 & 1471  \\ %& Moby Dick and Huckleberry Finn \\
 \hline                   
 LAMB & $\delta = .10$ & 3 & 1383  \\ %& Count of Monte Cristo \\
 \hline
 LAMB & $\delta = .10$ & 4 & 992  \\ %&  Little Women \\
 \hline
 LAMB & $\delta = .10$ & 5 & 645  \\ %& Picture of Dorian Gray \\
 \hline
 LAMB & $\delta = .10$ & 6 & 136  \\ %& comfort and atmosphere \\
 \hline
 LAMB & $\delta = .10$ & 7 & 301  \\ %& feelings and power structure \\
 \hline
 LAMB & $\delta = .05$ & 1 & 975  \\ %& Pride \& Prejudice \\
 \hline
 LAMB & $\delta = .05$ & 2 & 1485  \\ %& Shakespeare Works \\
 \hline                   
 LAMB & $\delta = .05$ & 3 & 1215  \\ %& Moby Dick \\
 \hline
 LAMB & $\delta = .05$ & 4 & 918  \\ %& Huckleberry Finn and Tom Sawyer \\
 \hline
 LAMB & $\delta = .05$ & 5 & 872  \\ %& Little Women \\
 \hline
 LAMB & $\delta = .05$ & 6 & 122  \\ %& Alice in Wonderland and Looking Glass \\
 \hline
 LAMB & $\delta = .05$ & 7 & 131  \\ %& comfort and atmosphere \\
 \hline
 LAMB & $\delta = .05$ & 8 & 290  \\ %& feelings and power structure \\
 \hline
 LAMB & $\delta = .01$ & 1 & 1276  \\ %&  Shakespeare Works \\
 \hline
 LAMB & $\delta = .01$ & 2 & 988  \\ %& Moby Dick \\
 \hline                   
 LAMB & $\delta = .01$ & 3 & 955  \\ %& Count of Monte Cristo \\
 \hline
 LAMB & $\delta = .01$ & 4 & 781  \\ %& Huckleberry Finn \\
 \hline
 LAMB & $\delta = .01$ & 5 & 432  \\ %& Picture of Dorian Gray \\
 \hline
 LAMB & $\delta = .01$ & 6 & 174  \\ %& Great Expectations \\
 \hline
 LAMB & $\delta = .01$ & 7 & 108  \\ %&  As You Like It, Alice in Wonderland,\\ & & & & and Looking Glass \\
 \hline
 LAMB & $\delta = .01$ & 8 & 119  \\ %& comfort and atmosphere \\
 \hline
 LAMB & $\delta = .01$ & 9 & 79  \\ %& sailing and worth \\
 \hline
 LAMB & $\delta = .01$ & 10 & 207  \\ %& Shakespeare Works and Great Expectations \\
 \hline
 LAMB & $\delta = .01$ & 11 & 264  \\ %& Tom Sawyer \\
 \hline
 \end{tabular}
\caption{Basic data for the different estimated coherent
sets produced by the LAMB method when it was applied to 
the text dataset in Section~\ref{sec: text}. 
%%The related texts
%%or themes were determined by comparing the terms in the estimated
%%coherent sets and basic knowledge about the source documents.
}
\label{tab: lamb text coh sets group 1}
\end{table}

\begin{figure}
\centering
\includegraphics[width = \textwidth]{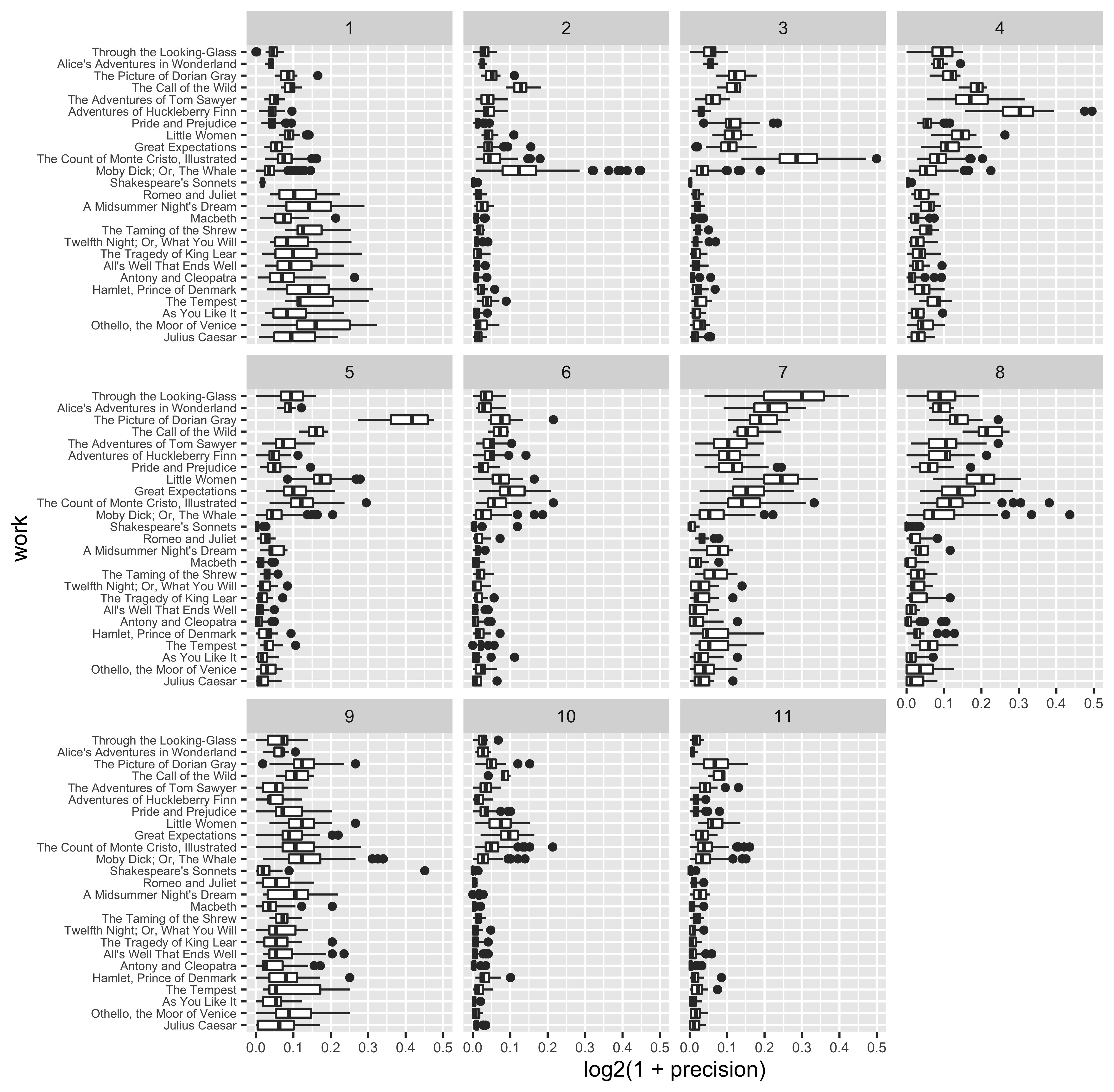}

\caption{Box plots of precision per work for each estimated coherent set 
obtained from the LAMB method
with $\delta = .01$ on the text dataset in Section~\ref{sec: text}.
Each data point within a box plot row corresponds to a chapter or scene in a given
work. Higher values of precision correspond to the estimated
coherent sets containing more terms in the work.}
\label{fig: lamb coh set box plots alpha 01}
\end{figure}

\begin{table}[H]
\centering
\begin{tabular}{ || c || c | c | c ||} % c || } 
 \hline
 \textbf{Method} & \textbf{Parameter} & \textbf{Set} & \textbf{Set Size}  \\ %& \textbf{Related Documents or Themes} \\          
 \hline \hline
LAMB & $\delta = .001$ & 1 & 744  \\ %&  Moby Dick \\
 \hline
 LAMB & $\delta = .001$ & 2 & 1031  \\ %& Shakespeare Works \\
 \hline                   
 LAMB & $\delta = .001$ & 3 & 616  \\ %& Pride and Prejudice \\
 \hline
 LAMB & $\delta = .001$ & 4 & 511  \\ %& Little Women \\
 \hline
 LAMB & $\delta = .001$ & 5 & 117  \\ %& As You Like It  and Looking Glass \\
 \hline
 LAMB & $\delta = .001$ & 6 & 646  \\ %& Huckleberry Finn \\
 \hline
 LAMB & $\delta = .001$ & 7 & 298  \\ %& The Picture of Dorian Gray \\
 \hline
 LAMB & $\delta = .001$ & 8 & 140  \\ %& Great Expectations \\
 \hline
 LAMB & $\delta = .001$ & 9 & 92  \\ %&  Alice in Wonderland \\
 \hline
 LAMB & $\delta = .001$ & 10 & 102  \\ %& comfort and atmosphere \\
 \hline                   
 LAMB & $\delta = .001$ & 11 & 107  \\ %& Great Expectations \\
 \hline
 LAMB & $\delta = .001$ & 12 & 103  \\ %& physical or mental state and location \\
 \hline
 LAMB & $\delta = .001$ & 13 & 145  \\ %& assistance, resistance, or judgement \\
 \hline
 LAMB & $\delta = .001$ & 14 & 113  \\ %& Great Expectations \\
 \hline
 LAMB & $\delta = .001$ & 15 & 196  \\ %& Great Expectations \\
 \hline
 LAMB & $\delta = .001$ & 16 & 172  \\ %& Hamlet and Great Expectations \\
 \hline
 LAMB & $\delta = .001$ & 17 & 158  \\ %& Great Expectations \\
 \hline
 LAMB & $\delta = .001$ & 18 & 222  \\ %& Tom Sawyer \\
 \hline
 LAMB & $\delta = .001$ & 19 & 76  \\ %& sailing and worth \\
 \hline
 LAMB & $\delta = .001$ & 20 & 78  \\ %& pleasure \\
 \hline
\end{tabular}
\caption{Basic data for the different estimated coherent
sets produced by the LAMB method applied to 
the text dataset in Section~\ref{sec: text}. 
%%The related texts
%%or themes were determined by comparing the terms in the estimated
%%coherent sets and basic knowledge about the source documents.
}
\label{tab: lamb text coh sets group 2}
\end{table}

\begin{table}[H]
\centering
\begin{tabular}{ || c || c | c | c || } 
 \hline
 \textbf{Method} & \textbf{Parameter} & \textbf{Set Filtering} & \textbf{Effective Number} \\          
 \hline \hline
 NMF & 5 topics & top 50 terms & 3.44 \\
 \hline
 NMF & 5 topics & top 100 terms & 3.25 \\
 \hline
 NMF & 5 topics & top 500 terms & 2.85 \\
 \hline
 NMF & 5 topics & top 1000 terms & 2.71 \\
 \hline
 NMF & 10 topics & top 50 terms & 5.72 \\
 \hline
 NMF & 10 topics & top 100 terms & 5.43 \\
 \hline
 NMF & 10 topics & top 500 terms & 4.50 \\
 \hline
 NMF & 10 topics & top 1000 terms & 4.12 \\
 \hline
 NMF & 15 topics & top 50 terms & 8.12 \\
 \hline
 NMF & 15 topics & top 100 terms & 7.32 \\
 \hline
 NMF & 15 topics & top 500 terms & 5.92 \\
 \hline
 NMF & 15 topics & top 1000 terms & 5.38 \\
 \hline
 NMF & 20 topics & top 50 terms & 10.30 \\
 \hline
 NMF & 20 topics & top 100 terms & 9.27 \\
 \hline
 NMF & 20 topics & top 500 terms & 7.17 \\
 \hline
 NMF & 20 topics & top 1000 terms & 6.33 \\
 \hline
 NMF & 25 topics & top 50 terms & 12.42 \\
 \hline
 NMF & 25 topics & top 100 terms & 10.67 \\
 \hline
 NMF & 25 topics & top 500 terms & 8.17 \\
 \hline
 NMF & 25 topics & top 1000 terms & 7.11 \\
 \hline
 NMF & 30 topics & top 50 terms & 13.84 \\
 \hline
 NMF & 30 topics & top 100 terms & 12.04 \\
 \hline
 NMF & 30 topics & top 500 terms & 9.16 \\
 \hline
 NMF & 30 topics & top 1000 terms & 7.91 \\
 \hline
\end{tabular}
\caption{Effective number of distinct term sets from the NMF method's soft clustering of terms. Multiple
sizes of soft clusters were considered to demonstrate the overlap between different soft clusters.}
\label{tab: nmf text effective numbers}
\end{table}

\begin{table}[H]
\centering
\begin{tabular}{ || c || c | c | c | c || } 
 \hline
 \textbf{Method} & \textbf{Parameter} & \textbf{Set Filtering} & \textbf{Ranking} & \textbf{Effective Number} \\          
 \hline \hline
 LDA & 5 topics & top 50 terms & probabilities & 3.40 \\
 \hline
 LDA & 5 topics & top 100 terms & probabilities & 3.37 \\
 \hline
 LDA & 5 topics & top 500 terms & probabilities & 2.96 \\
 \hline
 LDA & 5 topics & top 1000 terms & probabilities & 2.78 \\
 \hline
 LDA & 10 topics & top 50 terms & probabilities & 5.94 \\
 \hline
 LDA & 10 topics & top 100 terms & probabilities & 5.44 \\
 \hline
 LDA & 10 topics & top 500 terms & probabilities & 4.33 \\
 \hline
 LDA & 10 topics & top 1000 terms & probabilities & 4.00 \\
 \hline
 LDA & 15 topics & top 50 terms & probabilities & 8.30 \\
 \hline
 LDA & 15 topics & top 100 terms & probabilities & 7.30 \\
 \hline
 LDA & 15 topics & top 500 terms & probabilities & 5.60 \\
 \hline
 LDA & 15 topics & top 1000 terms & probabilities & 5.10 \\
 \hline
 LDA & 20 topics & top 50 terms & probabilities & 9.88 \\
 \hline
 LDA & 20 topics & top 100 terms & probabilities & 8.72 \\
 \hline
 LDA & 20 topics & top 500 terms & probabilities & 6.39 \\
 \hline
 LDA & 20 topics & top 1000 terms & probabilities & 5.73 \\
 \hline
 LDA & 25 topics & top 50 terms & probabilities & 11.12 \\
 \hline
 LDA & 25 topics & top 100 terms & probabilities & 9.85 \\
 \hline
 LDA & 25 topics & top 500 terms  & probabilities & 7.22 \\
 \hline
 LDA & 25 topics & top 1000 terms & probabilities & 6.40 \\
 \hline
 LDA & 30 topics & top 50 terms & probabilities & 12.70 \\
 \hline
 LDA & 30 topics & top 100 terms & probabilities & 11.18 \\
 \hline
 LDA & 30 topics & top 500 terms & probabilities & 8.17 \\
 \hline
 LDA & 30 topics & top 1000 terms & probabilities & 7.06 \\
 \hline
\end{tabular}
\caption{Effective number of distinct term sets from the LDA method's soft clustering of terms. Here the top terms
were determined by the estimated probabilities. Multiple
sizes of soft clusters were considered to demonstrate the overlap between different soft clusters.}
\label{tab: lda probs text effective numbers}
\end{table}

\begin{table}[H]
\centering
\begin{tabular}{ || c || c | c | c | c || } 
 \hline
 \textbf{Method} & \textbf{Parameter} & \textbf{Set Filtering} & \textbf{Ranking} & \textbf{Effective Number} \\          
 \hline \hline
 LDA & 5 topics & top 50 terms & term scores & 4.24 \\
 \hline
 LDA & 5 topics & top 100 terms & term scores & 4.08 \\
 \hline
 LDA & 5 topics & top 500 terms & term scores & 3.33 \\
 \hline
 LDA & 5 topics & top 1000 terms & term scores & 2.88 \\
 \hline
 LDA & 10 topics & top 50 terms & term scores & 6.20 \\
 \hline
 LDA & 10 topics & top 100 terms & term scores & 5.69 \\
 \hline
 LDA & 10 topics & top 500 terms  & term scores & 3.37 \\
 \hline
 LDA & 10 topics & top 1000 terms & term scores & 2.07 \\
 \hline
 LDA & 15 topics & top 50 terms & term scores & 7.42 \\
 \hline
 LDA & 15 topics & top 100 terms & term scores & 6.36 \\
 \hline
 LDA & 15 topics & top 500 terms & term scores & 2.64 \\
 \hline
 LDA & 15 topics & top 1000 terms & term scores & 1.40 \\
 \hline
 LDA & 20 topics & top 50 terms & term scores & 7.42 \\
 \hline
 LDA & 20 topics & top 100 terms & term scores & 5.59 \\
 \hline
 LDA & 20 topics & top 500 terms & term scores & 1.77 \\
 \hline
 LDA & 20 topics & top 1000 terms & term scores & 1.29 \\
 \hline
 LDA & 25 topics & top 50 terms & term scores & 6.94 \\
 \hline
 LDA & 25 topics & top 100 terms & term scores & 4.91 \\
 \hline
 LDA & 25 topics & top 500 terms & term scores & 1.30 \\
 \hline
 LDA & 25 topics & top 1000 terms & term scores & 1.27 \\
 \hline
 LDA & 30 topics & top 50 terms & term scores & 5.50 \\
 \hline
 LDA & 30 topics & top 100 terms & term scores & 3.57 \\
 \hline
 LDA & 30 topics & top 500 terms & term scores & 1.26 \\
 \hline
 LDA & 30 topics & top 1000 terms & term scores & 1.26 \\
 \hline
\end{tabular}
\caption{Effective number of distinct term sets from the LDA method's soft clustering of terms. Here the top terms
were determined by the term scores of \citet{bleiTopicModels-2009}. Multiple
sizes of soft clusters were considered to demonstrate the overlap between different soft clusters.}
\label{tab: lda term scores text effective numbers}
\end{table}

\newpage

\subsection{Additional last.fm Data Analysis}
\label{app: lastfm add results}

In this appendix we provide additional results
from the LAMB, NMF, and LDA methods
applied to the Last.fm dataset discussed in 
Section~\ref{sec: lastfm}. 
The artist sets discovered by the LAMB method,
which uses only binary co-occurence data,
include various genres of bands.
The NMF and LDA methods
were applied to count-valued artist-user matrices.
Tables~\ref{tab: nmf lastfm effective numbers}, 
\ref{tab: lda probs lastfm effective numbers},
and \ref{tab: lda term scores lastfm effective numbers} 
demonstrate that the sets of associated artists and bands 
discovered by the LAMB method
are reasonably discriminative relative to 
the results obtained from the NMF and LDA methods.

\begin{table}[H]
\centering
\begin{tabular}{ || c || c | c | c || } 
 \hline
 \textbf{Method} & \textbf{Parameter} & \textbf{Set Filtering} & \textbf{Effective Number} \\          
 \hline \hline
 LAMB & $\delta = .10$ & $\geq 25$ artists & 4.000 \\
 \hline
 LAMB & $\delta = .05$ & $\geq 25$ artistis & 4.710 \\
 \hline
 LAMB & $\delta = .01$ & $\geq 25$ artists  & 5.166 \\
 \hline
 LAMB & $\delta = .001$ & $\geq 25$ artists & 6.760 \\
 \hline
 NMF & 6 topics & top 50 artists & 4.820 \\
 \hline
 NMF & 6 topics & top 1000 artists & 3.391 \\
 \hline
 NMF & 9 topics & top 50 artists & 6.800 \\
 \hline
 NMF & 9 topics & top 1000 artists  & 4.009 \\
 \hline
  NMF & 12 topics & top 50 artists & 8.840 \\
 \hline
 NMF & 12 topics & top 1000 artists & 4.784 \\
 \hline
 LDA & 6 topics & top 50 artists (probabilities) & 5.020 \\
 \hline
 LDA & 6 topics & top 1000 artists (probabilities) & 4.058 \\
 \hline
 LDA & 9 topics & top 50 artists (probabilities) & 7.160 \\
 \hline
 LDA & 9 topics & top 1000 artists (probabilities) & 5.246 \\
 \hline
 LDA & 12 topics & top 50 artists (probabilities) & 8.820 \\
 \hline
 LDA & 12 topics & top 1000 artists (probabilities) & 6.157 \\
 \hline
\end{tabular}
\caption{Effective number of distinct artist sets
from applying the LAMB, NMF, and LDA methods
to the Last.fm dataset in Section~\ref{sec: lastfm}. The LAMB method outputs estimated coherent sets
while the NMF and LDA methods can produce soft clusters of artists and bands.}
\label{tab: lamb lastfm effective numbers extra}
\end{table}

\begin{table}[H]
\centering
\begin{tabular}{ || c || c | c | c | c || } 
 \hline
 \textbf{Method} & \textbf{Parameter} & \textbf{Set} & \textbf{Set Size} \\          
 \hline \hline
 LAMB & $\delta = .10$ & 1 & 2070  \\
 \hline
 LAMB & $\delta = .10$ & 2 & 1844  \\
 \hline                   
 LAMB & $\delta = .10$ & 3 & 2357  \\
 \hline
 LAMB & $\delta = .10$ & 4 & 2042   \\
 \hline
 LAMB & $\delta = .10$ & 5 & 37 \\
 \hline
 LAMB & $\delta = .05$ & 1 & 1939 \\
 \hline
 LAMB & $\delta = .05$ & 2 &1665   \\
 \hline                   
 LAMB & $\delta = .05$ & 3 & 1740 \\
 \hline
 LAMB & $\delta = .05$ & 4 & 37   \\
 \hline
 LAMB & $\delta = .05$ & 5 & 1490 \\
 \hline
 LAMB & $\delta = .01$ & 1 & 1122   \\
 \hline
 LAMB & $\delta = .01$ & 2 & 1606  \\
 \hline                   
 LAMB & $\delta = .01$ & 3 & 1371  \\
 \hline
 LAMB & $\delta = .01$ & 4 & 1861 \\
 \hline
 LAMB & $\delta = .01$ & 5 & 37  \\
 \hline
 LAMB & $\delta = .01$ & 6 & 82  \\
 \hline
 LAMB & $\delta = .001$ & 1 & 926   \\
 \hline
 LAMB & $\delta = .001$ & 2 & 1611  \\
 \hline                   
 LAMB & $\delta = .001$ & 3 & 1321  \\
 \hline
 LAMB & $\delta = .001$ & 4 & 78  \\
 \hline
 LAMB & $\delta = .001$ & 5 & 54  \\
 \hline
 LAMB & $\delta = .001$ & 6 & 37  \\
 \hline
 LAMB & $\delta = .001$ & 7 & 282  \\
 \hline
 \end{tabular}
\caption{Basic data for the different estimated coherent
sets produced by the LAMB method when it was applied to 
the Last.fm dataset in Section~\ref{sec: lastfm}.}
\label{tab: lamb lastfm coh sets}
\end{table}

\begin{table}[H]
\centering
\begin{tabular}{ || c || c | c | c | c || } 
 \hline
 \textbf{Method} & \textbf{Parameter} & \textbf{Set} & \textbf{Set Size} & \textbf{Sample of Artists and Bands} \\          
 \hline \hline
 LAMB & $\delta = .01$ & 1 & 1122 & Pearl Jam, Linkin Park, Sum 41, \\
  & & & & Breaking Benjamin, Alice Cooper, Creed \\
 \hline
 LAMB & $\delta = .01$ & 2 & 1606 & The Shins, Wilco, The Wombats, \\
  & & & & Phoenix, Beck, Modest Mouse, \\
  & & & & Passion Pit, Neil Young, The Beach Boys \\
 \hline                   
 LAMB & $\delta = .01$ & 3 & 1371 & Kanye West, Lupe Fiasco, Snoop Dogg,\\
  & & & & Rihanna,  Mary J. Blige, Madonna,\\
  & & & & Backstreet Boys, Justin Timberlake, Coldplay, \\
  & & & & My Chemical Romance, Green Day, \\
  & & & & The All-American Rejects \\
 \hline
 LAMB & $\delta = .01$ & 4 & 1861 & Tycho, RJD2, Wax Tailor, \\
  & & & & Basshunter, Fatboy Slim, The Presets \\
 \hline
 LAMB & $\delta = .01$ & 5 & 37 & Maxine Nightingale, Anita Ward, Gloria Gaynor, \\
  & & & & A Taste of Honey, George McCrae, Sia \\
 \hline
 LAMB & $\delta = .01$ & 6 & 82 & Joy Division, Tom Petty and the Heartbreakers, \\
  & & & & Judas Priest, Beastie Boys\\
 \hline
 \end{tabular}
\caption{A small sample of artists and bands for some of the LAMB method's estimated
coherent sets when it was applied to the Last.fm data.}
\label{tab: example lamb lastfm artist coh sets}
\end{table}

\begin{table}[H]
\centering
\begin{tabular}{ || c || c | c | c || } 
 \hline
 \textbf{Method} & \textbf{Parameter} & \textbf{Set Filtering} & \textbf{Effective Number} \\          
 \hline \hline
 NMF & 3 topics & top 50 artists & 2.780 \\
 \hline
 NMF & 3 topics & top 100 artists & 2.710 \\
 \hline
 NMF & 3 topics & top 500 artists & 2.430 \\
 \hline
 NMF & 3 topics & top 1000 artists & 2.317 \\
 \hline
 NMF & 6 topics & top 50 artists & 4.820 \\
 \hline
 NMF & 6 topics & top 100 artists &  4.380 \\
 \hline
 NMF & 6 topics & top 500 artists & 3.650 \\
 \hline
 NMF & 6 topics & top 1000 artists & 3.391 \\
 \hline
 NMF & 9 topics & top 50 artists & 6.800 \\
 \hline
 NMF & 9 topics & top 100 artists & 6.160 \\
 \hline
 NMF & 9 topics & top 500 artists & 4.632 \\
 \hline
 NMF & 9 topics & top 1000 artists & 4.009 \\
 \hline
 NMF & 12 topics & top 50 artists & 8.840 \\
 \hline
 NMF & 12 topics & top 100 artists & 7.830 \\
 \hline
 NMF & 12 topics & top 500 artists & 5.740 \\
 \hline
 NMF & 12 topics & top 1000 artists & 4.784 \\
 \hline
\end{tabular}
\caption{Effective number of distinct artist sets from the NMF method's soft clustering of artists and bands. Multiple
sizes of soft clusters were considered to demonstrate the overlap between different soft clusters.}
\label{tab: nmf lastfm effective numbers}
\end{table}

\begin{table}[H]
\centering
\begin{tabular}{ || c || c | c | c | c || } 
 \hline
 \textbf{Method} & \textbf{Parameter} & \textbf{Set Filtering} & \textbf{Ranking} & \textbf{Effective Number} \\          
 \hline \hline
 LDA & 3 topics & top 50 artists & probabilities & 2.720 \\
 \hline
 LDA & 3 topics & top 100 artists & probabilities &  2.560 \\
 \hline
 LDA & 3 topics & top 500 artists & probabilities & 2.416 \\
 \hline
 LDA & 3 topics & top 1000 artists & probabilities & 2.369 \\
 \hline
 LDA & 6 topics & top 50 artists & probabilities & 5.020 \\
 \hline
 LDA & 6 topics & top 100 artists & probabilities & 4.810 \\
 \hline
 LDA & 6 topics & top 500 artists & probabilities & 4.232 \\
 \hline
 LDA & 6 topics & top 1000 artists & probabilities & 4.058 \\
 \hline
 LDA & 9 topics & top 50 artists & probabilities & 7.160 \\
 \hline
 LDA & 9 topics & top 100 artists & probabilities & 6.740 \\
 \hline
 LDA & 9 topics & top 500 artists & probabilities & 5.670 \\
 \hline
 LDA & 9 topics & top 1000 artists & probabilities & 5.246 \\
 \hline
 LDA & 12 topics & top 50 artists & probabilities & 8.820 \\
 \hline
 LDA & 12 topics & top 100 artists & probabilities & 8.360 \\
 \hline
 LDA & 12 topics & top 500 artists & probabilities & 6.932 \\
 \hline
 LDA & 12 topics & top 1000 artists & probabilities & 6.157 \\
 \hline
\end{tabular}
\caption{Effective number of distinct artist sets from the LDA method's soft clustering of artists and bands. 
Here the top terms were determined by the estimated probabilities. Multiple
sizes of soft clusters were considered to demonstrate the overlap between different soft clusters.}
\label{tab: lda probs lastfm effective numbers}
\end{table}

\begin{table}[H]
\centering
\begin{tabular}{ || c || c | c | c | c || } 
 \hline
 \textbf{Method} & \textbf{Parameter} & \textbf{Set Filtering} & \textbf{Ranking} & \textbf{Effective Number} \\          
 \hline \hline
 LDA & 3 topics & top 50 artists & term scores & 2.840 \\
 \hline
 LDA & 3 topics & top 100 artists & term scores & 2.760 \\
 \hline
 LDA & 3 topics & top 500 artists & term scores & 2.624 \\
 \hline
 LDA & 3 topics & top 1000 artists & term scores & 2.485 \\
 \hline
 LDA & 6 topics & top 50 artists & term scores & 5.520 \\
 \hline
 LDA & 6 topics & top 100 artists & term scores & 5.060 \\
 \hline
 LDA & 6 topics & top 500 artists & term scores & 3.300 \\
 \hline
 LDA & 6 topics & top 1000 artists & term scores & 2.271 \\
 \hline
 LDA & 9 topics & top 50 artists & term scores & 7.020 \\
 \hline
 LDA & 9 topics & top 100 artists & term scores & 5.590 \\
 \hline
 LDA & 9 topics & top 500 artists & term scores & 2.390 \\
 \hline
 LDA & 9 topics & top 1000 artists & term scores & 1.438 \\
 \hline
 LDA & 12 topics & top 50 artists & term scores & 7.760 \\
 \hline
 LDA & 12 topics & top 100 artists & term scores & 5.820 \\
 \hline
 LDA & 12 topics & top 500 artists & term scores & 1.792 \\
 \hline
 LDA & 12 topics & top 1000 artists & term scores & 1.419 \\
 \hline
\end{tabular}
\caption{Effective number of distinct artist sets from the LDA method's soft clustering of artists and bands. 
Here the top artists or bands were determined by the term-scores of \citet{bleiTopicModels-2009}. Multiple
sizes of soft clusters were considered to demonstrate the overlap between different soft clusters.}
\label{tab: lda term scores lastfm effective numbers}
\end{table}

\end{document}